\documentclass[cernpreprint, orcidlogo, atlasdraft=false, texlive=2023, american, texmf, orcidlogo]{atlasdoc}
 
\usepackage{atlaspackage}
\usepackage{atlasbiblatex}
 
\usepackage{atlasphysics}
\usepackage{standalone}
\usepackage{multirow}
\usepackage{makecell}
\usepackage{float}
\usepackage{feynmf}
\usepackage{pgfplots}
\usepackage{subfig}
 
\graphicspath{{logos/}{figures/}}
 
\usepackage{ANA-HIGG-2021-20-PAPER-defs}

\AtlasTitle{Measurements of Higgs boson production by gluon-gluon fusion and vector-boson fusion using $H\rightarrow W W^* \rightarrow e\nu \mu\nu$ decays in $pp$ collisions at $\sqrt{s}=13$~TeV with the ATLAS detector}
 
\AtlasAbstract{
Higgs boson production via gluon-gluon fusion and vector-boson fusion in proton-proton collisions is measured in the $H\rightarrow W W^* \rightarrow e\nu \mu\nu$ decay channel.
The Large Hadron Collider delivered proton-proton collisions at a center-of-mass energy of 13~TeV between 2015 and 2018, which were recorded by the ATLAS detector, corresponding to an integrated luminosity of 139~fb$^{-1}$.
The total cross sections for Higgs boson production by gluon-gluon fusion and vector-boson fusion times the $H\rightarrow W W^*$ branching ratio are measured to be
$12.0\pm1.4$
and
$0.75\;^{+0.19}_{-0.16}~\mathrm{pb}$, respectively,
in agreement with the Standard Model predictions of $10.4\pm 0.6$ and $0.81\pm 0.02$~pb.
Higgs boson production is further characterized through measurements of Simplified Template Cross Sections in a total of 11 kinematic fiducial regions.
}
 
\author{The ATLAS Collaboration}
 
\HepDataRecord{128972}
 
\AtlasJournal{Phys. Rev. D}
\AtlasJournalRef{Phys. Rev. D 108 (2023) 032005}
\AtlasDOI{10.1103/PhysRevD.108.032005}
 
\AtlasRefCode{HIGG-2021-20} 
 
\PreprintIdNumber{CERN-EP-2022-078}

\hypersetup{pdftitle={ATLAS document},pdfauthor={The ATLAS Collaboration}}
 
\begin{document}
 
\maketitle


\section{Introduction}
\label{sec:Intro}


The Higgs boson is a neutral scalar particle associated with a field whose nonzero vacuum expectation value results in the breaking of electroweak (EW) symmetry in the Standard Model (SM) and gives mass to the $W$ and $Z$ bosons~\cite{Englert:1964et,Higgs:1964ia,Higgs:1964pj,Guralnik:1964eu}.
Observation of a new particle consistent with being the Higgs boson was reported by the ATLAS and CMS collaborations in 2012~\cite{HIGG-2012-27,CMS-HIG-12-028}.
The Higgs boson has a rich set of properties that can be verified experimentally. Measurements of these properties are a powerful test of the SM and can be used to constrain theories of physics beyond the SM (BSM). BSM physics can also alter the kinematics of the Higgs boson production and decay. The large data sample delivered by the Large Hadron Collider (LHC)~\cite{Evans:2008zzb} at CERN makes it possible to measure the Higgs boson cross section in different kinematic regions in order to probe for these effects.
 
This paper describes measurements of Higgs boson production by gluon-gluon fusion (ggF) and vector-boson fusion (VBF) using $\hwwenmn$ decays in proton-proton ($pp$) collisions at a center-of-mass energy of 13~\TeV. The data were recorded by the ATLAS detector~\cite{PERF-2007-01} during Run~2 (2015--2018) of the LHC and correspond to an integrated luminosity of \RunTwoLumi. The chosen decay channel takes advantage of the large branching ratio for \hww decay and the relatively low background from other SM processes due to having two charged leptons of different flavors in the final state. The measured cross section in the ggF production mode probes the Higgs boson couplings
to heavy quarks, while the VBF production mode directly probes the couplings to $W$ and~$Z$ bosons.
Previous studies of the $\hwwenmn$ decay channel have been reported by the CMS Collaboration using its $137$~\ifb\ full Run~2 dataset~\cite{CMS:2020dvg}
and by the ATLAS Collaboration using a partial Run~2 dataset corresponding to an integrated luminosity of approximately 36~\fb~\cite{HIGG-2016-07}.
Compared to the previous ATLAS Run~2 analysis, several improvements have been made in addition to using the larger dataset---most notably, a measurement of the ggF production mode in the final state with two or more reconstructed jets and measurements of cross sections in kinematic fiducial regions defined in the Simplified Template Cross Section (STXS) framework~\cite{deFlorian:2016spz,Badger:2016bpw}.
 
The outline of this paper is as follows. Section~\ref{sec:AnalysisOverview} provides an overview of the signal characteristics and the analysis strategy. Section~\ref{sec:DataAndSimSamples} describes the data and the simulated samples. Section~\ref{sec:EventReconstruction} describes the event reconstruction. Section~\ref{sec:EventSelection} details the various selections used to define the signal and control regions in the analysis. Section~\ref{sec:Background} discusses how the backgrounds are estimated. Section~\ref{sec:Systematics} provides commentary on the systematic uncertainties. Section~\ref{sec:FitProcedure} defines the likelihood fit procedure. Finally, the results of the analysis are presented in Sec.~\ref{sec:Results} and summarized in Sec.~\ref{sec:Conclusions}.


\section{Analysis overview}
\label{sec:AnalysisOverview}


The $\hwwenmn$ decay is characterized by two charged leptons and two undetected neutrinos in the final state.
The opening angle between the two charged leptons tends to be small due to the spin-0~nature of the Higgs boson and the
chiral structure of the weak force
in the decay of the two $W$~bosons~\cite{Dittmar:1996ss}.
This feature of the decay is exploited to separate the Higgs boson signal from the main backgrounds such as continuum production of $WW$, where the charged leptons are more likely to have a large opening angle.
 
In addition to the decay products of the Higgs boson, the final state can be populated by jets either from the quarks participating in the VBF production mode or from initial-state radiation from quarks or gluons (in both the ggF and VBF production modes).
The composition of the background processes changes significantly depending on the number of jets (\Njets) in the final state.
Therefore, the analysis is performed separately in the \ZeroJet, \OneJet, and \TwoJet channels.
The analysis is divided into four categories: one each for the \ZeroJet and \OneJet channels, which solely target the ggF signal production mode,
and two for the \TwoJet channel, which separately target the VBF and ggF production modes.
 
For each analysis category, a set of selections are applied in order to enhance the signal contribution in a sample of events referred to as a signal region (SR), and a final fit to these events is performed. For the categories targeting the ggF production mode, the fit variable discriminating between signal and SM background processes is the dilepton transverse mass, defined as $\mt = \sqrt{ {\left(E_\mathrm{{T}}^{\ell\ell} + E_\mathrm{{T}}^{\mathrm{miss}}\right)}^2 - {\left| \vpTll + \vmet \right| }^2 }$
with $E_\mathrm{{T}}^{\ell\ell} = \sqrt{|\vpTll |^2 + m_{\ell\ell}^2 }$, where $m_{\ell\ell}$ is the dilepton invariant mass, $\vpTll$ is the vector sum of the lepton transverse momenta, and $\vmet$ (with magnitude \met) is the missing transverse momentum.
For the \TwoJet category targeting the VBF production mode, the output of a deep neural network (DNN) trained to identify the VBF topology is used as the discriminating variable in the fit.
The analysis likelihood function combines all SRs and determines the best-fit values for a set of parameters of interest (POIs).
Cross sections times branching ratios
are measured for the ggF and VBF production modes and their combination in the inclusive jet multiplicity.

Cross-section measurements are also conducted in the Stage-1.2 STXS category scheme, which, relative to the 1.1 scheme~\cite{Berger:2019wnu}, refines the granularity of bins for ggF events with a Higgs boson produced with large transverse momentum.
Selected events are categorized according to the requirements placed on the transverse momentum of the reconstructed Higgs boson candidate
($\pTH$) and on potential additional hadronic jets.
The ggF STXS template process (referred to subsequently as $ggH$) is defined to be the Born-level $gg \to H$ process plus higher-order QCD and EW corrections. This includes real EW radiation, in particular the $gg \to Z(\to q\bar{q})H$ process.
The VBF STXS template process (referred to subsequently as EW $qqH$) is defined to include the $V(\to q\bar{q})H$ topology in addition to the usual VBF topology.
After merging certain STXS bins to ensure sensitivity for all the measured POIs, a total of 11 fiducial cross sections corresponding to different STXS-bound kinematic regions are measured: 6 for $ggH$ production and 5 for EW $qqH$ production.


\section{Data and simulation samples}
\label{sec:DataAndSimSamples}

 
\subsection{Detector and data samples}
\label{sec:DetectorAndData}
 
\newcommand{\AtlasCoordFootnote}{
ATLAS uses a right-handed coordinate system with its origin at the nominal interaction point (IP)
in the center of the detector and the \(z\) axis along the beam pipe.
The \(x\) axis points from the IP to the center of the LHC ring,
and the \(y\) axis points upward.
Cylindrical coordinates \((r,\phi)\) are used in the transverse plane,
\(\phi\) being the azimuthal angle around the \(z\) axis.
The pseudorapidity is defined in terms of the polar angle \(\theta\) as \(\eta = -\ln \tan(\theta/2)\).
Angular distance is measured in units of \(\Delta R \equiv \sqrt{(\Delta\eta)^{2} + (\Delta\phi)^{2}}\).}
 
The ATLAS detector at the LHC covers nearly the entire solid angle around the collision point.\footnote{\AtlasCoordFootnote}
It consists of an inner tracking detector surrounded by a thin superconducting solenoid, electromagnetic and hadronic calorimeters,
and a muon spectrometer incorporating three large superconducting toroidal magnets.
The inner-detector system (ID) is immersed in a \SI{2}{\tesla} axial magnetic field
and provides charged-particle tracking in the range \(|\eta| < 2.5\).
 
The high-granularity silicon pixel detector covers the vertex region and typically provides four measurements per track,
the first hit normally being in the insertable B-layer installed before Run~2~\cite{ATLAS-TDR-2010-19,PIX-2018-001}.
It is followed by the silicon microstrip tracker, which usually provides eight measurements per track.
These silicon detectors are complemented by the transition radiation tracker (TRT),
which enables radially extended track reconstruction up to \(|\eta| = 2.0\).
The TRT also provides electron identification information
based on the fraction of hits (typically 30 in total) above a higher energy-deposit threshold corresponding to transition radiation.
 
The calorimeter system covers the pseudorapidity range \(|\eta| < 4.9\).
Within the region \(|\eta|< 3.2\), electromagnetic calorimetry is provided by barrel and
end cap high-granularity lead/liquid-argon (LAr) calorimeters,
with an additional thin LAr presampler covering \(|\eta| < 1.8\)
to correct for energy loss in material upstream of the calorimeters.
Hadronic calorimetry is provided by the steel/scintillator-tile calorimeter,
segmented into three barrel structures within \(|\eta| < 1.7\), and two copper/LAr hadronic end cap calorimeters.
The solid angle coverage is completed with forward copper/LAr and tungsten/LAr calorimeter modules
optimized for electromagnetic and hadronic measurements respectively.
 
The muon spectrometer comprises separate trigger and
high-precision tracking chambers measuring the deflection of muons in a magnetic field generated by the superconducting air-core toroids.
The field integral of the toroids ranges between \num{2.0} and \SI{6.0}{\tesla\metre} (Tesla*meter)
across most of the detector.
A set of precision chambers covers the region \(|\eta| < 2.7\) with three layers of monitored drift tubes,
complemented by cathode-strip chambers in the forward region, where the background is highest.
The muon trigger system covers the range \(|\eta| < 2.4\) with resistive-plate chambers in the barrel and thin-gap chambers in the end caps.
 
Interesting events are selected to be recorded by the first-level trigger system implemented in custom hardware,
followed by selections made by algorithms implemented in software in the high-level trigger~\cite{TRIG-2016-01}.
The first-level trigger accepts events from the \SI{40}{\MHz} bunch crossings at a rate below \SI{100}{\kHz},
which the high-level trigger reduces in order to record events to disk at about \SI{1}{\kHz}.
A combination of unprescaled single-lepton triggers and one $e$--$\mu$ dilepton trigger~\cite{TRIG-2018-05,TRIG-2018-01} is employed in this analysis so as to maximize the total trigger efficiency.
The transverse momentum ($\pt$) threshold for single-electron(muon) triggers was 24(20)~\GeV\ for the first year of data taking and increased to 26~\GeV\ for both lepton flavors during the remainder of Run~2.
The $e$--$\mu$ trigger had a $\pt$ threshold of 17~\GeV\ for electrons and 14~\GeV\ for muons.
The full ATLAS Run~2 dataset is used for this analysis, consisting of $pp$ collision data produced at $\sqrt{s}=13~\TeV$ and recorded between 2015 and 2018.
The data are subjected to quality requirements~\cite{DAPR-2018-01}, including the removal of events recorded when relevant detector components were not operating correctly.
The total integrated luminosity after this cleaning of the data corresponds to $\mathrm{139~fb^{-1}}$~\cite{ATLAS-CONF-2019-021}.
 
An extensive software suite~\cite{ATL-SOFT-PUB-2021-001} is used in the reconstruction and analysis of real and simulated data, in detector operations, and in the trigger and data acquisition systems of the experiment.
 
\subsection{Simulated event samples}
\label{sec:MonteCarloSamples}

Higgs boson production and decay into pairs of $W$ bosons or leptonically decaying $\tau$-leptons were simulated for each of the four main production modes: ggF and VBF, as well as $WH$ and $ZH$ (production in association with a vector boson, collectively referred to as $VH$).
 
The ggF production was simulated at next-to-next-to-leading-order (NNLO) accuracy in QCD\footnote{When higher orders (NLO and NNLO) are specified, QCD is implied if not explicitly stated.} using the \POWHEG~NNLOPS program~\cite{Nason:2004rx, Frixione:2007vw, Alioli:2010xd, Hamilton:2013fea, Hamilton:2015nsa}, interfaced with \PYTHIA[8.212]~\cite{Sjostrand:2014zea} for parton showering and nonperturbative effects.
The simulation achieves NNLO accuracy for arbitrary inclusive ${gg\to H}$ observables by reweighting the Higgs boson rapidity\footnote{The rapidity is defined in terms of a particle's energy $E$ and momentum in the direction of the beam pipe $p_z$ as $y = \frac{1}{2}\ln\frac{E+p_z}{E-p_z}$.} spectrum in Higgs plus one jet (Hj)-MiNLO~\cite{Hamilton:2012np, Campbell:2012am, Hamilton:2012rf} to that of HNNLO~\cite{Catani:2007vq}.
The gluon-gluon fusion prediction from the Monte Carlo (MC) samples is normalized to the next-to-next-to-next-to-leading-order cross section in QCD plus electroweak corrections at next-to-leading order (NLO)~\cite{deFlorian:2016spz, Anastasiou:2016cez, Anastasiou:2015vya, Dulat:2018rbf, Harlander:2009mq, Harlander:2009bw, Harlander:2009my, Pak:2009dg, Actis:2008ug, Actis:2008ts, Bonetti:2018ukf}.
 
VBF events were generated with \POWHEGBOX~\cite{Nason:2004rx, Frixione:2007vw, Alioli:2010xd, Nason:2009ai}, interfaced with \PYTHIA[8.230]~\cite{Sjostrand:2014zea} with the dipole recoil option enabled to model the parton shower, hadronization and underlying event.
The \POWHEGBOX\ prediction is accurate to NLO in QCD corrections and tuned to match calculations with effects due to finite heavy-quark masses and soft-gluon resummations up to next-to-next-to-leading-logarithm (NNLL) accuracy.
The MC prediction is normalized to an approximate-NNLO QCD cross section with NLO electroweak corrections~\cite{Ciccolini:2007jr, Ciccolini:2007ec, Bolzoni:2010xr}.
 
The uncertainties due to the parton shower and hadronization model for the ggF and VBF Higgs boson signal samples are
evaluated using the events in the nominal sample generated with \POWHEGBOX\ but interfaced to an alternative showering program
\HERWIG[7]~\cite{Bahr:2008pv,Bellm:2015jjp} instead of \PYTHIA[8]. To estimate the uncertainty related to the matching between the matrix element and the parton shower for ggF and VBF production, MC events produced with the \MGFiveNLO~\cite{Alwall:2014hca} generator (where MG5 denotes $\MADGRAPH5$) and interfaced to \HERWIG[7] are used. They are accurate to NLO in QCD and utilize the \textsc{NNPDF30\_nlo\_as\_0118}~\cite{Ball:2014uwa} parton distribution function (PDF) set.
In both cases, the \textsc{H7UE} set of tuned parameters (tune)~\cite{Bellm:2015jjp} and the \MMHT[lo] PDF set~\cite{Harland-Lang:2014zoa} were used for the underlying event.
 
The $VH$ production was simulated using \POWHEGBOX[v2]~\cite{Nason:2009ai,Alioli:2010xd,Nason:2004rx,Frixione:2007vw} and interfaced with \PYTHIA[8.212] for parton showering and nonperturbative effects.
The \POWHEGBOX\ prediction is accurate to NLO for the production of $VH$ plus one jet. Samples for the loop-induced process $gg \to ZH$ were generated with \POWHEGBOX[v2] interfaced to \PYTHIA[8.235].
The MC prediction is normalized to cross sections calculated at NNLO in QCD (including the $gg \to ZH$ contribution) and at NLO in electroweak corrections~\cite{Ciccolini:2003jy, Brein:2003wg, Brein:2011vx, Denner:2014cla, Brein:2012ne}.
 
The ggF, VBF, and $VH$ Higgs boson samples use the \PDFforLHC[15]~\cite{Butterworth:2015oua} PDF set and the AZNLO tune~\cite{STDM-2012-23} of \PYTHIA[8]. The sample normalizations account for the decay branching ratios calculated with \textsc{HDECAY}~\cite{Djouadi:1997yw, Spira:1997dg, Djouadi:2006bz} and \PROPHECY~\cite{Bredenstein:2006ha, Bredenstein:2006rh, Bredenstein:2006nk} assuming a Higgs boson mass of $\SI{125.09}{\GeV}$~\cite{HIGG-2014-14}. An uncertainty of 2.16\%~\cite{deFlorian:2016spz} is assigned to the \hww branching ratio, which includes the uncertainty in the Higgs boson mass. All Higgs boson samples are generated with a Higgs boson mass of 125~\GeV, with the uncertainty in the Higgs boson mass being negligible for kinematic distributions.
 
The production of \ttH\ events was modeled using the \POWHEGBOX[v2]~\cite{Frixione:2007nw,Nason:2004rx,Frixione:2007vw,Alioli:2010xd,Hartanto:2015uka} generator at NLO with the \NNPDF[3.0nlo]~\cite{Ball:2014uwa} PDF set.
The events were interfaced to \PYTHIA[8.230] using the A14 tune~\cite{ATL-PHYS-PUB-2014-021}.
The prediction is normalized to the cross section computed at NLO QCD and NLO EW accuracy~\cite{deFlorian:2016spz}.
The sample is inclusive in Higgs decay modes and the cross section is computed assuming a Higgs boson mass of $\SI{125.09}{\GeV}$.

To model the SM background, quark-initiated production of $WW$, $WZ$, $V\gamma^{\ast}$, and $ZZ$ involving the strong interaction was simulated with the \SHERPA[2.2.2]~\cite{Bothmann:2019yzt} generator.
Fully leptonic final states were generated using matrix elements at NLO accuracy in QCD for up to one additional parton and at leading-order (LO) accuracy for up to three additional parton emissions.
For $V\gamma^{\ast}$, the $\gamma^{\ast}$ mass was generated with a lower bound of 4~\GeV.
Samples for the loop-induced processes $gg \to WW$ and $ZZ$ were generated using LO-accurate matrix elements for up to one additional parton emission.
 
For the quark-initiated $WW$ background, systematic uncerainties are evaluated
via samples simulated with alternative settings of the \SHERPA[2.2.2]
generator. The uncertainty in the matching procedure is assessed by varying the
parameter $Q_\text{cut}$, which determines the transition between the matrix-element
and parton-shower domain~\cite{Hoeche:2009rj}. Specifically, alternative
samples with $Q_\text{cut}=15$ and $\SI{30}{\GeV}$ instead of the nominal value of
$Q_\text{cut}=\SI{20}{\GeV}$ are considered. The uncertainties in the shower
model are estimated via samples, in which either the resummation scale,
$\mu_\text{q}$, is increased or decreased by a factor of 2, or the alternative
recoil scheme described in Refs.~\cite{Schumann:2007mg,Hoeche:2009xc} is used.
 
The production of $V\gamma$ final states was simulated with the \SHERPA[2.2.8]~\cite{Bothmann:2019yzt} generator.
Matrix elements are at NLO QCD accuracy for up to one additional parton and at LO accuracy for up to three additional parton emissions.
 
Triboson production ($VVV$) was simulated with the
\SHERPA[2.2.2] generator using factorized gauge-boson
decays. The matrix elements are accurate to NLO for the inclusive process and
to LO for up to two additional parton emissions.
 
For all nominal multiboson samples generated with \SHERPA, the matrix-element calculations were matched and merged with the \SHERPA parton shower based on Catani-Seymour dipole factorization~\cite{Gleisberg:2008fv,Schumann:2007mg} using the MEPS@NLO prescription~\cite{Hoeche:2011fd,Hoeche:2012yf,Catani:2001cc,Hoeche:2009rj}.
The virtual QCD corrections were provided by the \openloops\ library~\cite{Cascioli:2011va,Denner:2016kdg}.
The \NNPDF[3.0nnlo]~\cite{Ball:2014uwa} set of PDFs was used, along with the dedicated set of tuned parton-shower parameters developed by the \SHERPA authors.
 
Electroweak $WW$ production in association with two jets ($WWjj$) was
generated by \MGFiveNLO\ with LO matrix elements using the \NNPDF[3.0nlo] PDF set.
For the nominal sample, \MGFiveNLO\ was interfaced with \PYTHIA[8.244], using the
A14 tune to model the parton shower, hadronization, and underlying event. An
alternative sample is utilized to evaluate the shower model uncertainty, for
which \MGFiveNLO\ was instead interfaced with \HERWIG[7].
 
The quark- or gluon-initiated production of \Zjets was simulated with the \SHERPA[2.2.1]~\cite{Bothmann:2019yzt} generator using NLO matrix elements for up to two partons, and LO matrix elements for up to four partons, calculated with the \textsc{Comix} and \openloops libraries.
They were matched with the \SHERPA parton shower~\cite{Schumann:2007mg} using the MEPS@NLO prescription.
The \NNPDF[3.0nnlo] set of PDFs was used, and the samples are normalized to a NNLO prediction~\cite{Anastasiou:2003ds}.
 
An additional sample of \Zjets events was made with the
\MGFiveNLO\ generator to evaluate the matching
uncertainty. The matrix elements are accurate to LO for up to four
final-state partons. The \NNPDF[2.3lo]~\cite{Ball:2012cx} PDF set was
used. Events were interfaced to \PYTHIA[8.186]~\cite{Sjostrand:2007gs} with the
A14 tune to model the parton shower,
hadronization, and underlying event.

Electroweak production of $\ell\ell jj$ final states was simulated with \SHERPA[2.2.1], but using
LO matrix elements with up to two additional parton emissions.

The production of \ttbar\ events was modeled using the \POWHEGBOX[v2] generator at NLO with the \NNPDF[3.0nlo] PDF set and the \hdamp\ parameter\footnote{The \hdamp\ parameter is a resummation damping factor and one of the parameters that controls the matching of \Powheg\ matrix elements to the parton shower and thus effectively regulates the high-\pt\ radiation against which the \ttbar\ system recoils.} set to $1.5\times$\mtop~\cite{ATL-PHYS-PUB-2016-020}.
In order to correct for a known mismodeling of the leading-lepton \pt due to missing higher-order corrections, an NNLO reweighting was applied to the sample~\cite{NNLOReweighting}.
The events were interfaced to \PYTHIA[8.230] to model the parton shower, hadronization, and underlying event, with parameters set according to the A14 tune and using the \NNPDF[2.3lo] set of PDFs.
 
The associated production of top quarks and $W$ bosons (mainly $Wt$) was modeled using the \POWHEGBOX[v2] generator at NLO in QCD using the five-flavor scheme and the \NNPDF[3.0nlo] set of PDFs.
The diagram removal scheme~\cite{Frixione:2008yi} was used to remove interference and overlap with \ttbar\ production.
The events were interfaced to \PYTHIA[8.230] using the A14 tune and the \NNPDF[2.3lo] set of PDFs.
In all samples generated with \POWHEGBOX[v2], the decays of bottom and charm hadrons were performed by \EVTGEN[1.6.0]~\cite{Lange:2001uf}.
 
The \Wjets and multijet backgrounds are estimated from data. Generated samples of \Wjets and $Z$+jets events are used to validate the estimate and to determine the flavor composition uncertainties. These MC samples were generated using \POWHEGBOX interfaced with \PYTHIA[8.186], with \SHERPA[2.2.1], and with \MGFiveNLO~\cite{Alwall:2014hca,Frederix:2012ps} interfaced with \PYTHIA[8.186].
 
The MC generators, PDFs, and programs used for the underlying event and parton shower (UEPS) are summarized in Table~\ref{tab:mcsamples}.
The alternative generators or UEPS models used to estimate systematic uncertainties are also listed in parentheses.
Finally, the orders of the perturbative prediction for each sample are reported.
 
For all MC samples, the events were processed through the ATLAS detector simulation~\cite{SOFT-2010-01} based on
\GEANT~\cite{Agostinelli:2002hh}. The effect of pileup was modeled by overlaying the hard-scattering event with simulated inelastic $pp$ events generated with \PYTHIA[8.186] using the \NNPDF[2.3lo] set of PDFs and the A3 tune~\cite{ATL-PHYS-PUB-2016-017}.

\begin{table}[h]
\centering
\caption{
Overview of simulation tools used to generate signal and background processes, as well as to model the UEPS. The PDF sets are also summarized.
Alternative event generators or quantities varied to estimate systematic uncertainties are shown in parentheses.}
\label{tab:mcsamples}
\scalebox{0.66}{
\begin{tabular}{l l l l l}
\hline\hline
Process              & Matrix element                                              & PDF set                 & UEPS model                                           & Prediction order\\
& (alternative)                                               &                         & (alternative model)                                  &  for total cross section\\
\hline
ggF $H$              & \POWHEGBOX[v2]~\cite{Hamilton:2013fea,Hamilton:2015nsa,Alioli:2010xd,Nason:2004rx,Frixione:2007vw}
& \multirow{2}{*}{\PDFforLHC[15nnlo]~\cite{Butterworth:2015oua}} &\multirow{2}{*}{\PYTHIA[8]~\cite{Sjostrand:2014zea}} & \multirow{2}{*}{N$^{3}$LO QCD + NLO EW~\cite{deFlorian:2016spz,Anastasiou:2016cez,Anastasiou:2015vya,Dulat:2018rbf,Harlander:2009mq,Harlander:2009bw,Harlander:2009my,Pak:2009dg,Actis:2008ug,Actis:2008ts,Bonetti:2018ukf}} \\
& NNLOPS~\cite{Nason:2009ai,Hamilton:2013fea,Campbell:2012am} &                         &    & \\
& (\MGFiveNLO)~\cite{Alwall:2014hca,Frederix:2012ps}          &                         & (\HERWIG[7])~\cite{Bellm:2015jjp} & \\
VBF $H$              & \POWHEGBOX[v2]~\cite{Nason:2009ai,Alioli:2010xd,Nason:2004rx,Frixione:2007vw}
& \PDFforLHC[15nlo]  & \PYTHIA[8]        & NNLO QCD + NLO EW~\cite{Ciccolini:2007jr,Ciccolini:2007ec,Bolzoni:2010xr} \\
& (\MGFiveNLO)                                                &                         & (\HERWIG[7])                                        & \\
$VH$ excluding\ $gg\to ZH$ & \POWHEGBOX[v2]                                            & \PDFforLHC[15nlo]  & \PYTHIA[8]  & NNLO QCD + NLO EW~\cite{Ciccolini:2003jy,Brein:2003wg,Brein:2011vx,Denner:2014cla,Brein:2012ne} \\
\ttH                 & \POWHEGBOX[v2]                                             & \NNPDF[3.0nlo]    & \PYTHIA[8]               & NLO~\cite{deFlorian:2016spz} \\
$gg\to ZH$           & \POWHEGBOX[v2]                                             & \PDFforLHC[15nlo]  & \PYTHIA[8]               & NNLO QCD + NLO EW~\cite{Altenkamp:2012sx,Harlander:2014wda} \\
 
$qq \to WW$          & \SHERPA[2.2.2]~\cite{Bothmann:2019yzt}                     & \NNPDF[3.0nnlo]~\cite{Ball:2014uwa} & \SHERPA[2.2.2]~\cite{Gleisberg:2008fv,Schumann:2007mg,Hoeche:2011fd,Hoeche:2012yf,Catani:2001cc,Hoeche:2009rj} & NLO~\cite{Buccioni:2019sur,Cascioli:2011va,Denner:2016kdg} \\
& ($Q_\text{cut}$)                                            &                         & (\SHERPA[2.2.2]~\cite{Schumann:2007mg,Hoeche:2009xc}; $\mu_\text{q}$)  \\
$qq \to WWqq$        & \MGFiveNLO~\cite{Alwall:2014hca}                             & \NNPDF[3.0nlo]    & \PYTHIA[8]                                      & LO \\
&                                                             &                         & (\HERWIG[7])                                        & \\
$gg \to WW/ZZ$         & \SHERPA[2.2.2]                                             & \NNPDF[3.0nnlo]   & \SHERPA[2.2.2]                                      & NLO~\cite{Caola:2015rqy}  \\
$WZ/V\gamma^{\ast}/ZZ$ & \SHERPA[2.2.2]                                             & \NNPDF[3.0nnlo]   & \SHERPA[2.2.2]                                      & NLO~\cite{Cascioli:2013gfa} \\
$V\gamma$            & \SHERPA[2.2.8]~\cite{Bothmann:2019yzt}                     & \NNPDF[3.0nnlo]   &\SHERPA[2.2.8]                                       & NLO~\cite{Cascioli:2013gfa} \\
$VVV$                  & \SHERPA[2.2.2]                                             & \NNPDF[3.0nnlo]   &\SHERPA[2.2.2]                                       & NLO  \\
$t\bar{t}$           & \POWHEGBOX[v2]                                             & \NNPDF[3.0nlo]    & \PYTHIA[8]                                          & NNLO+NNLL~\cite{Beneke:2011mq,Cacciari:2011hy,Baernreuther:2012ws,Czakon:2012zr,Czakon:2012pz,Czakon:2013goa,Czakon:2011xx} \\  
& (\MGFiveNLO)                                                &                         & (\HERWIG[7])                                        & \\
$Wt$                  &\POWHEGBOX[v2]                                              & \NNPDF[3.0nlo]    & \PYTHIA[8]                         & NNLO~\cite{Kidonakis:2010ux,Kidonakis:2013zqa} \\
& (\MGFiveNLO)                                                &                         & (\HERWIG[7])                                          & \\
$Z/\gamma^{\ast}$    & \SHERPA[2.2.1]                                             & \NNPDF[3.0nnlo]   & \SHERPA[2.2.1]                                      & NNLO~\cite{Anastasiou:2003ds} \\
& (\MGFiveNLO)                                                & \\
\hline\hline
\end{tabular}
}
\end{table}


\section{Event reconstruction}
\label{sec:EventReconstruction}


Primary vertices in the event are reconstructed from tracks in the ID with $\pT > 500~\MeV$.
Events are required to have at least one primary vertex with at least two associated tracks.
The hard-scatter vertex is selected as the vertex with the highest $\sum \pT^2$, where the sum is over all the tracks associated with that particular vertex.
 
Electron candidates are reconstructed by matching energy clusters in the electromagnetic calorimeter to well-reconstructed tracks that are extrapolated to the calorimeter~\cite{EGAM-2018-01}.
All candidate electron tracks are fitted using a Gaussian sum filter~\cite{ATLAS-CONF-2012-047} to account for bremsstrahlung energy losses.
Electron candidates are required to satisfy $|\eta| < 2.47$, excluding the transition region $1.37 < |\eta| < 1.52$ between the barrel and end caps of the LAr calorimeter.
Muon candidates are reconstructed from a global fit of matching tracks from the inner detector and muon spectrometer~\cite{ATLAS:2020auj}.
They are required to satisfy $|\eta| < 2.5$.
In order to reject particles misidentified as prompt leptons, several identification requirements as well as isolation and impact parameter criteria~\cite{EGAM-2018-01, ATLAS:2020auj} are applied.
For electrons, a likelihood-based identification method~\cite{EGAM-2018-01} is employed, which takes into account a number of discriminating variables such as electromagnetic shower shapes, track properties, transition radiation response, and the quality of the cluster-to-track matching.
Electron candidates with $15~\GeV < \et < 25~\GeV$ must satisfy the ``tight'' likelihood working point, which has an efficiency of approximately 70\% for these electrons.
For $\et > 25~\GeV$, where misidentification backgrounds (discussed further in Sec.~\ref{sec:bkg_misid}) are less important, electron candidates must satisfy the ``medium'' likelihood working point, which has an efficiency of approximately 85\% for an electron with $\pt\! \sim$40~\GeV.
For muons, a cut-based identification method~\cite{ATLAS:2020auj} is employed, using the ``tight'' working point with an efficiency of $\sim$95\% so as to maximize the sample purity.
The impact parameter requirements are $|z_0\sin\theta| < 0.5$~mm and $|d_0|/\sigma_{d_0} < 5\ (3)$ for electrons (muons).\footnote{$z_0$ and $d_0$ are the longitudinal and transverse impact parameters, respectively. $d_0$ is defined by the point of closest approach of the track to the beamline in the r-$\phi$ plane, while $z_0$ is the longitudinal distance to the hard-scatter primary vertex from this point.}
Leptons are required to be isolated from other activity in the event by placing upper bounds on both other transverse energy (using topological clusters in the calorimeter) within a cone of size $\Delta R = 0.2$ around the lepton and other \pT (using tracks) within a cone of variable size no larger than $\Delta R = 0.2$~(0.3) for electrons (muons).
At least one of the offline reconstructed leptons must be matched to an online object that triggered the recording of the event.
In the case where the $e$--$\mu$ trigger is solely responsible for the event being recorded, each lepton must correspond to one of the trigger objects.
This trigger matching scheme also requires the $\pt$ of the lepton to be at least 1~\GeV\ above the trigger-level threshold.
 
Jets are reconstructed using the anti-$k_t$ algorithm with a radius parameter of $R = 0.4$ and particle-flow objects as input~\cite{Cacciari:2008gp,Fastjet,PERF-2015-09}.
The four-momentum of the jets is corrected for the response of the noncompensating calorimeter, signal losses due to noise threshold effects, energy loss in inactive material, and contamination from pileup (defined as additional $pp$ interactions in the same and neighboring bunch crossings)~\cite{PERF-2016-04}.
For jets entering the analysis, a kinematic selection of $\pt > 20~\GeV$ and $|\eta| <  4.5$ is applied.
In the context of event categorization, only jets with $\pt > 30~\GeV$ are considered for jet counting.
Furthermore, a jet-vertex-tagger multivariate discriminant selection that reduces contamination from pileup~\cite{ATLAS-CONF-2014-018} is applied to jets with $20 < \pt < 60~\GeV$ and $|\eta| <  2.4$, utilizing calorimeter and tracking information to separate hard-scatter jets from pileup jets.
Jets with $\pT > 20~\GeV$ and $|\eta| < 2.5$ containing $b$-hadrons ($b$-jets) are identified using a neural-network discriminant based on a number of lower-level taggers which utilize relevant quantities such as the associated track impact parameters and information from secondary vertices.
The working point that is adopted has an average 85\% $b$-jet tagging efficiency, as estimated from simulated $t\bar{t}$ events~\cite{ATL-PHYS-PUB-2017-013,FTAG-2018-01}.
 
The following procedure is adopted in the case of overlapping objects.
If two electrons share an ID track, the lower-\et electron is removed.
If a muon shares an ID track with an electron, the electron is removed.
For electrons and jets, the jet is removed if $\Delta R(\mathrm{jet}, e) < 0.2$ and the jet is not tagged as a $b$-jet.
For any surviving jets, the electron is removed if $\Delta R(\mathrm{jet}, e) < 0.4$.
For muons and jets, the jet is removed if $\Delta R(\mathrm{jet}, \mu) < 0.2$, the jet has less than three associated tracks with $\pT > 500$~\MeV, and the jet is not tagged as a $b$-jet.
For any surviving jets, the muon is removed if $\Delta R(\mathrm{jet}, \mu) < 0.4$.
 
The quantity $\vmet$ is calculated as the negative vector sum of the $\pt$ of all the selected leptons and jets, together with reconstructed tracks that are not associated with these objects but are consistent with originating from the primary $pp$ collision~\cite{PERF-2016-07}.
A second definition of missing transverse momentum (in this case denoted by $\pTmiss$) uses tracks for the hadronic hard term as well, replacing the calorimeter-measured jets with their associated tracks instead. The $\pTmiss$ observable is used directly in the selection of events because of its ability to discriminate better against the \Ztt background, while the \met observable is used to build signal-sensitive variables such as \mT due to its superior resolution.


\section{Event selection and categorization}
\label{sec:EventSelection}


The initial sample of events is required to satisfy the data quality and trigger criteria, as well as to contain exactly two
leptons identified as discussed in the previous section, with different flavor and opposite charge.
In addition, the higher-$\pt$ (leading) lepton is required to have $\pt > 22~\GeV$ and the subleading lepton is required to have $\pt > 15~\GeV$.
Di-$\tau$-lepton backgrounds from low-mass Drell-Yan (DY) production and meson resonances are removed by requiring a dilepton invariant mass $\mll > 10~\GeV$.
In the analysis categories targeting the ggF production mode, a $\pTmiss > 20~\GeV$ selection is applied, which significantly reduces both the \Ztt background and the multijet backgrounds with misidentified leptons.
The above criteria define the event preselection.
Figure~\ref{fig:Plots:Njets} shows the jet multiplicity distribution at the preselection level.
All histograms in this paper include underflow and overflow events unless otherwise stated.
The different background compositions as a function of jet multiplicity motivate the division of the data sample into separate $N_\textrm{jet}$ categories.
Four main analysis categories are defined: the \ZeroJet category targeting the ggF production mode as described in Sec.~\ref{sec:ZeroJetCategory}, the \OneJet category targeting the ggF production mode as described in Sec.~\ref{sec:OneJetCategory}, the \TwoJet category targeting the VBF production mode as described in Sec.~\ref{sec:VBF-TwoJetCategory}, and the \TwoJet category targeting the ggF production mode as described in Sec.~\ref{sec:ggF-TwoJetCategory}.
\begin{figure}[!h]
\centering
\includegraphics[width=0.55\textwidth]{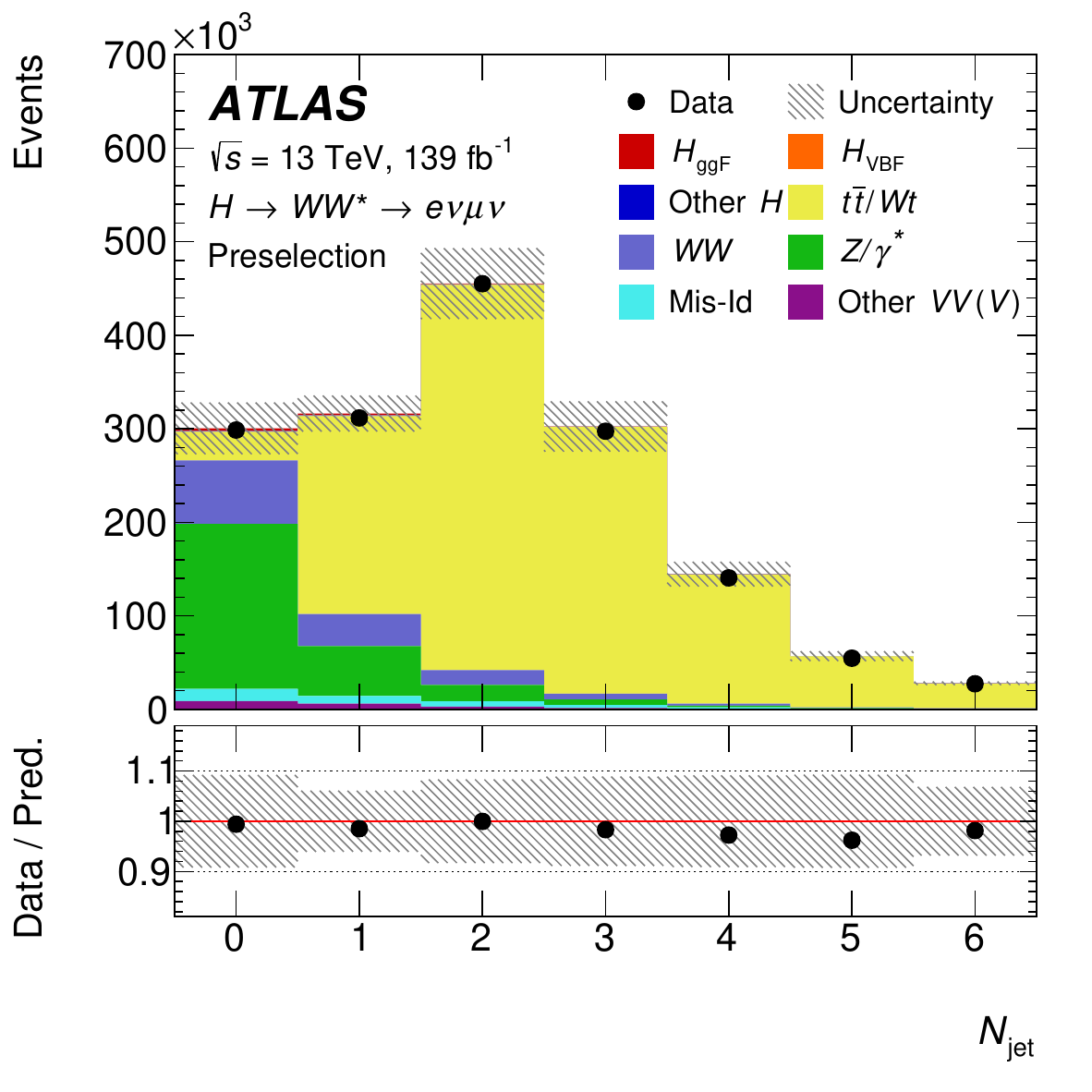}{\caption{
Jet multiplicity distribution, for jets with $\pT > 30$~GeV and $|\eta| < 4.5$, after applying the preselection criteria (prefit normalizations). The hatched band shows the normalization component of the total prefit uncertainty, assuming SM Higgs boson production. The various background components are discussed in more detail in Sec.~\ref{sec:Background}.
\label{fig:Plots:Njets}
}}
\end{figure}
To reject background from top-quark production, events containing $b$-jets with $\pt > 20~\GeV$ are vetoed in all analysis categories.
The remaining selections used to define the analysis SRs are described separately for each category of events below, while Table~\ref{tab:HWWselection} provides a summary of the full set of SR selections.
\begin{table}[!h]
\centering
\caption{
Event selection criteria used to define the SRs in the \hwwenmn\ analysis. The definitions of the variables can be found in the text of this paper.
\label{tab:HWWselection}
}
\scalebox{0.76}{
\renewcommand{\arraystretch}{1.4}
\begin{tabular}{c|| c| c| c| c}
\dbline
Category          & $\NjetThirty=0$ ggF & $\NjetThirty=1$ ggF & $\NjetThirty\ge2$ ggF & $\NjetThirty\ge2$ VBF \\
 
\hline\hline
\multirow{2}{*}{ Preselection}     &
\multicolumn{4}{c}{
\begin{tabular}{c}
Two isolated, different-flavor leptons ($\ell\,{=}\,e, \mu$) with opposite charge\\
$\pTlead\,{>}\,22~\GeV$ , $\pTsublead\,{>}\,15~\GeV$ \\
$\mll\,{>}\,10~\GeV$ \\
\end{tabular}
}\\
\cline{2-5}
& \multicolumn{3}{c|}{$\pTmiss\,{>}\,20~\GeV$}  & \\
\hline\hline
\multirow{3}{*}{Background rejection}  & \multicolumn{4}{c}{$\Nbjetsub=0$}       \\
\cline{2-5}
& $\dphillMET > \pi/2$  &  \multicolumn{3}{c}{$\mtt\,{<}\,m_Z - 25~\GeV$} \\\cline{3-5}
& $\pTll\,{>}\,30~\GeV$                 &  $\maxmtlep\,{>}\,50~\GeV$  & \multicolumn{2}{c}{ }    \\
\hline\hline
\multirow{6}{*}{\!\!\!\!\!
\begin{tabular}{c}
$\hwwenmn$  \\  topology
\end{tabular}
}                          & \multicolumn{3}{c|}{$\mll\,{<}\,55~\GeV$}    &  \\
& \multicolumn{3}{c|}{$\dphill\,{<}\,1.8$}    & \\\cline{2-4}
& \multicolumn{2}{c|}{ } & fail central jet veto  & \\
& \multicolumn{2}{c|}{ } & or & central jet veto \\
& \multicolumn{2}{c|}{ } & fail outside lepton veto & outside lepton veto\\\cline{4-4}
& \multicolumn{2}{c|}{ } & $| \mjj - 85 | > 15\ \GeV$ & $\mjj > 120\ \GeV$ \\
& \multicolumn{2}{c|}{ } & or & \\
& \multicolumn{2}{c|}{ } & $\dyjj > 1.2$ & \\
\hline\hline
Discriminating fit variable   & \multicolumn{3}{c|}{$\mt$}    & DNN   \\
\dbline
\end{tabular}
}
\end{table}
 
\subsection{\ZeroJet category}
\label{sec:ZeroJetCategory}
 
Events with a significant mismeasurement of the missing transverse momentum are suppressed by requiring $\vmet$ to point away from the dilepton transverse momentum ($\dphillMET > \pi/2$).
In the absence of a jet to balance the dilepton system, the magnitude of the dilepton momentum \pTll is expected to be small in DY events.
A requirement of $\pTll\,{>}\,30~\GeV$ further reduces the DY contribution while retaining the majority of the signal events.
 
Continuum $WW$ production and resonant Higgs boson production can be separated by exploiting the spin-0 property of the Higgs boson, which, when combined with the $V - A$ nature of the $W$ boson decay, leads to a small opening angle between the charged leptons.
A requirement of $\mll\,{<}\,55~\GeV$, which combines the small lepton opening angle with the kinematics of a low-mass Higgs boson ($m_H = 125~\GeV$), significantly reduces both the $WW$ and DY backgrounds.
A requirement of $\dphill\,{<}\,1.8$ significantly reduces the remaining DY background while retaining most of the signal.
The \mll and \dphill selections in the \ZeroJet category are indicated by dashed lines in Figs.~\ref{fig:ggf:Plots:selections}(a) and~\ref{fig:ggf:Plots:selections}(b), with an arrow at the top pointing to the region retained.
The \ZeroJet SR is further split into four subregions with boundaries in \mll at $\mll \le30~\GeV$ and $\mll > 30~\GeV$ as well as in the \pt of the subleading lepton at $\pTsublead \le20~\GeV$ and $\pTsublead > 20~\GeV$.
 
\begin{figure}[!h]
\centering
\subfloat[]{
\includegraphics[width=0.49\textwidth]{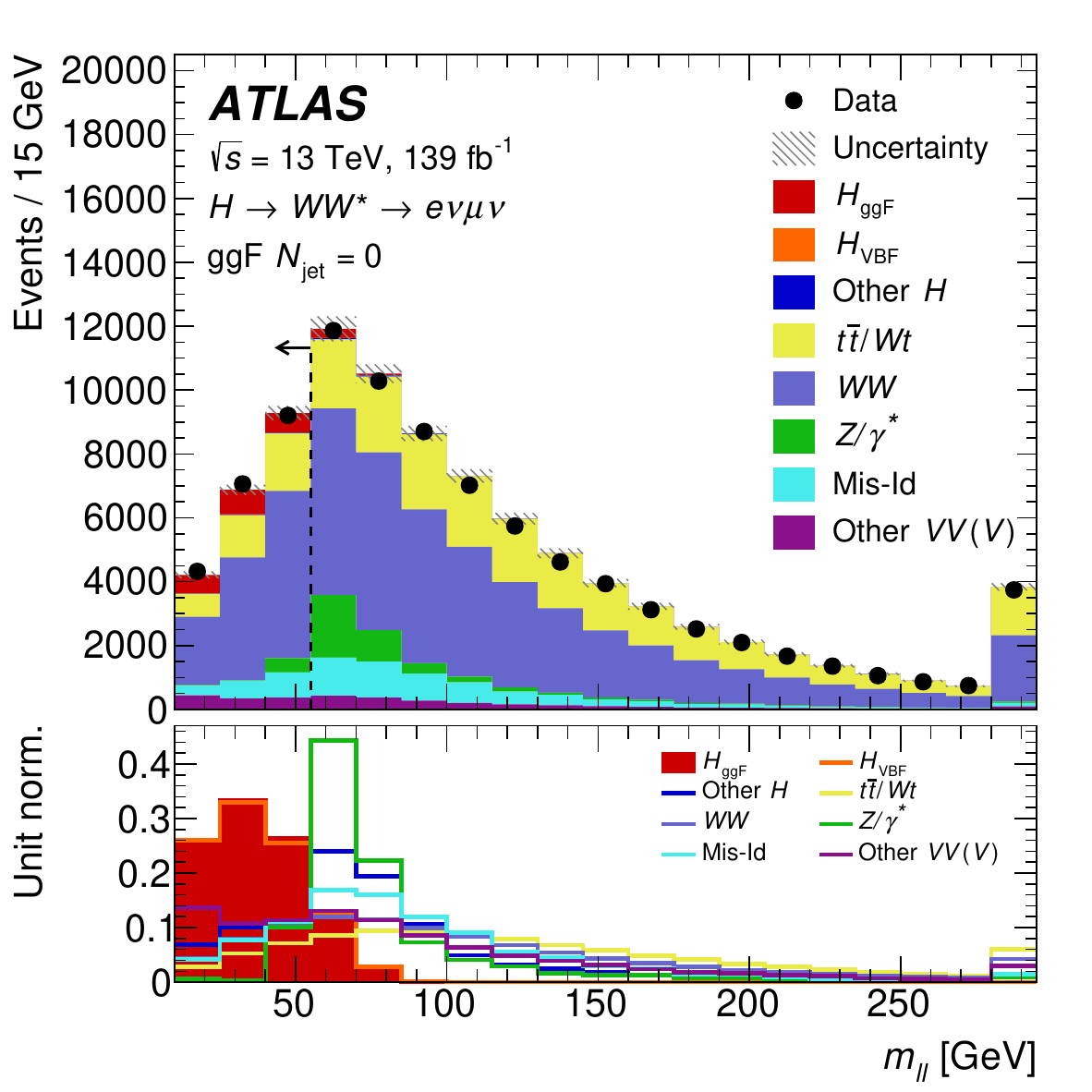}
}
\subfloat[]{
\includegraphics[width=0.49\textwidth]{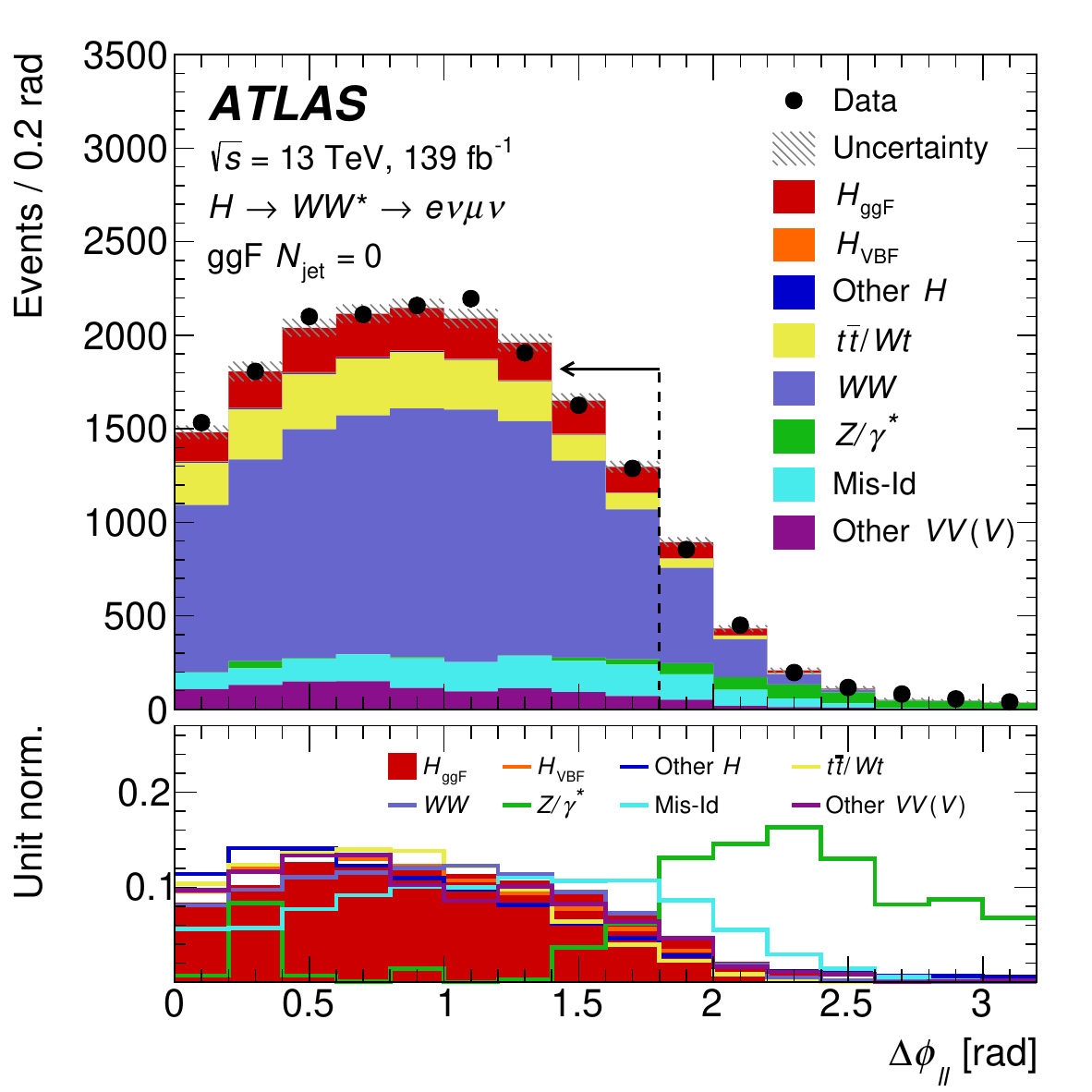}
} \\
\subfloat[]{
\includegraphics[width=0.49\textwidth]{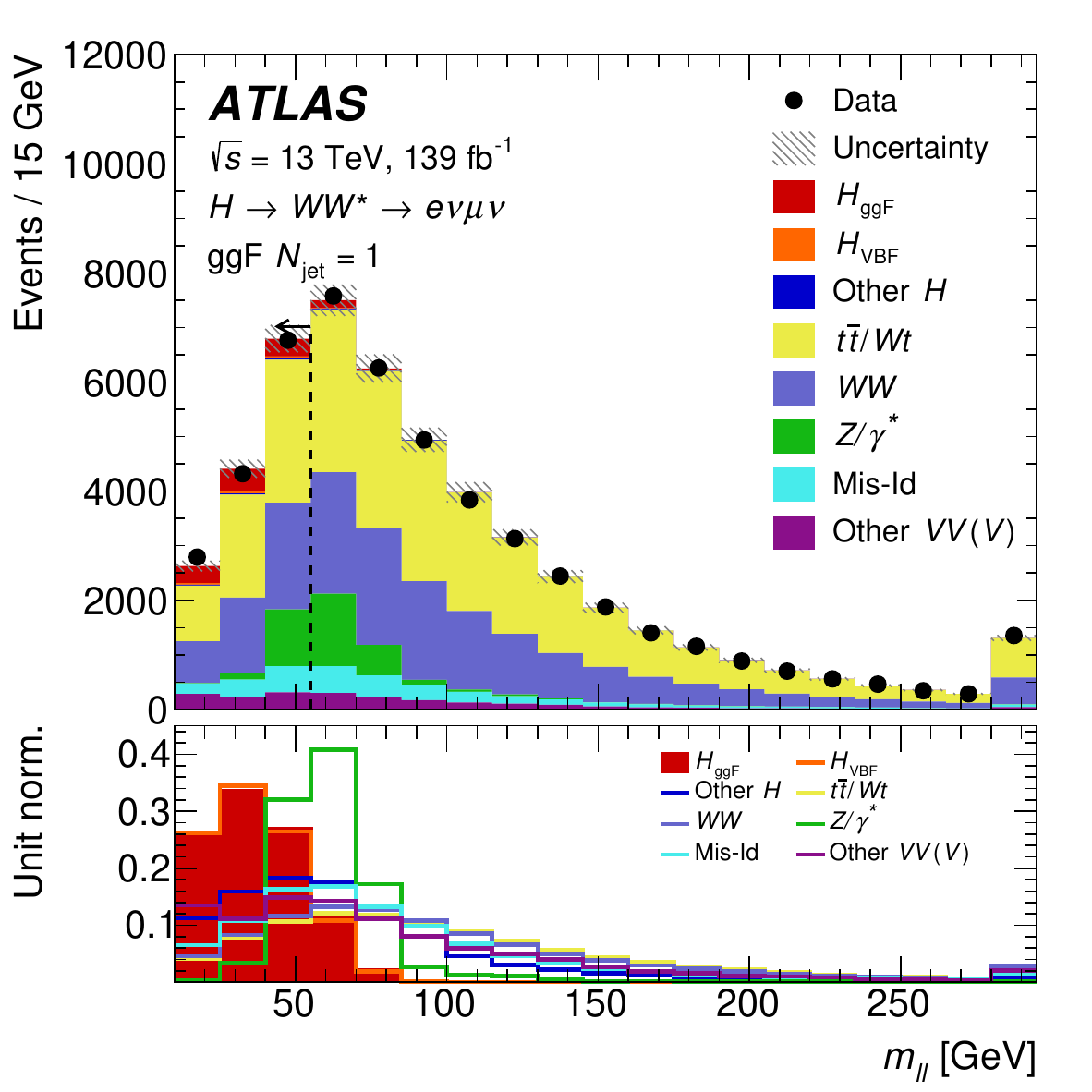}
}
\subfloat[]{
\includegraphics[width=0.49\textwidth]{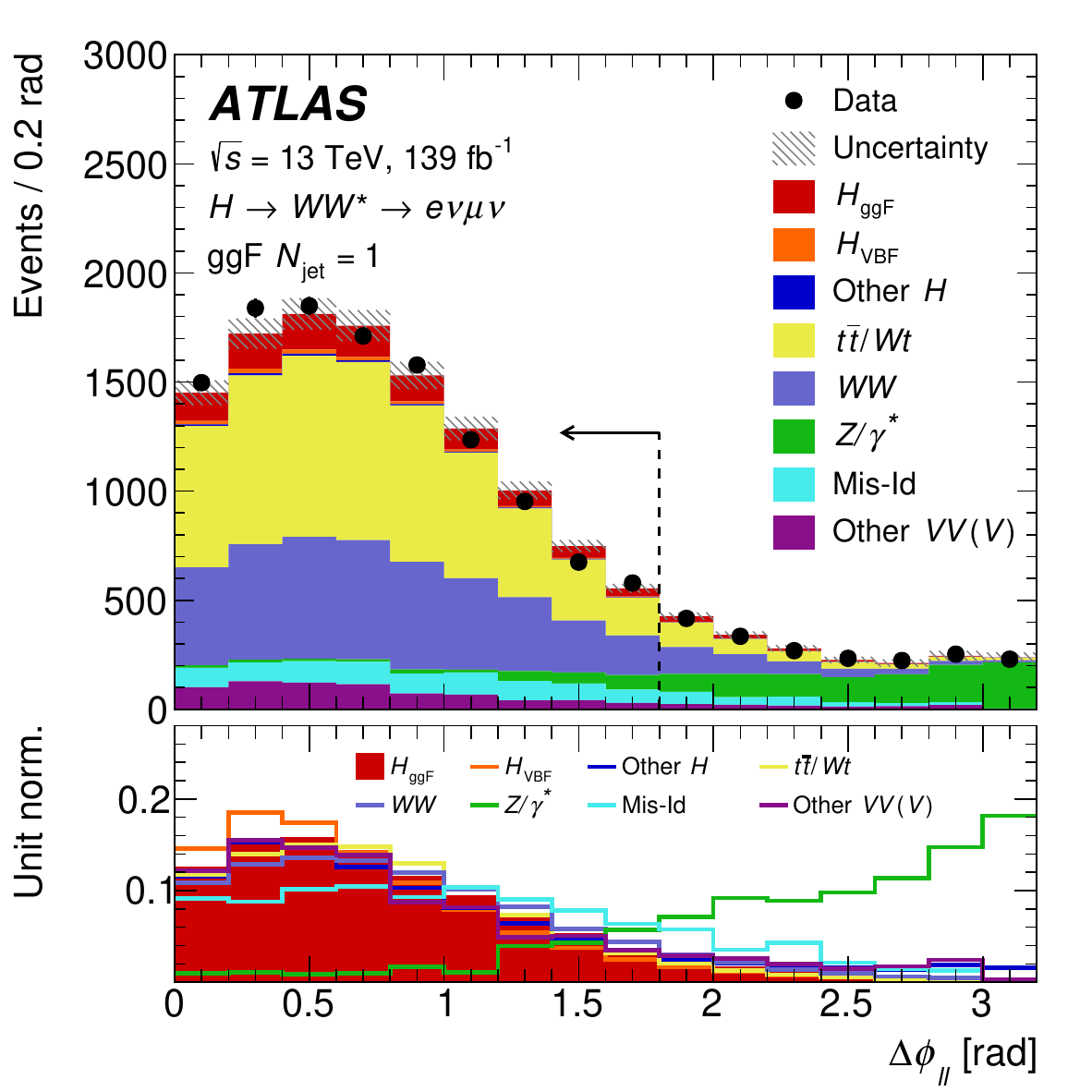}
}
\caption{
Distributions of (a) \mll and (b) \dphill in the \ZeroJet category as well as (c) \mll and (d) \dphill in the \OneJet category, after the preselection and background rejection steps, and also after the selection on \mll for the \dphill plots. The dashed lines indicate where the selection on the observable is made. The distributions are normalized to
their nominal yields, before the final fit to all SRs and CRs (prefit normalizations). The hatched band shows the normalization component of the total prefit uncertainty, assuming SM Higgs boson production. The bottom panels show the normalized distributions for the signal and backgrounds, from which it can be inferred which background processes are primarily removed by the indicated selections. \label{fig:ggf:Plots:selections}
}
\end{figure}

\subsection{\OneJet category}
\label{sec:OneJetCategory}
 
A requirement is applied to the maximum transverse mass defined as the maximum value of $\mT^{\ell_i}$, $\maxmtlep$, where
\begin{equation*}
\mT^{\ell_i} = \sqrt{2\,\pt^{\ell_i}\cdot\met\cdot\left(1-\cos{\Delta\phi\left({\ell_i},\met\right)} \right)},
\end{equation*}
and $\ell_i$ can be either the leading or the subleading lepton. This quantity tends to have small values for the DY background and large values for the signal process.
It also has small values for multijet production, where misidentified leptons are frequently measured with energy lower than the jets from which they originate. Therefore, these backgrounds are substantially reduced with a requirement of $\maxmtlep\,{>}\,50~\GeV$.
The one-jet requirement improves the rejection of \Ztt background. Using the direction and magnitude of the measured missing transverse momentum and projecting it along the directions defined by the two reconstructed charged leptons, the mass of the leptonically decaying $\tau$-lepton pair, \mtautau, can be reconstructed using the so-called collinear approximation~\cite{Plehn:1999xi}.
A requirement of $\mtt\,{<}\,m_Z - 25~\GeV$\footnote{For this selection, $m_Z$ is set to 91.1876~\GeV.} significantly reduces the remaining DY contribution and is applied in all categories with \OneTwoJetSimple.
The same \dphill and \mll selections as described in Sec.~\ref{sec:ZeroJetCategory} are also applied in the \OneJet category and are illustrated in Figs.~\ref{fig:ggf:Plots:selections}(c) and~\ref{fig:ggf:Plots:selections}(d), respectively.
The \OneJet SR is further split into four subregions with the same boundaries as defined for the \ZeroJet category.
 
\subsection{VBF-enriched \TwoJet category}
\label{sec:VBF-TwoJetCategory}
 
The VBF process is characterized by the kinematics of the two leading jets in the event, which are predominantly emitted in the forward region, and by the relatively low levels of hadronic activity between them due to the mediating weak bosons that do not exchange color.
In order to construct a SR enriched in this VBF topology, events are rejected if they contain additional jets with $\pt > 30~\GeV$ that lie in the pseudorapidity gap between the two leading jets (central jet veto) or if either lepton lies outside the pseudorapidity gap between the two leading jets (outside lepton veto).
Furthermore, the invariant mass of the two leading jets, \mjj, is required to be above 120~\GeV\ to ensure orthogonality to analyses targeting the $V(\rightarrow qq)H$ production mode.
 
The events in this category are analyzed using a DNN that is implemented through Keras~\cite{chollet2015keras} and TensorFlow~\cite{tensorflow}, considering VBF Higgs boson production as signal and the other processes as background (including ggF Higgs boson production).
The hyperparameters are optimized to find the best-performing set.
The architecture of the DNN exhibits seven dense hidden layers, with the first hidden layer consisting of 256 nodes and each successive layer decreasing in size.
Nodes in the hidden layers use rectified linear units as activation functions, while the output node utilizes a sigmoid activation function.
Cross-entropy is used to calculate the loss, and dropout is used as a regularization method to prevent overtraining.
A total of 15 kinematic variables built from the leptons ($\ell$), jets ($j$), and $\vmet$ in the event are used as inputs to the DNN.
The following variables are chosen to provide discrimination based on the VBF topology: \mjj; the difference between the two jet rapidities (\dyjj); the lepton $\eta$-centrality ($\lepetacent$, where $C_\ell = | 2\eta_\ell - \sum \eta_{j} | / \Delta \eta_{jj}$), which quantifies the positions of the leptons relative to the leading jets in pseudorapidity~\cite{BARGER1995106}; the \pt of the three leading jets (\pTjone, \pTjtwo, \pTjthree, where \pTjthree is set to $0~\GeV$ if there is no third jet in the event); and the invariant masses of all four possible lepton-jet pairs between the leptons and the two leading jets (\mlonejone, \mlonejtwo, \mltwojone, \mltwojtwo).
The variables \mll, \dphill, and \mt are also utilized, targeting the features of the \hww decay.
Two additional variables are also included:
the total transverse momentum (\pttot), defined as the magnitude of the vectorial sum of the \pt of all selected objects, and the \METSig, which provides separation between events with real undetected high-\pt particles and events where the $\met$ is the result of resolution effects~\cite{ATLAS-CONF-2018-038}.
The observables providing the best discrimination between signal and background are \mjj and \dyjj, and their distributions in the $\TwoJet$ VBF SR are shown in Fig.~\ref{fig:vbf:Plots:SR}.
\begin{figure}[!h]
\centering
\subfloat[]{
\includegraphics[width=0.49\textwidth]{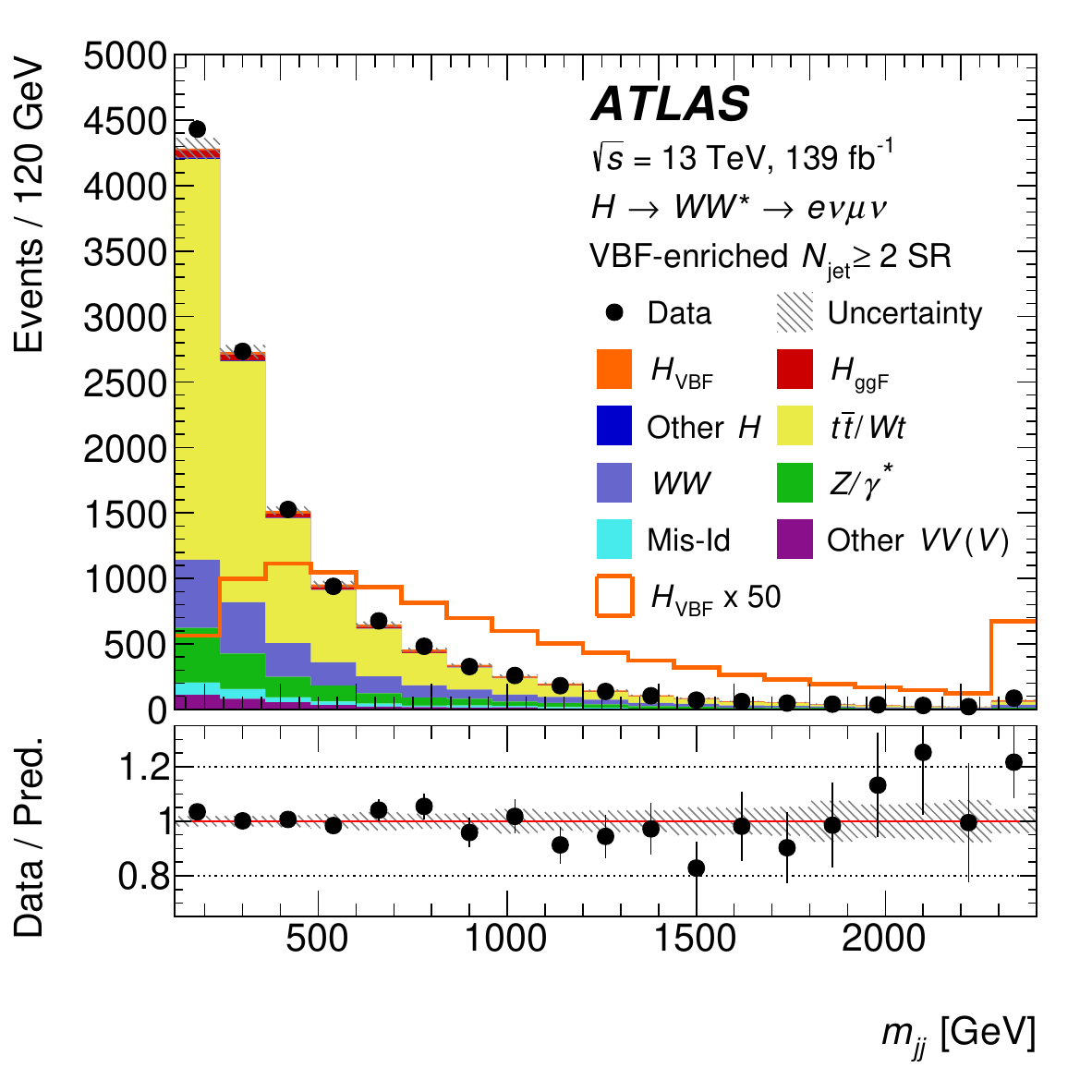}
}
\subfloat[]{
\includegraphics[width=0.49\textwidth]{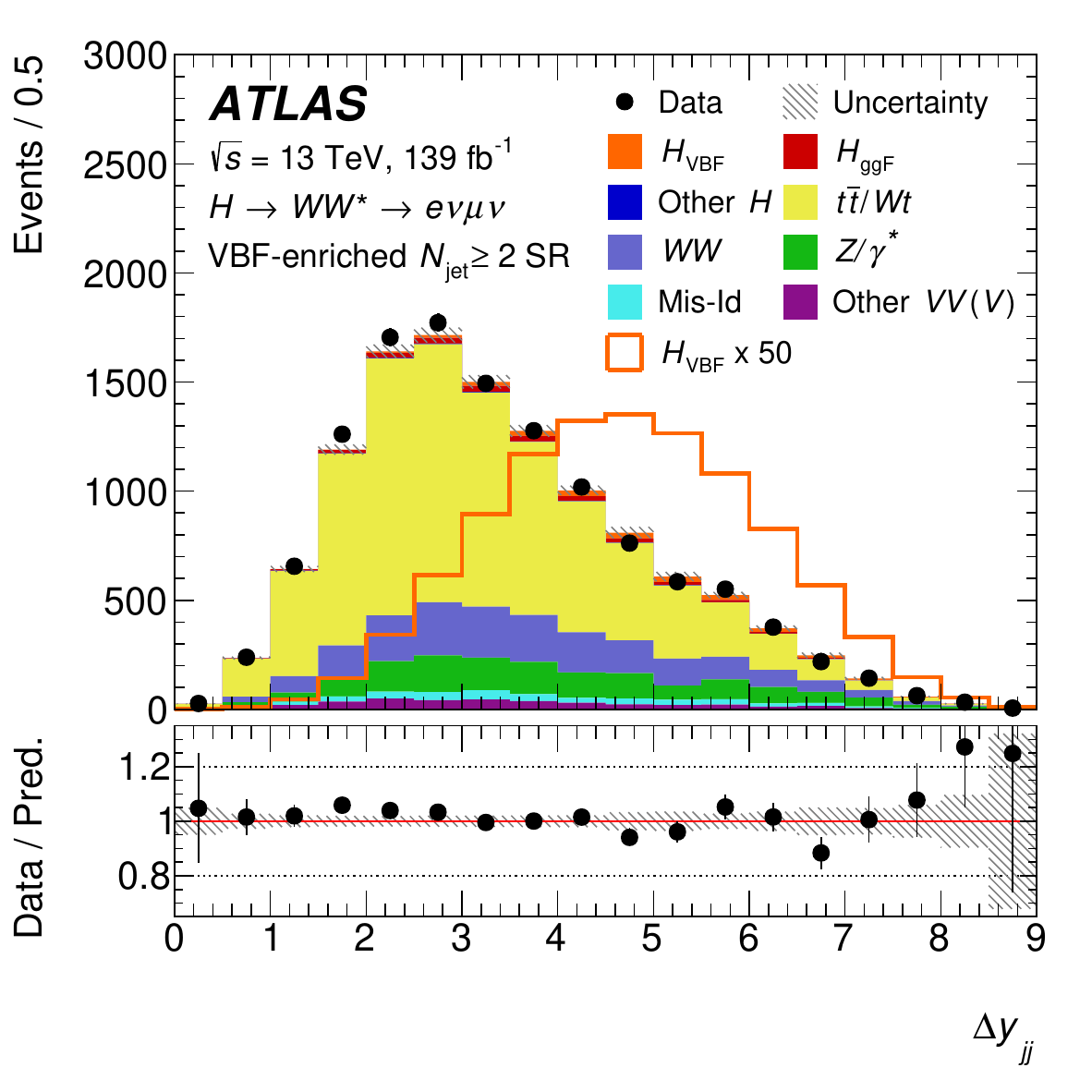}
}
\caption{
Distributions of (a) $\mjj$ and (b) $\dyjj$ in the $\TwoJet$ VBF SR. The solid line shows the expected VBF signal scaled by a factor of 50. The signal and background yields are normalized to the output from the final fit to all SRs and CRs. The hatched band shows the normalization component of the total postfit uncertainty, assuming SM Higgs boson production. The last bin of each distribution is inclusive (includes the overflow). \label{fig:vbf:Plots:SR}
}
\end{figure}
The DNN output reflects how ``VBF-like'' the event's kinematics are, and thus is used as a classifier, with the signal purity improving as the output value increases.
The DNN bin boundaries in the VBF-sensitive range are chosen with an algorithm that aims to make the bins as narrow as possible, while also requiring at least ten expected signal and background events each per bin as well as at most 20\% statistical uncertainty in the background.
This method helps to mitigate the risk of strong statistical fluctuations in the fit, and yields narrower bins on average for larger values of the DNN output, giving a total of seven bins: $[0$--$0.25,0.25$--$0.52,0.52$--$0.68,0.68$--$0.77,0.77$--$0.83,0.83$--$0.87,0.87$--$1.00]$.
In the bin with the highest DNN output, the expected VBF signal-to-background ratio is approximately 2 to 1.
 
\subsection{ggF-enriched \TwoJet category}
\label{sec:ggF-TwoJetCategory}
 
The ggF-enriched \TwoJet category is forced to be mutually exclusive to the VBF-enriched \TwoJet category by requiring events to fail either the central jet veto or outside lepton veto.
Furthermore, $V(\rightarrow qq)H$ production is suppressed by rejecting events in a region defined by $|\mjj - 85|\,\le15\,\GeV$ and $\dyjj\,\le1.2$.
The same \dphill and \mll selections as described in Sec.~\ref{sec:ZeroJetCategory} are also applied in the ggF-enriched \TwoJet category and are shown in Figs.~\ref{fig:ggf2j:Plots:selections}(a) and~\ref{fig:ggf2j:Plots:selections}(b), respectively, together with the \mtt selection in Fig.~\ref{fig:ggf2j:Plots:selections}(c).
The ggF-enriched \TwoJet SR is further split into two subregions with a boundary at $\mll = 30~\GeV$.
\begin{figure}[!h]
\centering
\subfloat[]{
\includegraphics[width=0.49\textwidth]{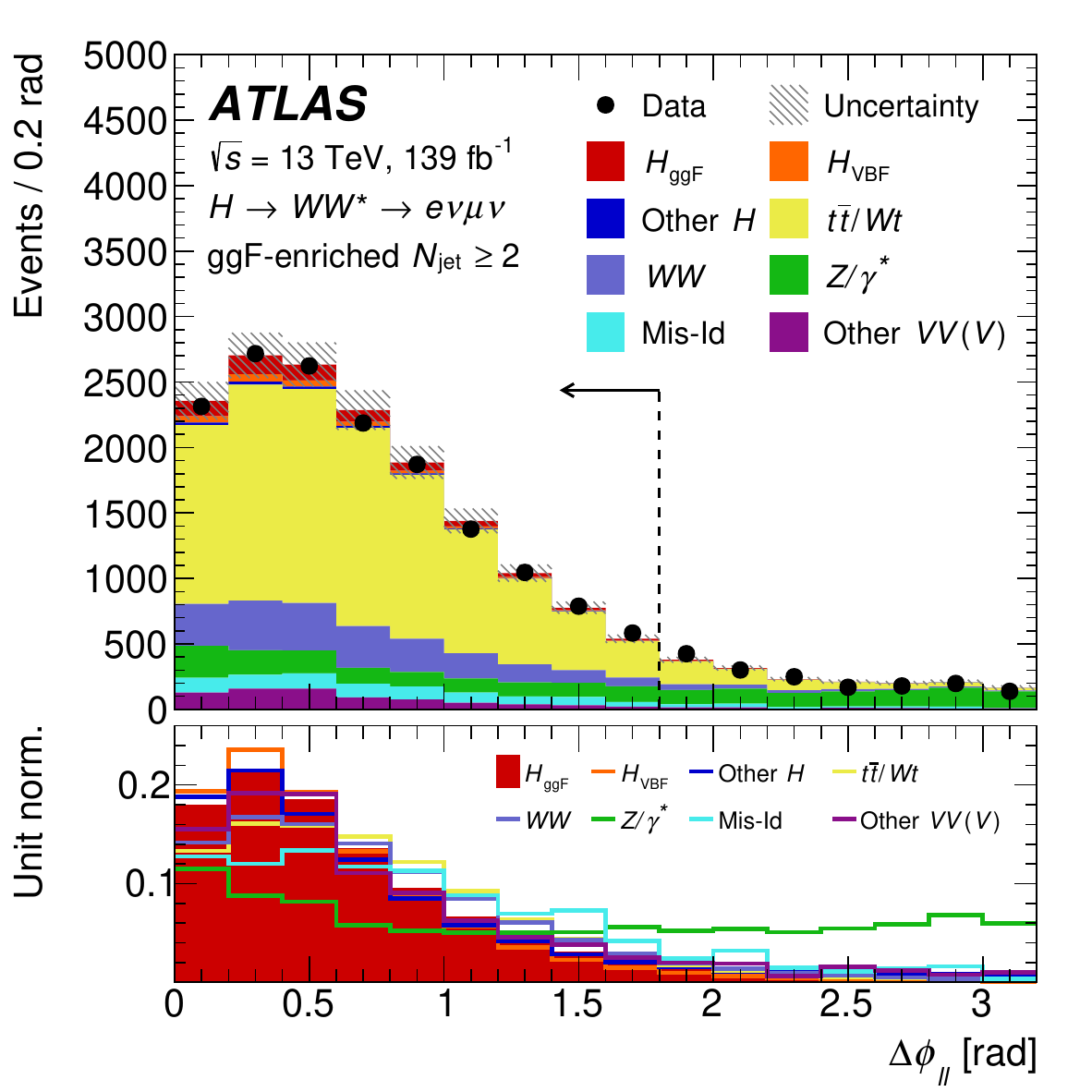}
}
\subfloat[]{
\includegraphics[width=0.49\textwidth]{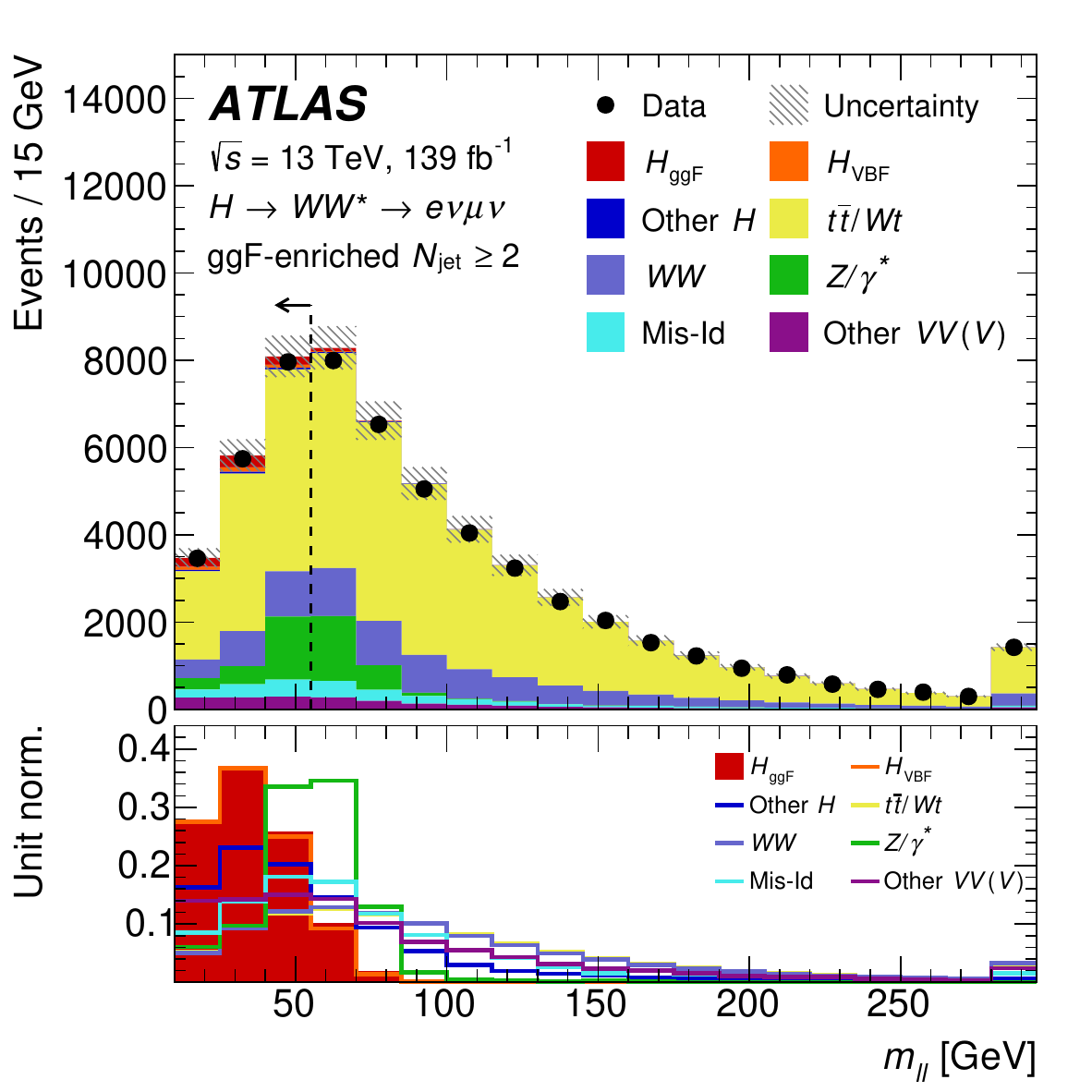}
} \\
\subfloat[]{
\includegraphics[width=0.49\textwidth]{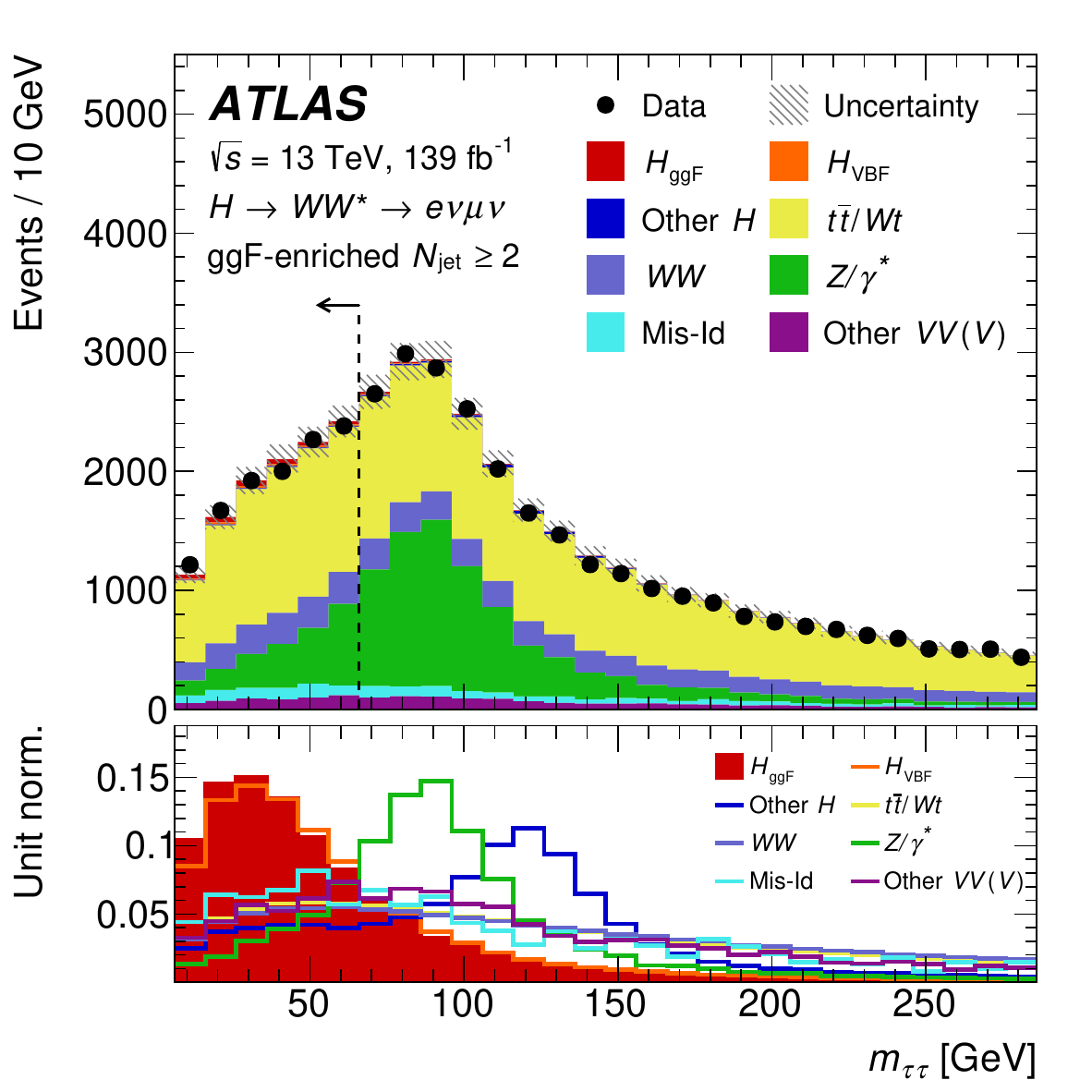}
}
\caption{
Distributions of (a) \dphill, (b) \mll, and (c) \mtt in the ggF-enriched \TwoJet category, after requiring all selections up to the corresponding observable. Underflow and overflow events are not included in (c). The dashed lines indicate where the selection on the observable is made. The distributions are normalized to
their nominal yields, before the final fit to all SRs and CRs (prefit normalizations). The hatched band shows the normalization component of the total prefit uncertainty, assuming SM Higgs boson production. The bottom panels show the normalized distributions for the signal and backgrounds, from which it can be inferred which background processes are primarily removed by the indicated selections. \label{fig:ggf2j:Plots:selections}
}
\end{figure}
 
\subsection{STXS categorization}
\label{sec:STXS-Categorization}

In order to optimize the measurements in bins aligned with those of the Stage-1.2 STXS framework, several STXS kinematic fiducial regions are merged and the separation of the selected events into SRs differs slightly from the description above.
The STXS bin merging strategy, referred to as Reduced Stage-1.2, and the reconstructed SRs are illustrated in Fig.~\ref{fig:STXS_Diagram}.
\begin{figure}[!h]
\centering
\includegraphics[width=0.99\textwidth]{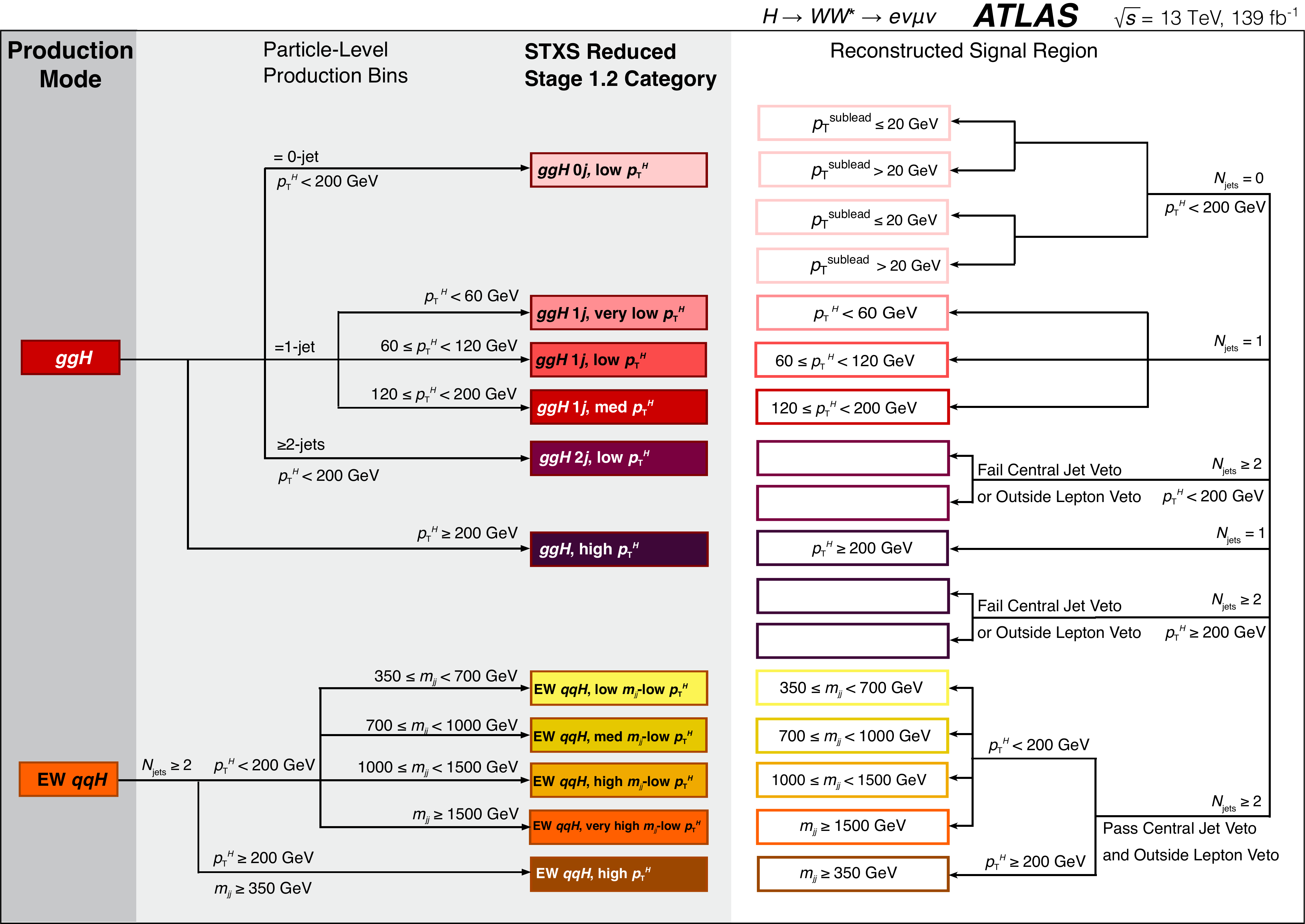}
\caption{
Two sets (Production Mode Stage and Reduced Stage 1.2) of exclusive phase-space regions (production bins) defined at parton level for the measurement of the Higgs boson production cross sections (left and middle-left shaded panels), and the corresponding reconstructed signal regions (right panel).
The description of the production bins as well as the corresponding reconstructed signal regions is given in Sec.~\ref{sec:STXS-Categorization}.
The colors of each reconstructed signal region box indicate the STXS bin which provides the largest relative contribution.
\label{fig:STXS_Diagram}
}
\end{figure}
For the exclusive \ZeroJet $ggH$ category, only a single STXS cross section is measured, with the same SR splitting as defined in Sec.~\ref{sec:ZeroJetCategory} and applying a $\pTH < 200~\GeV$ selection.
For the exclusive \OneJet $ggH$ category, all three Stage-1.2 measurements are retained, with the SR split along the same \pTH bin boundaries.
For the exclusive \TwoJet $ggH$ category with $\pTH < 200~\GeV$, only a single STXS cross section is measured, with the same SR splitting as defined in Sec.~\ref{sec:ggF-TwoJetCategory} but also including a $\pTH < 200~\GeV$ selection.
For the \Njets inclusive $ggH$ category with $\pTH \ge200~\GeV$, only a single STXS cross section is measured, using both the \OneJet SR with an added $\pTH \ge200~\GeV$ selection and the same ggF-enriched \TwoJet SR with splitting as defined in Sec.~\ref{sec:ggF-TwoJetCategory} but also including a $\pTH \ge200~\GeV$ selection. No events with \ZeroJet at the event-generation level are reconstructed in the region with $\pTH \ge200~\GeV$.
For the exclusive \TwoJet EW $qqH$ categories targeting VBF production, the bins separated by \pTHjj value\footnote{\pTHjj is defined at the reconstruction level as the transverse momentum of the system composed of the two leptons + \met + two leading jets in the event.} are merged, and so are the bins separated by \mjj for $\pTH \ge200~\GeV$, resulting in a total of five measured cross sections.
The same SR described in Sec.~\ref{sec:VBF-TwoJetCategory} is used, but split into five subregions with the same boundaries that define the measured EW $qqH$ STXS cross sections. In addition, the same DNN and training is used for the final fit's discriminating variable.
The exclusive \TwoJet EW $qqH$ category with $\mjj < 350~\GeV$ that targets $V(\rightarrow q\bar{q})H$ production and also the $V(\to \textrm{leptons})H$ and $t\bar{t}H$ categories to which this analysis is not sensitive are fixed to their expected yields in the fit.
Figure~\ref{fig:STXS_Composition} shows the relative contributions of the different merged STXS bins in all reconstructed SRs.
In each case, the target categories provide the largest contribution in the corresponding SRs which aim to select them.

\begin{figure}[!h]
\centering
\includegraphics[width=0.90\textwidth]{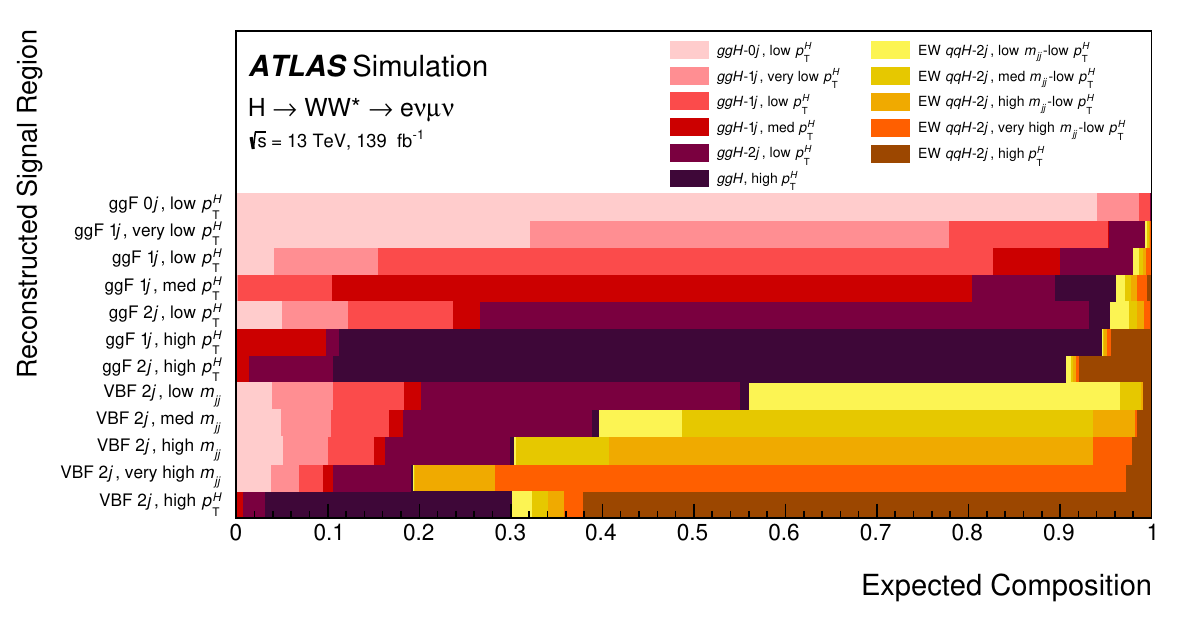}
\caption{
Relative SM signal composition in terms of the measured STXS bin for each reconstructed signal region.
\label{fig:STXS_Composition}
}
\end{figure}


\section{Background estimation}
\label{sec:Background}


The background contamination in the SRs originates from various processes: nonresonant $WW$,
top-quark pair ($t\bar{t}$) and single-top-quark ($Wt$), diboson ($WZ$, $ZZ$, $W\gamma$, $W\gamma^{\ast}$, and $Z\gamma$) and Drell-Yan (mainly $\Ztt$, hereafter denoted by \DY) production.
Other background contributions arise from \Wjets\ and multijet production with misidentified leptons,
which are either nonprompt leptons from decays of heavy-flavor hadrons or jets misidentified as prompt leptons. The backgrounds
with misidentified leptons are estimated using a data-driven technique.
Dedicated data regions with low expected signal, hereafter called control regions (CRs), are used to normalize the predictions of the $WW$, top-quark, and \Ztt backgrounds. Table~\ref{tab:CRsSelection} summarizes the selections used to define the CRs, which start from the preselection defined in Table~\ref{tab:HWWselection}. The background estimates for the remaining background processes, most notably the diboson processes other than $WW$, are obtained from simulated samples normalized to the theoretical cross sections for these processes.
 
\begin{table}[!h]
\caption{
Event selection criteria used to define the control regions in the \hwwenmn\ analysis.
Every control region selection starts from the selection labeled ``Preselection'' in Table~\ref{tab:HWWselection}, and
$\Nbjetbetween$ represents the number of $b$-jets with $20 < \pT < 30~\GeV$.}
\label{tab:CRsSelection}
\centering
\scalebox{0.77}{
\renewcommand{\arraystretch}{1.4}
\begin{tabular}{c|| c | c | c | c}
\dbline
CR                         & $\NjetThirty=0$ ggF & $\NjetThirty=1$ ggF & $\NjetThirty\ge2$ ggF & $\NjetThirty\ge2$ VBF \\
\hline\hline
\multirow{8}{*}{$qq\rightarrow WW$}      & \multicolumn{3}{c|}{$\Nbjetsub=0$}                                                   &   \\ \cline{2-4}
& $\dphillMET > \pi/2$  & \multicolumn{2}{c|}{$\mll\,{>}\,80~\GeV$}              &   \\ \cline{3-4}
& $\pTll\,{>}\,30~\GeV$         & $|\mtt-m_Z|\,{>}\,25~\GeV$  & $\mtt<m_Z-25~\GeV$   &   \\
&  $55\,{<}\,\mll\,{<}\,110~\GeV$   & $\maxmtlep\,{>}\,50~\GeV$  & $\mTtwo\,{>}\,165~\GeV$ &   \\ \cline{4-4}
&  $\dphill\,{<}\,2.6$  &           & fail central jet veto                                &    \\
& &           & or fail outside lepton veto                          &    \\ \cline{4-4}
& &           & $|\mjj-85| > 15~\GeV$                               &    \\
& &           &  or\ \ \ \ $\dyjj > 1.2$                                     &    \\
\hline\hline
\multirow{10}{*}{$t\bar{t}$/$Wt$} & \multirow{2}{*}{$\Nbjetbetween>0$}    & $\Nbjetnom = 1$  & \multirow{2}{*}{$\Nbjetsub=0$}   & \multirow{2}{*}{$\Nbjetsub=1$}   \\
&                                       & $\Nbjetbetween  = 0$          &              &       \\ \cline{3-5}
& $\dphillMET > \pi/2$                  & \multicolumn{3}{c}{\phantom{ccccc}$\mtt\,{<}\,m_Z - 25~\GeV$} \\ \cline{3-5}
& $\pTll\, > \, 30~\GeV$                & $\maxmtlep\,{>}\,50~\GeV$                 & $\mll\,{>}\,80~\GeV$ &         \\
& $\dphill\,{<}\,2.8$                   &        &       $\dphill\,{<}\,1.8$      &    \\
&                                       &        &       $\mTtwo\,{<}\,165~\GeV$ &   \\ \cline{4-4}
& &           & fail central jet veto                                & central jet veto \\
& &           & or fail outside lepton veto                          & outside lepton veto  \\ \cline{4-4}
& &           & $|\mjj-85| > 15~\GeV$                               &    \\
& &          & or\ \ \ \ $\dyjj > 1.2$                                     &    \\
\hline\hline
\multirow{8}{*}{$Z/\gamma^{\ast}$}    &                        \multicolumn{4}{c}{\phantom{ccccc}$\Nbjetsub=0$}           \\ \cline{2-5}
& \multicolumn{2}{c|}{\phantom{ccccc}$\mll < 80~\GeV$} & $\mll\,{<}\,55~\GeV$ & $\mll\,{<}\,70~\GeV$    \\\
& \multicolumn{2}{c|}{no $\pTmiss$ requirement}  &  &   \\\cline{2-4}
& $\dphill > 2.8$        & \multicolumn{2}{c|}{$\mtt > m_Z - 25~\GeV$} &$|\mtt - m_Z| \leq 25~\GeV$  \\ \cline{3-4}
& &   $\maxmtlep\,{>}\,50~\GeV$       & fail central jet veto                                & central jet veto \\
& &           & or fail outside lepton veto                          & outside lepton veto  \\ \cline{4-4}
& &           & $|\mjj-85| > 15~\GeV$                               &    \\
& &           & or\ \ \ \ $\dyjj > 1.2$                                     &    \\
\dbline
\end{tabular}
}
\end{table}
 
\subsection{$WW$ background}
\label{sec:bkg_WW}
 
The nonresonant $WW$ background mainly originates from the quark-initiated process (labeled $qqWW$) with a small additional contribution from the gluon-initiated process proceeding via a box-diagram ($ggWW$). The $ggWW$ process produces approximately 10\% of the total $WW$ background contribution and is estimated from simulated samples normalized to the theoretical cross sections. The $qqWW$ process is normalized to the observed yields in dedicated CRs, defined separately for each analysis category. The CRs are orthogonal to the SRs and enriched in the $WW$ process.
For the \ZeroJet category,
the selected \mll region is modified to $55 < \mll < 110$~\GeV\ and the \dphill selection is relaxed to $\dphill < 2.6$ (from $\dphill < 1.8$ in the SRs). The upper bound on the \mll selection reduces the contamination from top-quark processes in the $WW$ CR, whereas the \dphill selection removes most of the \Ztt contamination. The $WW$ CR in the \OneJet category differs from the SR by requiring $\mll > 80$~\GeV\ and $|\mtt-m_Z|\,{>}\,25~\GeV$. For the \TwoJet categories, it is difficult to find a region with high $WW$ process purity because of the overwhelming background from top-quark processes. For the ggF-enriched \TwoJet category, the $qqWW$ process is normalized to the yield in the CR, whereas the $qqWW$ background in the VBF-enriched \TwoJet category is estimated from simulated samples normalized to the theoretical cross section. The $WW$ CR for the ggF-enriched \TwoJet category is defined by requiring $\mll > 80$~\GeV\ and $\mTtwo > 165$~\GeV. The \mTtwo variable~\cite{Lester:1999tx}
is defined as
\begin{equation*}
\mTtwo^2 = \underset{{p\mkern-9.5mu/}_1+{p\mkern-9.5mu/}_2={p\mkern-9.5mu/}_\text{T}}{\textrm{min}} \left[\textrm{max}\{\mt^2(p_\text{T}^a,{p\mkern-9.5mu/}_1),\mt^2(p_\text{T}^b,{p\mkern-9.5mu/}_2)\}\right] ,
\end{equation*}
where the minimization is over all possible two-momenta, ${p\mkern-9.5mu/}_{1,2}$, such that their sum gives the observed missing transverse momentum ${p\mkern-9.5mu/}_\text{T}$, and where each of $p_\text{T}^a$ and $p_\text{T}^b$ is the combined transverse momentum of a charged lepton and a jet.
The \mTtwo selection is indicated by a dashed line in Fig.~\ref{fig:WW_CR_MT2}, with an arrow at the top pointing to the region retained.
For all $WW$ CRs, a \bjet veto is maintained.
 
The $WW$ CRs have prefit $WW$ process purities of 67\% (\ZeroJet), 34\% (\OneJet), and 39\% (ggF-enriched \TwoJet). 
The postfit background normalization factors, from the fit described in Sec.~\ref{sec:FitProcedure} are summarized in Table~\ref{tab:CRs_NF}.
Figure~\ref{fig:WW_CRs} presents the postfit \mt distributions in the \ZeroJet, \OneJet, and ggF-enriched \TwoJet CRs.
 
\begin{figure}[htp]
\centering
\includegraphics[width=0.7\textwidth]{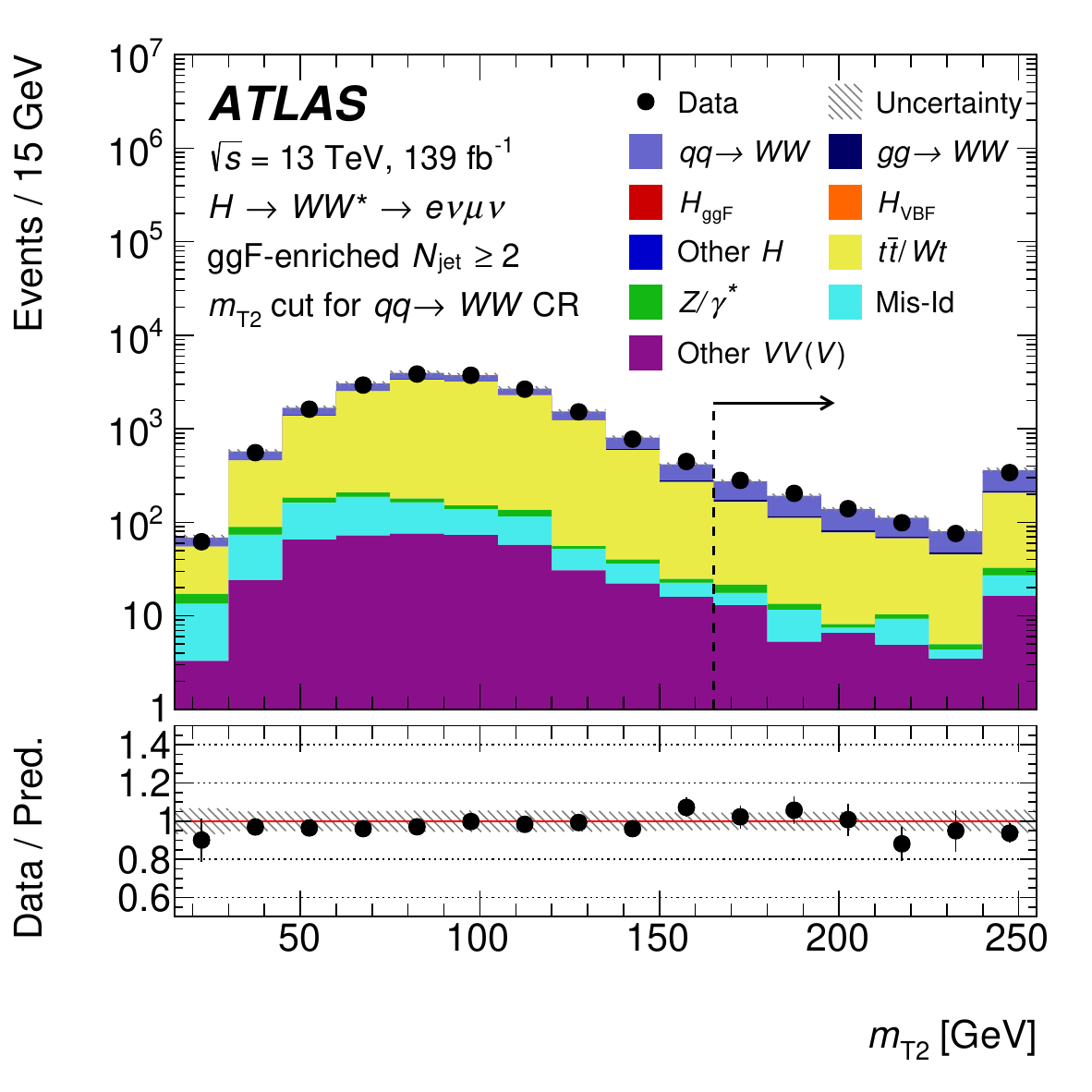}
\caption{
Distribution of the \mTtwo variable used in the definition of the $WW$ CR for the ggF-enriched \TwoJet category and after the prior selections summarized in Table~\ref{tab:CRsSelection} have been applied.
The dashed line indicates where the selection on the observable is made.
The distributions are normalized to their nominal yields, before the final fit to all SRs and CRs (prefit normalizations).
The hatched band shows the normalization component of the total prefit uncertainty.
\label{fig:WW_CR_MT2}
}
\end{figure}
 
\begin{figure}[htp]
\centering
\subfloat[]{
\includegraphics[width=0.5\textwidth]{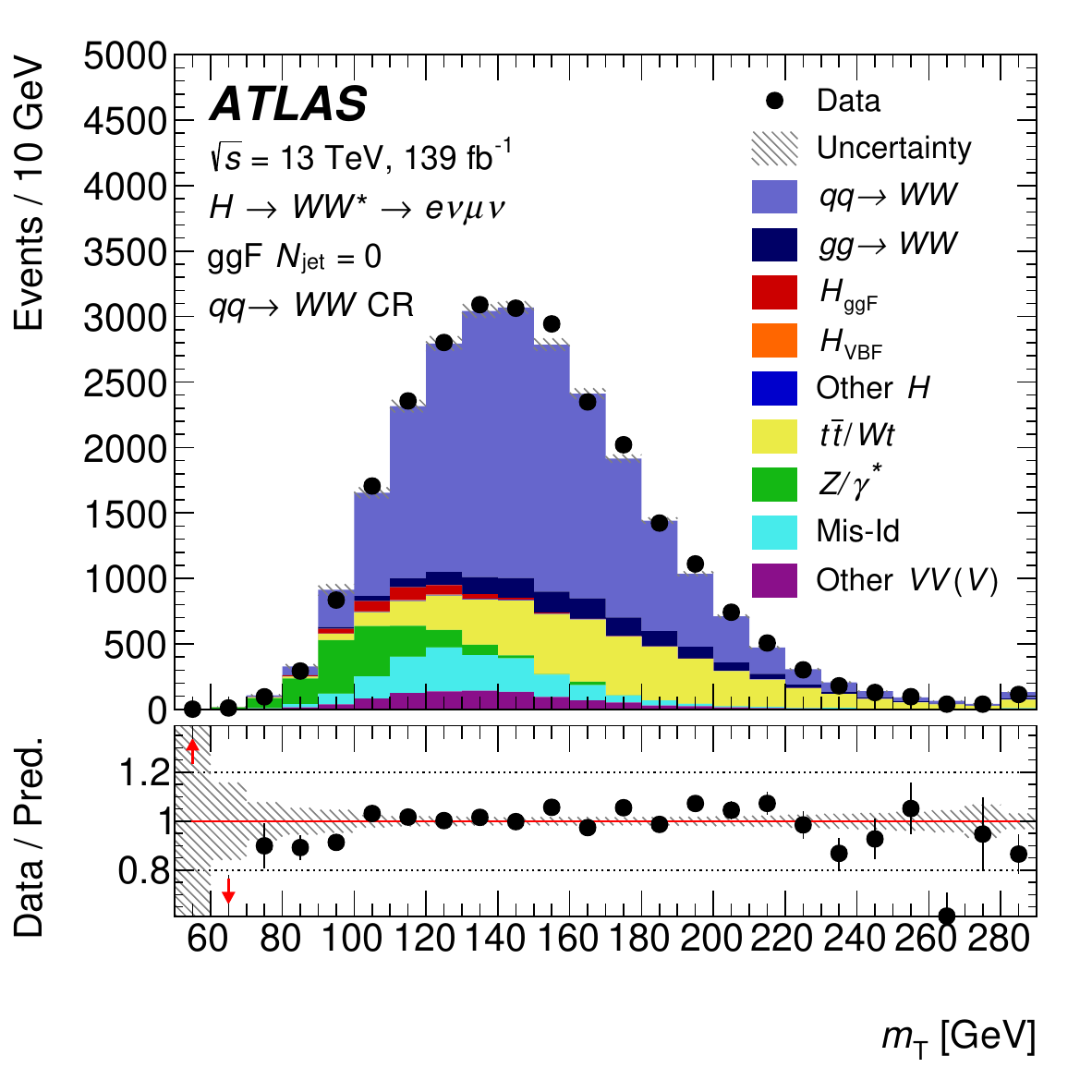}
}
\subfloat[]{
\includegraphics[width=0.5\textwidth]{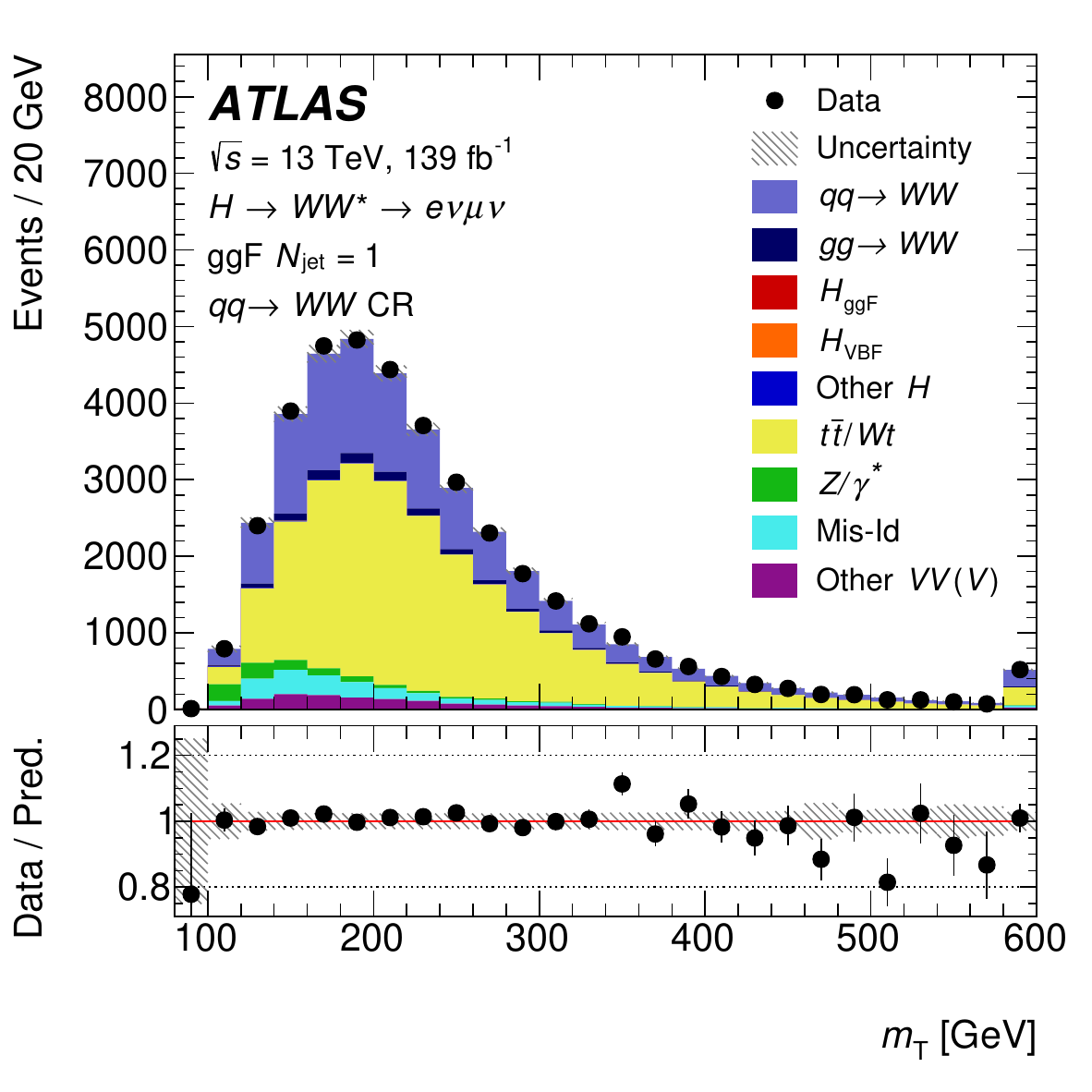}
} \\
\subfloat[]{
\includegraphics[width=0.5\textwidth]{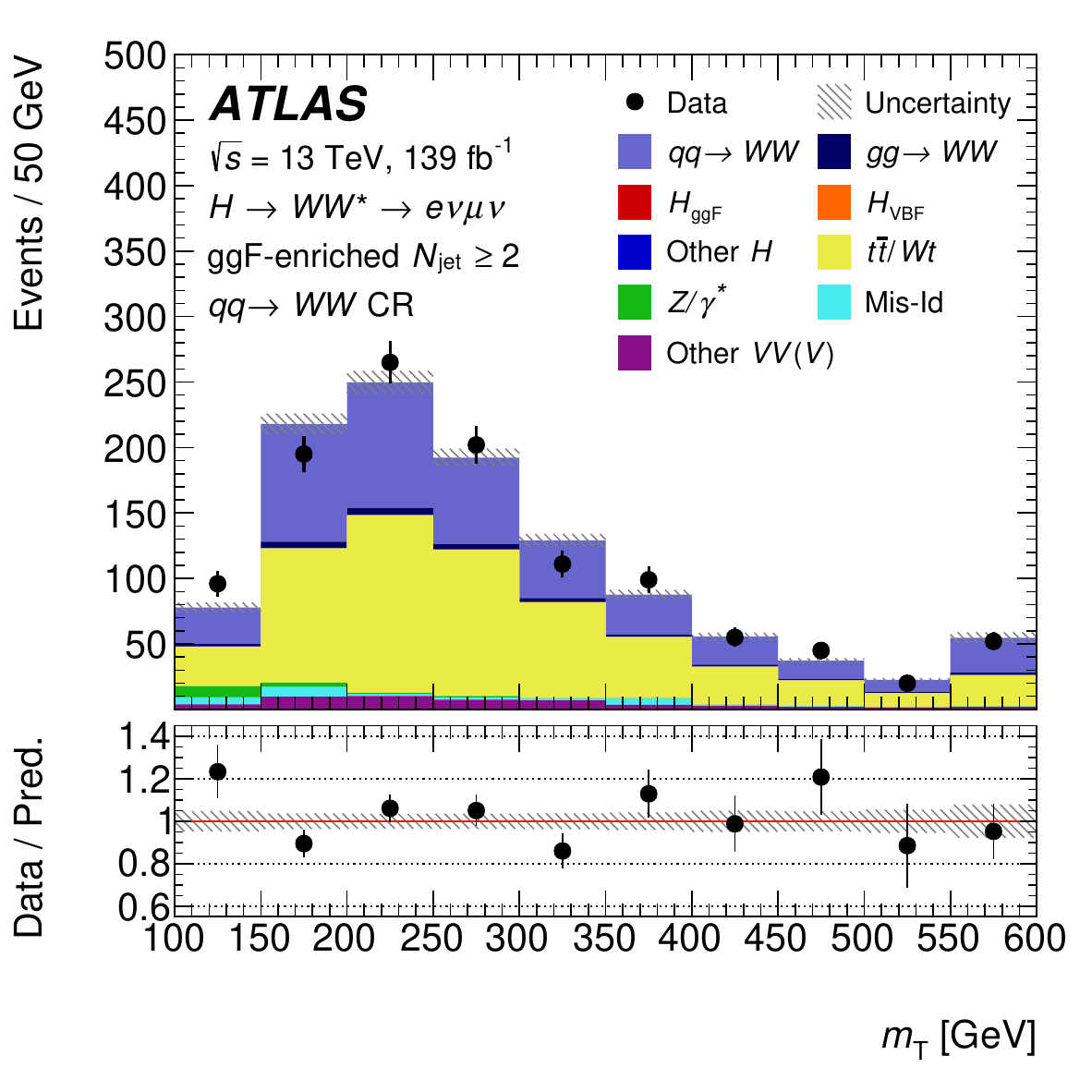}
}
\caption{
Postfit $\mT$ distributions in the (a) \ZeroJet, (b) \OneJet, and (c) ggF-enriched \TwoJet $WW$ CRs with signal (normalized to postfit measurement) and background modeled contributions.
The red arrow in the lower panel of (a) indicates that the central value of the data lies above the window. The last bin of each distribution is inclusive (includes the overflow).
The hatched band shows the total uncertainty, assuming SM Higgs boson production.
Some contributions are too small to be visible.
\label{fig:WW_CRs}
}
\end{figure}
 
\subsection{Top-quark backgrounds}
\label{sec:bkg_top}
 
The top-quark backgrounds affecting this analysis are associated with the $t\bar{t}$ and $Wt$ processes.
They are normalized to the observed combined top yields in CRs, defined separately for each analysis category.
The uncertainties in the relative contributions of $t\bar{t}$ and $Wt$ are accounted for by considering their relevant uncertainties separately, while their ratios are similar between respective CRs and SRs.
The CRs are orthogonal to the SRs, normally by inverting the \bjet veto.
The exception is in the ggF-enriched \TwoJet category, where the top-quark CR is defined with a \bjet veto and orthogonality to the SR and $WW$ CR is achieved by requiring $\mll > 80$~\GeV\ and $\mTtwo < 165$~\GeV, respectively.
This is possible due to the \TwoJet categories having high top-quark event purity even with a \bjet veto, and this definition reduces the uncertainties from the \bjet selection in this category. For the \ZeroJet category, the top-quark CR requires the presence of a reconstructed jet with $20 < \pt < 30$~\GeV\ which is identified as coming from a $b$-quark.
 
The top-quark CRs have prefit top-quark process purities of 89\% (\ZeroJet), 98\% (\OneJet), 71\% (ggF-enriched \TwoJet), and 97\% (VBF-enriched \TwoJet). The postfit background normalization factors are summarized in Table~\ref{tab:CRs_NF}.
Figure~\ref{fig:Top_CRs} presents the postfit \mt distributions in the \ZeroJet, \OneJet, and ggF-enriched \TwoJet CRs, as well as the postfit DNN output distribution in the VBF-enriched \TwoJet CR.
 
\begin{figure}[htp]
\centering
\subfloat[]{
\includegraphics[width=0.5\textwidth]{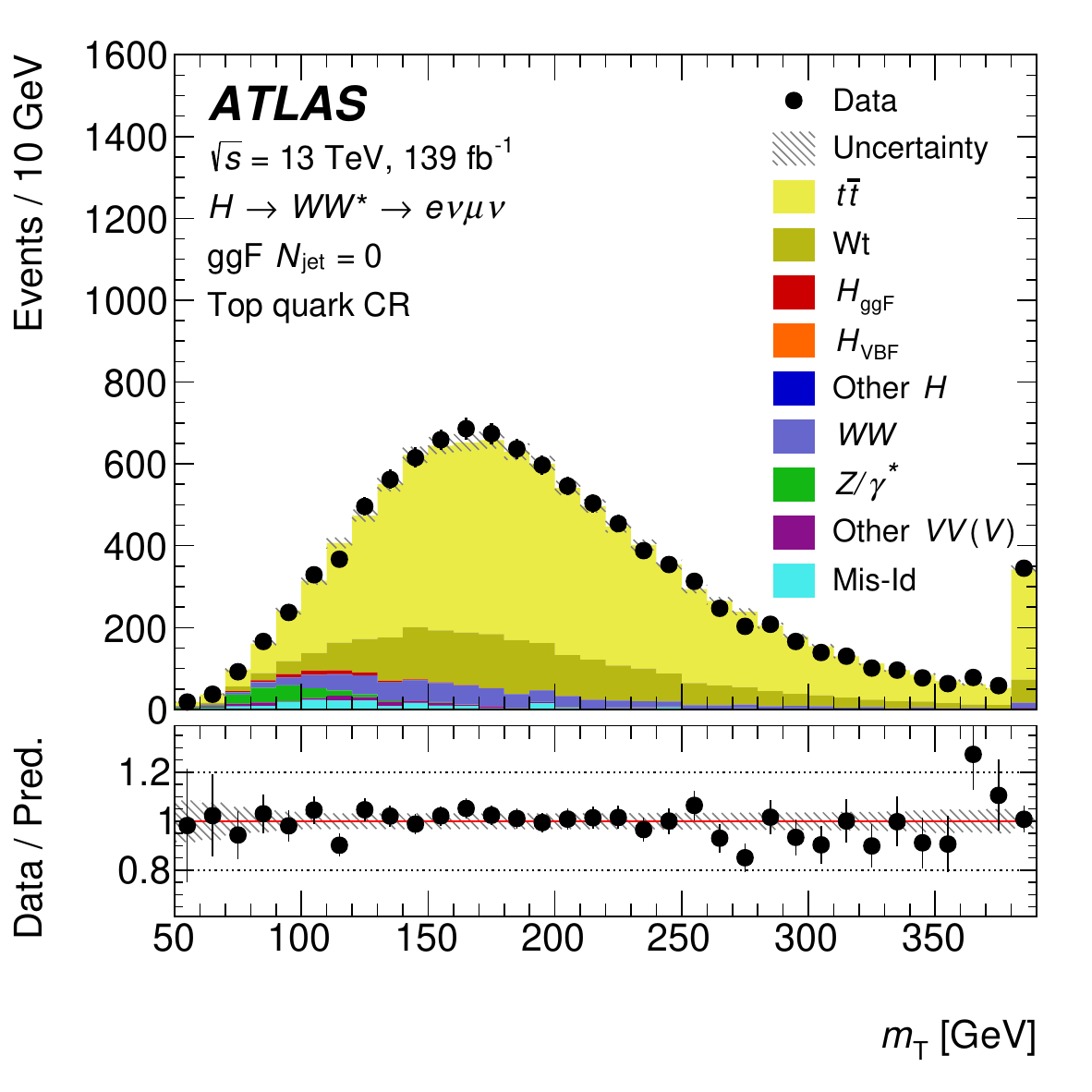}
}
\subfloat[]{
\includegraphics[width=0.5\textwidth]{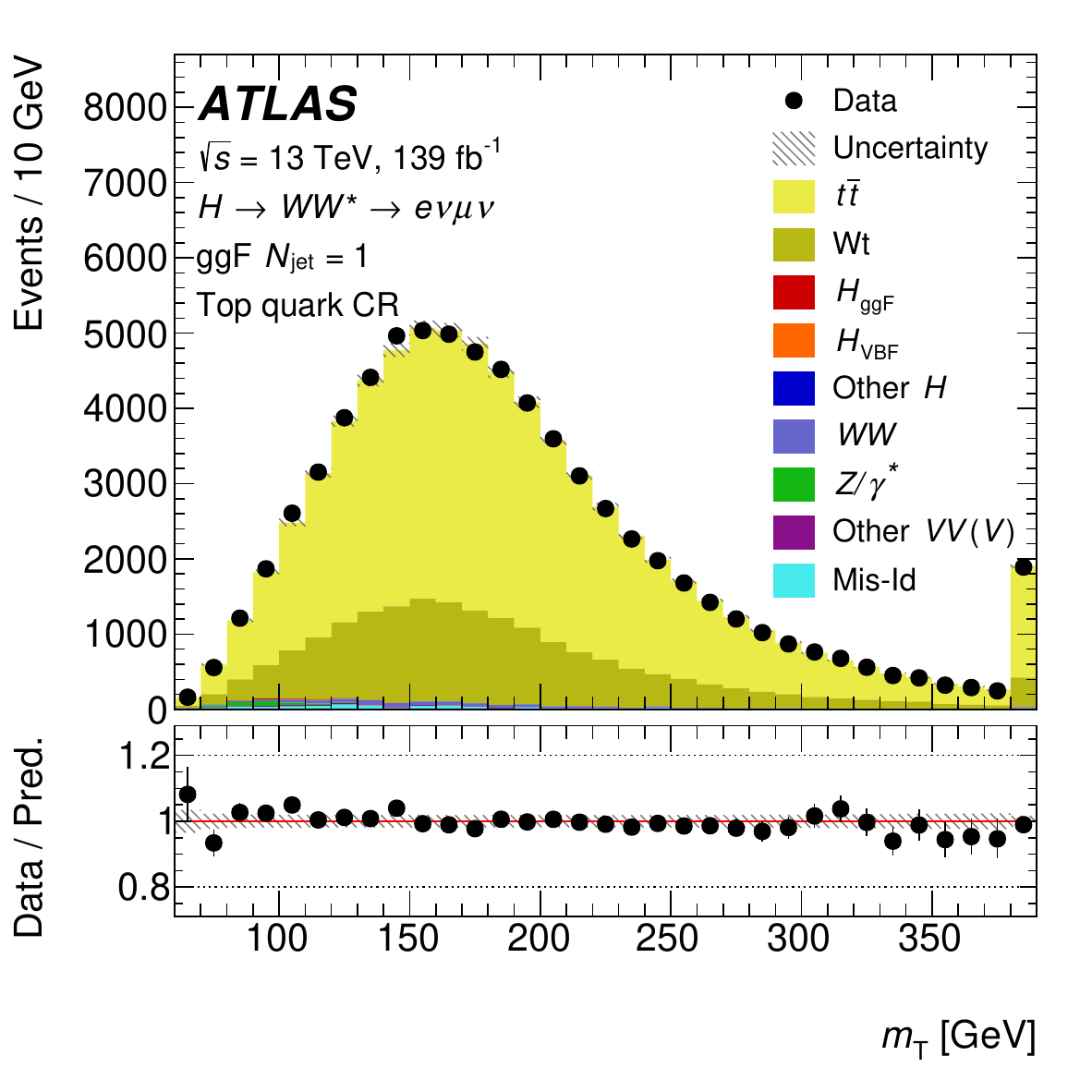}
} \\
\subfloat[]{
\includegraphics[width=0.5\textwidth]{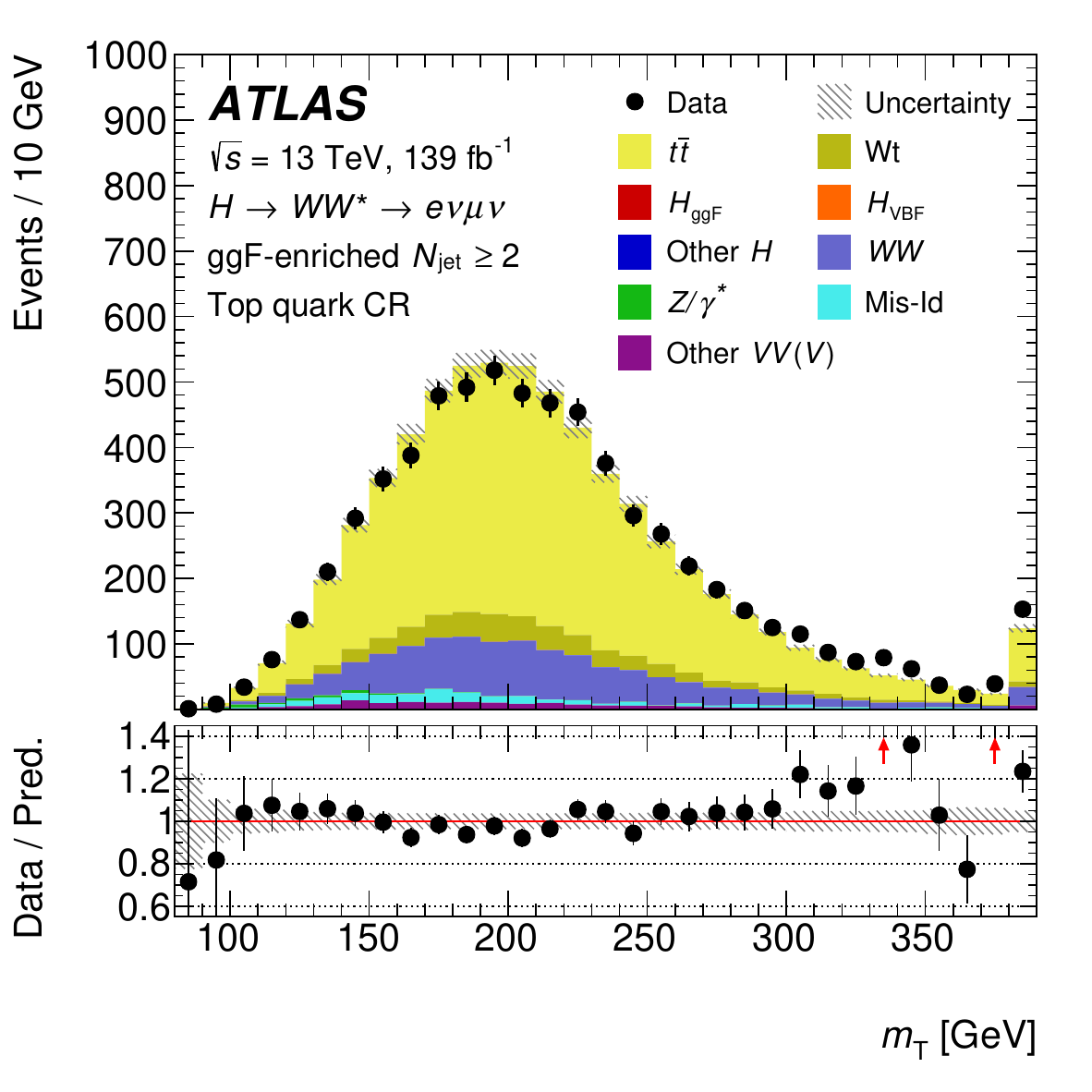}
}
\subfloat[]{
\includegraphics[width=0.5\textwidth]{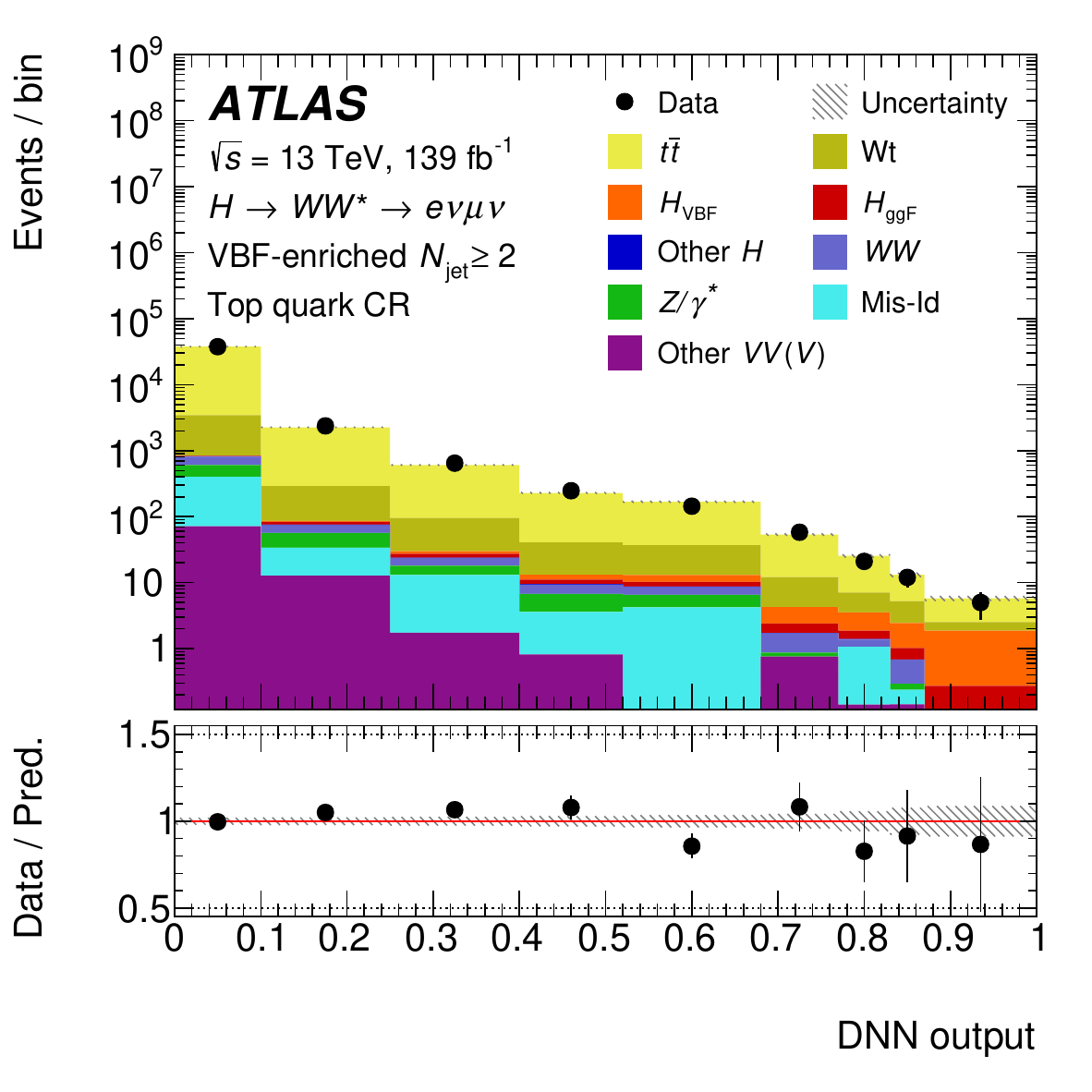}
}
\caption{
Postfit \mT distributions in the (a) \ZeroJet, (b) \OneJet, and (c) ggF-enriched \TwoJet top-quark CRs, as well as (d) the postfit DNN output distribution in the VBF-enriched \TwoJet top-quark CR, with signal (normalized to postfit measurement) and background modeled contributions.
The first two bins of the DNN output distribution used in the final fit, $[0$--$0.25,0.25$--$0.52]$, are split here into four bins, $[0$--$0.1,0.1$--$0.25,0.25$--$0.4,0.4$--$0.52]$, to illustrate the shape of the steeply falling background.
The red arrows in the lower panel of (c) indicate that the central value of the data lies above the window. The last bin of each \mT distribution is inclusive (includes the overflow).
The hatched band shows the total uncertainty, assuming SM Higgs boson production.
Some contributions are too small to be visible.
\label{fig:Top_CRs}
}
\end{figure}
 
\subsection{\Ztt background}
\label{sec:bkg_Ztt}
 
The \Ztt background is normalized in dedicated CRs, defined separately for each analysis category. For the \ZeroJet category, the reconstructed leptons are required to have a large opening angle, $\dphill > 2.8$. For the \OneJet and \TwoJet categories, the main selection that separates the \Ztt CR from the SR is that the \Ztt CR includes the region in \mtt around the nominal $Z$~boson mass. For all \Ztt CRs, a \bjet veto is maintained. These CRs have prefit \Ztt process purities of 94\% (\ZeroJet), 76\% (\OneJet), 76\% (ggF-enriched \TwoJet), and 77\% (VBF-enriched \TwoJet). The postfit background normalization factors are summarized in Table~\ref{tab:CRs_NF}.
Figure~\ref{fig:Ztt_CRs} presents the postfit \mt distributions in the \ZeroJet, \OneJet, and ggF-enriched \TwoJet CRs as well as the postfit DNN output distribution in the VBF-enriched \TwoJet CR.
 
\begin{figure}[htp]
\centering
\subfloat[]{
\includegraphics[width=0.5\textwidth]{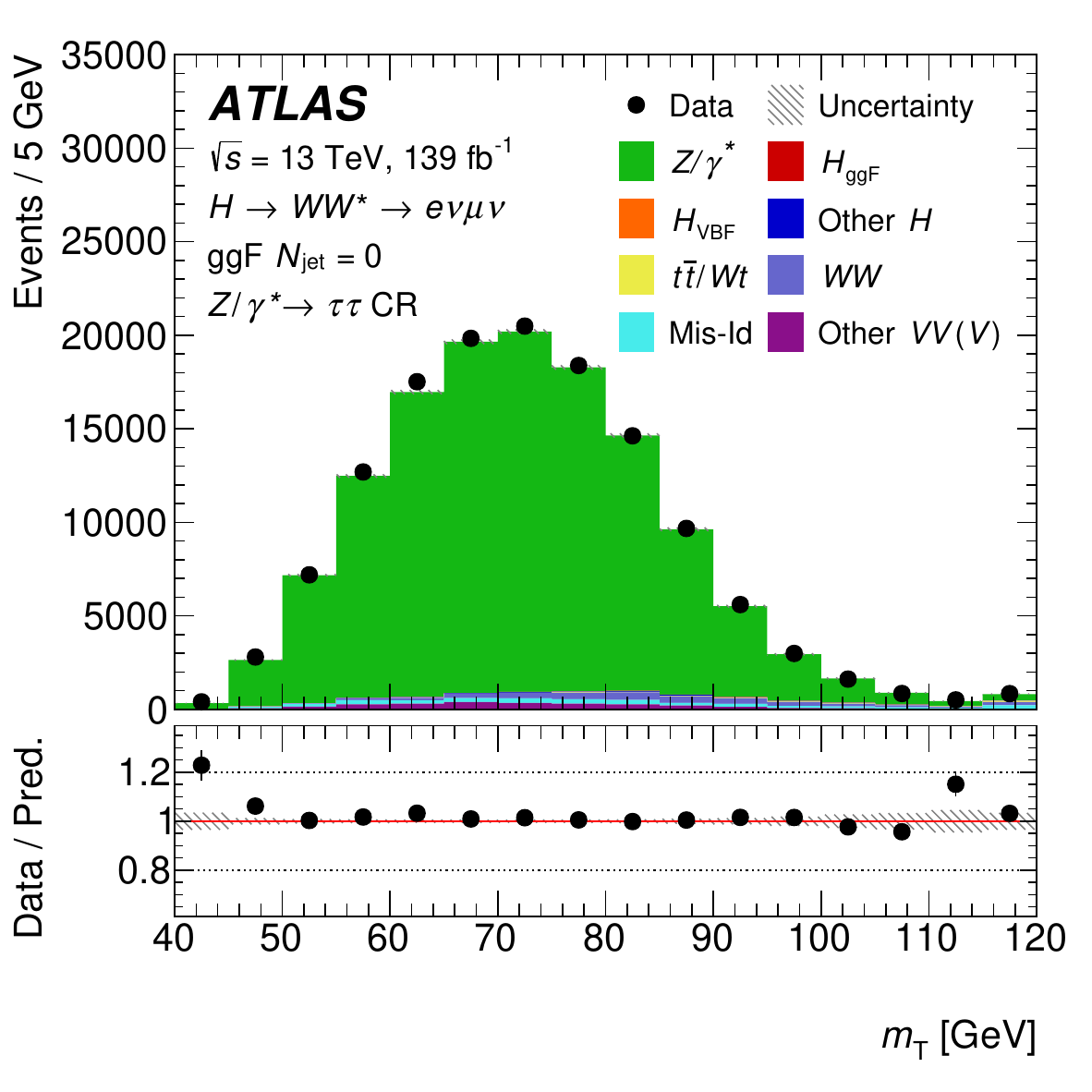}
}
\subfloat[]{
\includegraphics[width=0.5\textwidth]{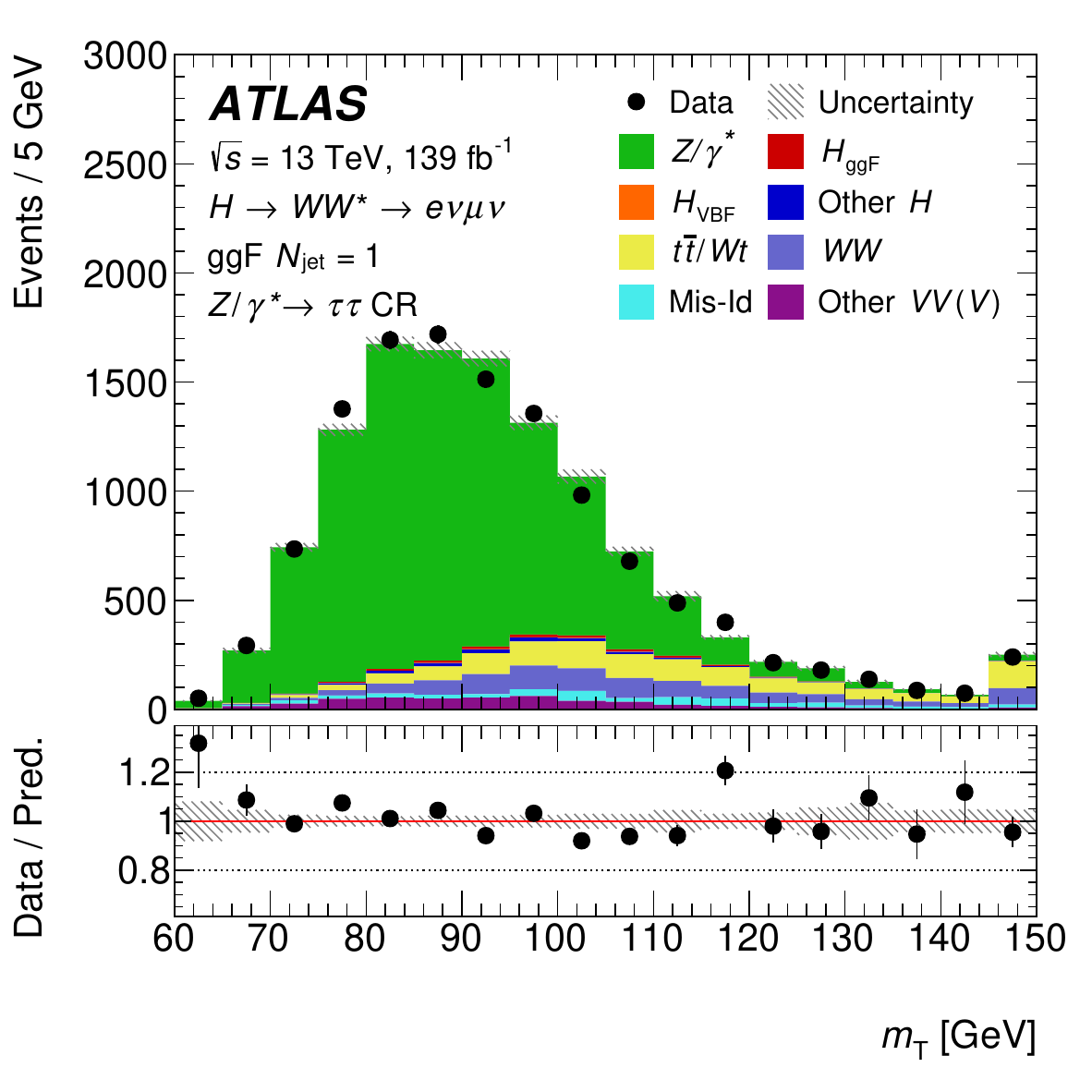}
} \\
\subfloat[]{
\includegraphics[width=0.5\textwidth]{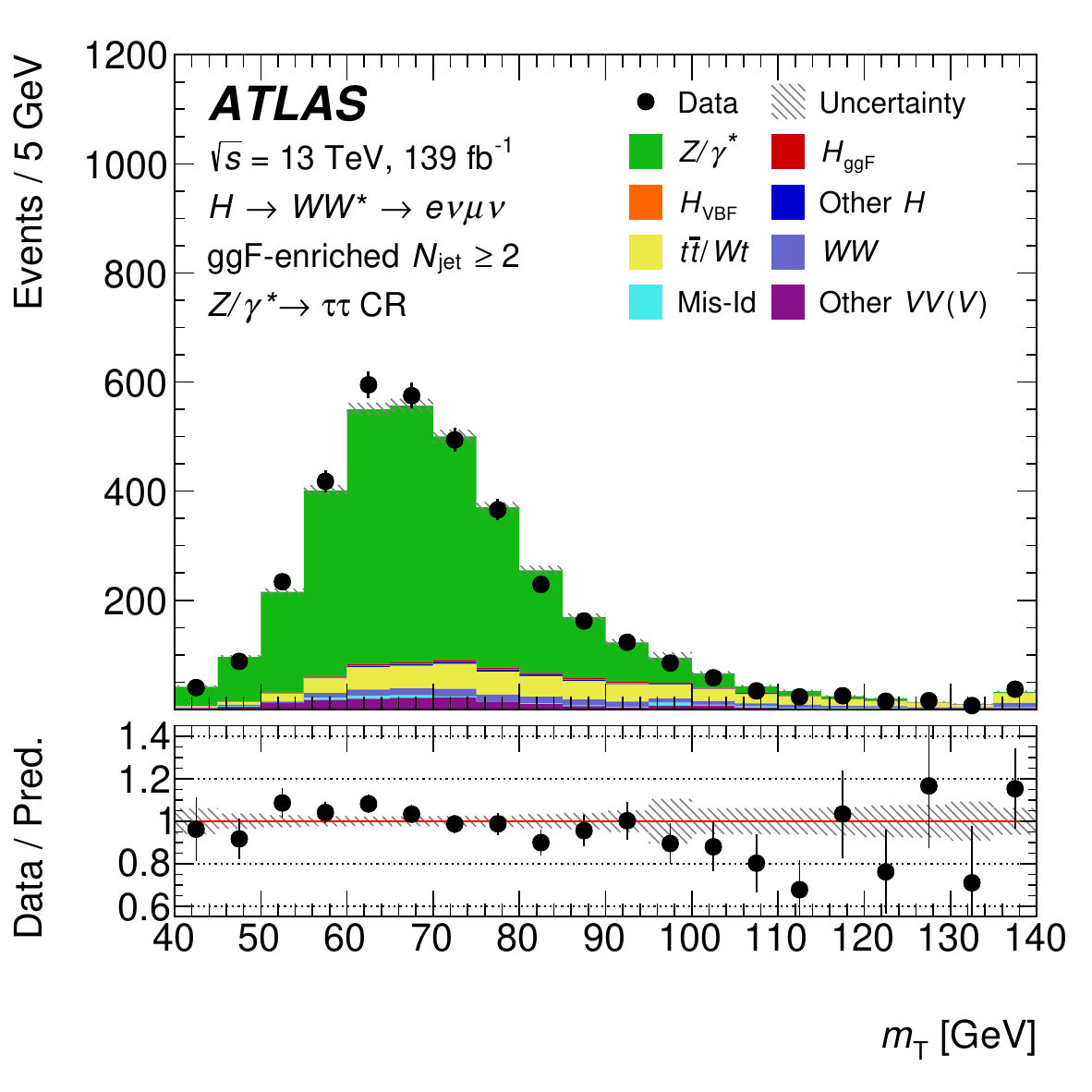}
}
\subfloat[]{
\includegraphics[width=0.5\textwidth]{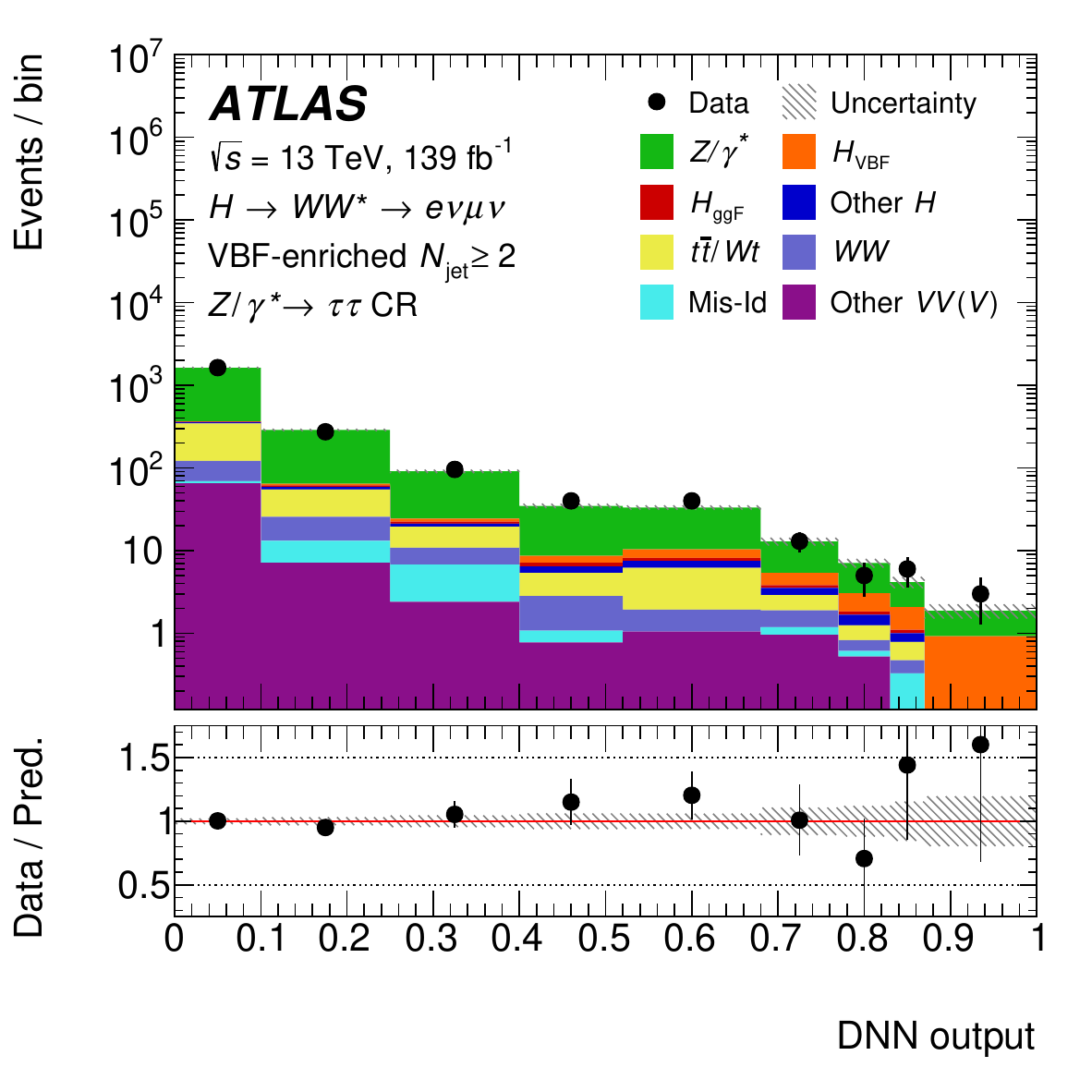}
}
\caption{
Postfit \mT distributions in the (a) \ZeroJet, (b) \OneJet, and (c) ggF-enriched \TwoJet \Ztt CRs, as well as (d) the postfit DNN output distribution in the VBF-enriched \TwoJet \Ztt CR, with signal (normalized to postfit measurement) and background modeled contributions.
The first two bins of the DNN output distribution used in the final fit, $[0$--$0.25,0.25$--$0.52]$, are split here into four bins, $[0$--$0.1,0.1$--$0.25,0.25$--$0.4,0.4$--$0.52]$, to illustrate the shape of the steeply falling background.
The last bin of each \mT distribution is inclusive (includes the overflow).
The hatched band shows the total uncertainty, assuming SM Higgs boson production.
Some contributions are too small to be visible.
\label{fig:Ztt_CRs}
}
\end{figure}
 
\begin{table}[!h]
\centering
\renewcommand{\arraystretch}{1.6}
\caption{
Postfit normalization factors which scale the corresponding estimated yields in the relevant signal region; the dash indicates where a MC-based normalization is used.
The quoted uncertainties include both the statistical and systematic contributions.
}
\label{tab:CRs_NF}
\scalebox{1.0}{
\begin{tabular}{c c c c}
\hline\hline
Category      & $WW$           &   $t\bar{t}/Wt$   & $Z/\gamma^{\ast}$   \\
\hline\hline
\ZeroJet ggF    &  $1.02^{+0.07}_{-0.07}$ &  $0.93^ {+0.22}_{-0.17}$ & $0.96^ {+0.07}_{-0.06}$ \\
\OneJet ggF    &  $0.85^{+0.16}_{-0.15}$ &  $1.05^ {+0.19}_{-0.16}$ & $0.98^ {+0.10}_{-0.09}$ \\
\TwoJet ggF  &  $0.81^{+0.34}_{-0.33}$ &  $0.96^ {+0.23}_{-0.18}$ & $0.98^ {+0.18}_{-0.17}$ \\
\TwoJet VBF  &                       &  $0.92^ {+0.33}_{-0.21}$ & $0.93^ {+0.23}_{-0.19}$ \\
\noalign{\vskip 1mm}
\hline\hline
\end{tabular}
}
\end{table}
 
\subsection{Backgrounds with misidentified leptons}
\label{sec:bkg_misid}
 
The backgrounds originating from either one or two misidentified (Mis-Id) leptons are primarily due to \Wjets and multijet processes, respectively. They are estimated using a data-driven technique\footnote{The method to estimate the Mis-ID backgrounds is based on previous ggF+VBF \hww analyses by ATLAS~\cite{ATLAS:2014aga,Aaboud:2018jqu}.} where control samples are established in which all nominal selections are applied with the exception that one of the two lepton candidates fails to meet all of the identification criteria defined in Sec.~\ref{sec:EventReconstruction}, but satisfies a looser set of identification criteria (referred to as an anti-identified lepton).
The expected Mis-Id background yields in the signal and control regions are extrapolated from the observed number of events in the corresponding samples with anti-identified leptons, after subtracting the expected contribution from processes with two prompt leptons. The method appropriately accounts for all processes with misidentified leptons, as long as they are represented in the sample with one anti-identified lepton and the extrapolation factor is similar to the nominal value.
The small contribution from multijet processes with two misidentified leptons is accounted for in the extrapolation by applying a correction term evaluated in a sample where both lepton candidates are anti-identified. The correction is largest in the VBF-enriched \TwoJet category, for which a direct \pTmiss selection is not applied. In this case, the multijet processes constitute approximately 25\% of the total misidentified lepton yield in the SR.
 
The extrapolation factor that is used to extrapolate the expected Mis-Id background yields in the control samples to the SRs is determined in a sample of $Z$+jets-enriched events, where a three-lepton selection is applied to target events with a leptonically decaying $Z$ boson plus a misidentified lepton candidate recoiling against the $Z$ boson.
It is defined as the ratio of the number of events in which the misidentified lepton candidate is identified to the number of events in which it is anti-identified and is measured in bins of lepton \pt (and $|\eta|$) in the case of electrons (muons).
A correction factor is used to account for the fact that the sources of misidentified leptons, such as hadrons, nonprompt leptons from heavy-flavor decays, and photons, contribute in different ratios
to the $Z$+jets-enriched sample in which the extrapolation factor is derived and the largest source of Mis-Id background in the SR (\Wjets events).
This sample composition correction factor is determined from the ratio of extrapolation factors measured in \Wjets and $Z$+jets MC simulation.
 
\subsection{Control regions for the STXS measurements}
\label{sec:STXS_CRs}
 
For the cross-section measurements in the STXS framework, the CRs defined above for the $WW$, top-quark, and \Ztt processes are further split into smaller CRs corresponding to the various STXS SRs. For the \ZeroJet CRs, no further splitting is needed and identical CRs are used. For the \OneJet category, the CRs for the $WW$, top-quark, and \Ztt processes are each divided into four regions defined by $\pTH < 60$~\GeV, $60 \le \pTH < 120$~\GeV, $120 \le \pTH < 200$~\GeV, and $\pTH \ge200$~\GeV. The CRs for the $WW$, top-quark, and \Ztt processes in the ggF-enriched \TwoJet category are further divided into regions with $\pTH < 200$~\GeV\ and $\pTH \ge200$~\GeV. The $\pTH \ge200$~\GeV\ STXS category targeting the ggF production mode is common to all jet multiplicities. For the VBF-enriched \TwoJet category, the top-quark and \Ztt CRs are split into three regions each.
Two regions are defined for $\pTH < 200$~\GeV\ by $350 \le \mjj < 700$~\GeV\ and $\mjj \ge 700$~\GeV, while one region is defined for $\pTH \ge200$~\GeV.


\section{Systematic uncertainties}
\label{sec:Systematics}

 
Uncertainties from both experimental and theoretical sources affect the results of the analysis.
This section describes the estimation of their effects on the signal and background normalizations as well as, where applicable, their effects on the shape of the final discriminant.
The relative impacts that various sources of systematic uncertainties have on the measured ggF and VBF cross sections are obtained from the likelihood fit described in Sec.~\ref{sec:FitProcedure} and are listed in Table~\ref{tab:UncertaintyBreakdown_2-POI}.
 
\subsection{Experimental uncertainties}
\label{sec:ExperimentalUncertainties}
 
The uncertainties related to the reconstruction of the objects used in the analysis are determined using data-driven methods
on high-statistics samples of processes such as $Z\to\ell\ell$.
Uncertainties associated with the selected leptons originate from the reconstruction and identification efficiency, the energy (or momentum) scale and resolution, and the isolation efficiency~\cite{EGAM-2018-01,ATLAS:2020auj}. For jets, uncertainties arise from the jet energy scale and resolution~\cite{PERF-2016-04}, the jet-vertex tagger's performance, and the $b$-jet identification~\cite{FTAG-2018-01}. Furthermore, uncertainties due to the trigger selection~\cite{TRIG-2018-05,TRIG-2018-01} and the soft term in the reconstruction of \met~\cite{PERF-2016-07} are estimated. The uncertainty in the modeling of pileup for simulated samples is estimated by varying the reweighting to the profile in data within its uncertainties. The uncertainty in the combined 2015--2018 integrated luminosity is 1.7\% \cite{ATLAS-CONF-2019-021}, obtained using the LUCID-2 detector~\cite{Avoni:2018iuv} for the primary luminosity measurements. The integrated luminosity uncertainty is only applied to the Higgs boson signal and to background processes that are normalized to theoretical predictions.
 
Three sources of uncertainty related to the extrapolation factor used in the data-driven Mis-Id background estimate are considered: the statistical uncertainty of the extrapolation factor itself, an uncertainty related to the subtraction of processes with two prompt leptons from the $Z$+jets-enriched sample used to derive the extrapolation factor, and an uncertainty in the sample composition correction factor. Together, they amount to a total uncertainty on the electron (muon) extrapolation factor ranging from $10\%$ ($12\%$) at low $\pt$ to $35\%$ ($75\%$) at high $\pt$ where there is a small number of Mis-Id leptons.
 
The largest experimental uncertainties affecting the ggF measurement come from the $b$-jet identification, the pileup modeling, the jet energy resolution, and the Mis-Id background estimate. For the VBF measurement, the largest experimental uncertainty comes from the \met reconstruction.

\subsection{Theoretical uncertainties}
\label{sec:TheoreticalUncertainties}
 
Uncertainties from the renormalization and factorization scale choices, underlying-event modeling, and choice of PDF are estimated for all processes.
For signal, top, and $Z/\gamma^{\ast}$ processes, the uncertainties from the parton shower and the matrix-element matching are estimated by comparing predictions from the nominal and alternative generators that are described in Sec.~\ref{sec:MonteCarloSamples}.
For the prediction of $qqWW$ and of $WZ$, $ZZ$, and $V\gamma^{\ast}$ production ($VV$), variations of the matching scale and nonperturbative effects are considered instead of an alternative program for estimating the matrix-element matching uncertainties.
The uncertainty from the resummation scale is estimated for the \SHERPA samples.
 
For signal processes, the approach described in Refs.~\cite{deFlorian:2016spz,Stewart:2011cf} is used to estimate the variations due to the impact of higher-order contributions not included in the calculations and of migration effects on the \Njets ggF cross sections.
In particular, the uncertainties from the choice of factorization and renormalization scales, the choice of resummation scales, and the ggF migrations between the 0-jet and 1-jet phase-space bins or between the 1-jet and $\ge$ 2-jet bins are considered~\cite{deFlorian:2016spz,Stewart:2013faa,Liu:2013hba,Boughezal:2013oha,Gangal:2013nxa}.
 
The $ggWW$ process is simulated at LO precision for up to one additional parton emission.
Therefore, a conservative $-50\%/+100\%$ normalization uncertainty is assigned to this process for the \TwoJet categories and to the STXS measurement in the region with $\pTH \ge200$~\GeV\ targeting the ggF production mode.
A similar $-50\%/+100\%$ normalization uncertainty is also assigned to the $V\gamma$ sample, due to a mismodeling of the $\gamma \to e$ misidentification rate which primarily affects events with $\mt \lesssim 80$~\GeV\ and can be seen in Fig.~\ref{fig:ggF_MT}(a).
The EW $WW$ process, which contributes most significantly in the highest VBF DNN bin, is assigned an additional normalization uncertainty of $15\%$ due to NLO EW corrections, as calculated using the leading-log approximation~\cite{Biedermann_2017,Denner_2001}.
For $Wt$, an additional uncertainty estimated by comparing samples with different diagram removal schemes~\cite{Frixione:2008yi} is applied.
A normalization uncertainty of $12\%$, as estimated in the sample with a three-lepton selection described in Sec.~\ref{sec:bkg_misid}, is also applied to the non-$WW$ and $VVV$ backgrounds.
For backgrounds which are normalized to CR yields, uncertainties are estimated for the CR-to-SR extrapolation factors.
Only uncertainties that change the ratios of SR yields to CR yields affect the extrapolation.
The uncertainties in the extrapolation factors are treated as uncorrelated between different jet multiplicities and between the CRs for the STXS measurement.
 
The uncertainties in the STXS measurements are estimated for each SR separately, where the 11 STXS bins are treated as different processes, and cover the migration of events between STXS SRs. Merged SRs are used to determine the uncertainties when only a small number of events are available in the simulated samples in a particular SR.
 
The largest theoretical uncertainties affecting the ggF signal come from the measurement of exclusive jet multiplicities and from the parton shower. For the VBF signal, the comparisons of different event generators for the matrix-element matching and for the parton shower result in the largest uncertainties in the measurement. For background processes, the theoretical uncertainties in the $WW$ and top-quark backgrounds result in the largest contributions to the overall uncertainty.


\section{Fit procedure}
\label{sec:FitProcedure}

 
Results are obtained from a profile likelihood fit~\cite{Cowan:2010js} to data in the signal and control regions. Uncertainties enter the fit as nuisance parameters in the likelihood function. Theoretical uncertainties affecting the signal and the experimental uncertainties affecting both signal and background are in general correlated between signal and control regions in all analysis categories. Theoretical uncertainties in the backgrounds and the background normalization factors are uncorrelated between different analysis categories.
 
The \mt distribution is used as the final fit discriminant in each of four regions defined by \mll and subleading lepton \pt in both the \ZeroJet and \OneJet categories, as described in Secs.~\ref{sec:ZeroJetCategory} and~\ref{sec:OneJetCategory}. The same binning of the \mt distribution is used in all regions: $[0$--$90,90$--$100,100$--$110,110$--$120,120$--$130,130$--$\infty]$. The SR in the ggF-enriched \TwoJet category is split into two bins of \mll, but there is no split in subleading lepton \pt. In both regions, the \mt distribution is divided into six bins with the same boundaries as for the \ZeroJet and \OneJet categories. For the VBF-enriched \TwoJet category, the DNN output is used as the discriminating fit variable.
The distribution is divided into seven bins with the boundaries defined in Sec.~\ref{sec:VBF-TwoJetCategory}.
 
For the STXS measurements, two modifications are made: the four STXS regions in the \OneJet category are no longer split into bins defined by \mll and subleading lepton \pt, while the STXS measurements targeting the VBF production mode define four bins for the DNN output with boundaries $[0$--$0.5,0.5$--$0.7,0.7$--$0.84,0.84$--$1.00]$.
 
The cross sections for the ggF and VBF production modes are determined in a simultaneous fit to all nominal SRs and CRs in the \ZeroJet, \OneJet, and \TwoJet categories. The ggF and VBF cross sections are the two unconstrained POIs in this fit. A second fit is performed using these same regions, but measuring a single POI for the combined ggF and VBF yield. In both fits, the other Higgs boson production modes are fixed to their expected yields. A third fit is made to all the STXS regions, where the 11 cross sections measured are POIs.
No nuisance parameters are significantly pulled or constrained in any of the fits.


\section{Signal region yields and results}
\label{sec:Results}


Table~\ref{tab:ggF_VBF_yields} shows the postfit SR yields for all of the four analysis categories defined in Sec.~\ref{sec:EventSelection}. The uncertainty in the total expected yield reflects incomplete knowledge of the observed yield in each analysis category and is not indicative of the precision of the analysis.
Furthermore, the relative error in the yields of the background processes for which dedicated CRs are defined is in many cases less than the relative error in the corresponding normalization factor displayed in Table~\ref{tab:CRs_NF} due to effects of anticorrelation with some nuisance parameters modeling theory uncertainties.

\begin{table}[htb]
\centering
\caption{
Postfit MC and data yields in the ggF and VBF SRs. Yields in the bin with the highest VBF DNN output are also presented.
The quoted uncertainties correspond to the statistical uncertainties, together with the experimental and theory modeling systematic uncertainties.
The sum of all the contributions may differ from the total value due to rounding.
Moreover, the uncertainty in the total yield differs from the sum in quadrature of the single-process uncertainties due to the effect of anticorrelations between the sources of their systematic uncertainties, which are larger than their MC statistical uncertainties.
\label{tab:ggF_VBF_yields}
}
\scalebox{0.8}{
\begin{tabular}{lS[table-format=5,table-number-alignment=right]@{$\,\pm\,$}
S[table-format=3,table-number-alignment=left]
S[table-format=5,table-number-alignment=right]@{$\,\pm\,$}
S[table-format=3,table-number-alignment=left]
S[table-format=5,table-number-alignment=right]@{$\,\pm\,$}
S[table-format=3,table-number-alignment=left]
S[table-format=5,table-number-alignment=right]@{$\,\pm\,$}
S[table-format=3,table-number-alignment=left]
S[table-format=2.2,table-number-alignment=right]@{$\,\pm\,$}
S[table-format=1.2,table-number-alignment=left]}
\dbline
Process & \multicolumn{2}{c}{\ZeroJet ggF} & \multicolumn{2}{c}{\OneJet ggF} & \multicolumn{2}{c}{\TwoJet ggF} & \multicolumn{4}{c}{\TwoJet VBF}  \tabularnewline
& \multicolumn{2}{c}{}  &  \multicolumn{2}{c}{}     &     \multicolumn{2}{c}{}     &  \multicolumn{2}{c}{} &\multicolumn{2}{c}{DNN:}\tabularnewline
& \multicolumn{2}{c}{}  &  \multicolumn{2}{c}{}     &     \multicolumn{2}{c}{}     &  \multicolumn{2}{c}{Inclusive} &\multicolumn{2}{c}{$[0.87,1.0]$}\tabularnewline
\sgline
$H_{\mathrm{ggF}}$  &    2100  & 220      &  1100 & 130     &  440  & 90    &  209    & 40     & 2.6   &  0.9   \tabularnewline
$H_{\mathrm{VBF}}$  &      23  & 9        &   103 & 30      &  46   & 12      &  180    & 40     & 28.8  &  5.5   \tabularnewline
\sgline
Other Higgs         &   40     & 20        & 55    & 28       &  55   & 27      &  29     & 15      & 0.04   &  0.02   \tabularnewline
$WW$                &   9700   & 350      & 3500 & 410     &  1500 & 470    &  2100   & 340    & 4.6   &  1.2  \tabularnewline
$t\bar{t}/Wt$       &   2200   & 210      & 5300  & 340     &  6100 & 500    &  7600   & 370   & 2.6   &  0.8   \tabularnewline
$Z/\gamma^{*}$      &   140    & 50       &  280  & 40      &  930  & 70     &  1300   & 300    & 0.6   &  0.1   \tabularnewline
Other $VV$          &   1400   & 130      &  840  & 100     &  470  & 90     &  380    & 80     & 0.6   &  0.1   \tabularnewline
Mis-Id              &   1200   & 130      &  720  & 90      &  470  & 50     &  330   & 40    & 1.7   &  0.2   \tabularnewline
\sgline
Total               &   16770  & 130       & 11940 &  110   &  10030 & 100    &  12200  & 180    & 42.0  &  5.1  \tabularnewline
Observed  &  \multicolumn{2}{c}{\num{16726}}  &  \multicolumn{2}{c}{\num{11917}} & \multicolumn{2}{c}{\ \ \num[group-minimum-digits=4]{9982}} &  \multicolumn{2}{c}{\num{12189}} & \multicolumn{2}{c}{\num{38}}  \tabularnewline
\dbline
\end{tabular}
}
\end{table}

The \mT distributions for the separate \ZeroJet, \OneJet, and ggF-enriched \TwoJet SRs, as well as the combination of ggF SRs, are shown in Fig.~\ref{fig:ggF_MT}.
The bottom panels of Fig.~\ref{fig:ggF_MT} display the difference between the data and the total estimated background compared to the \mT distribution of a SM Higgs boson with $\mH = 125.09~\GeV$.
The total signal observed in all categories (see Table~\ref{tab:ggF_VBF_yields}) is about 4000 events and agrees, in both shape and rate, with the expected SM signal.
The observed (expected) signal yield in the ggF-enriched \TwoJet category, with the VBF contribution fixed to the Standard Model prediction, reaches a significance of $2.2\sigma$ ($1.6\sigma$) above the background expectation.
 
\begin{figure}[htb]
\centering
\subfloat[]{
\includegraphics[width=0.355\textwidth]{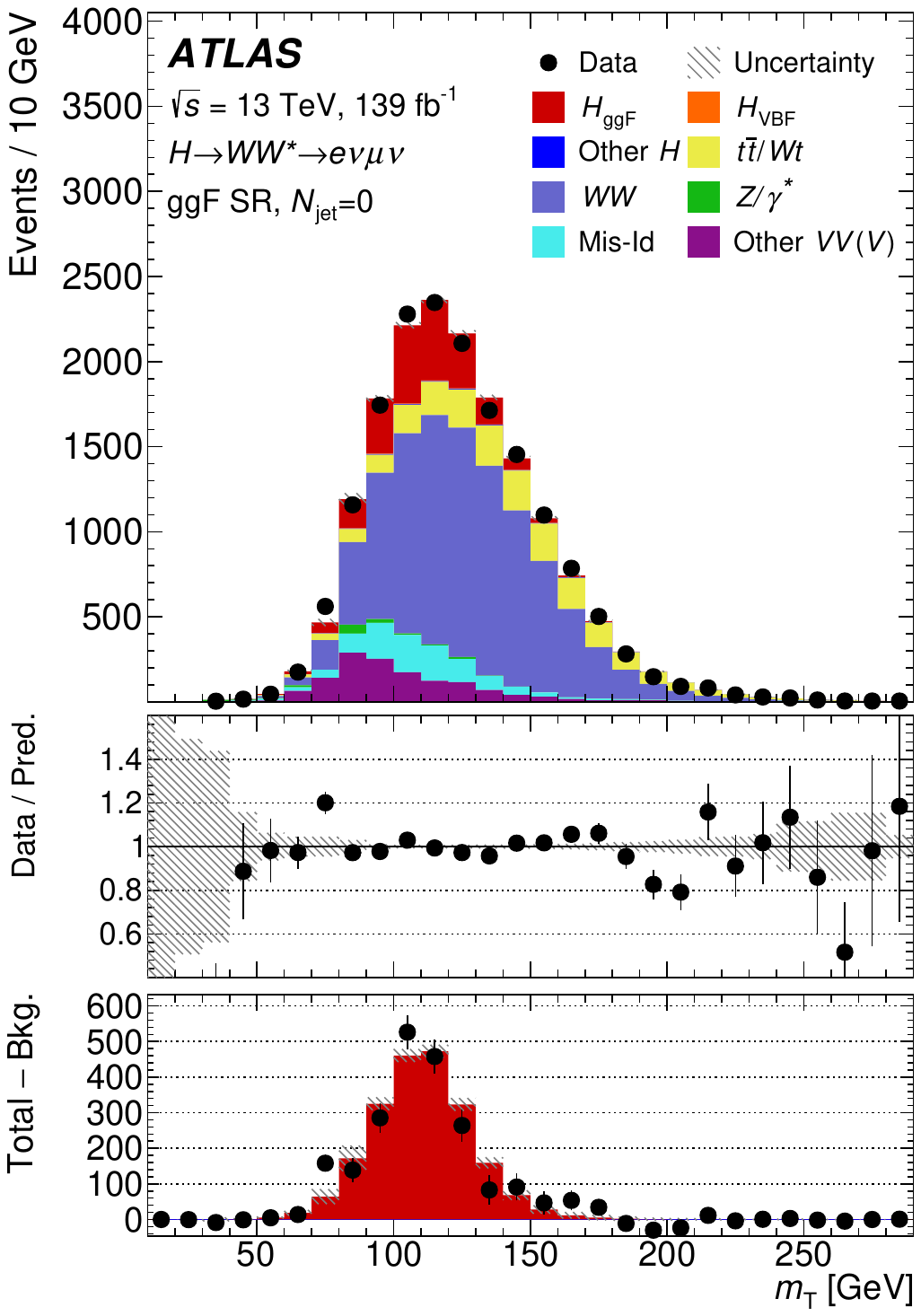}
}
\subfloat[]{
\includegraphics[width=0.355\textwidth]{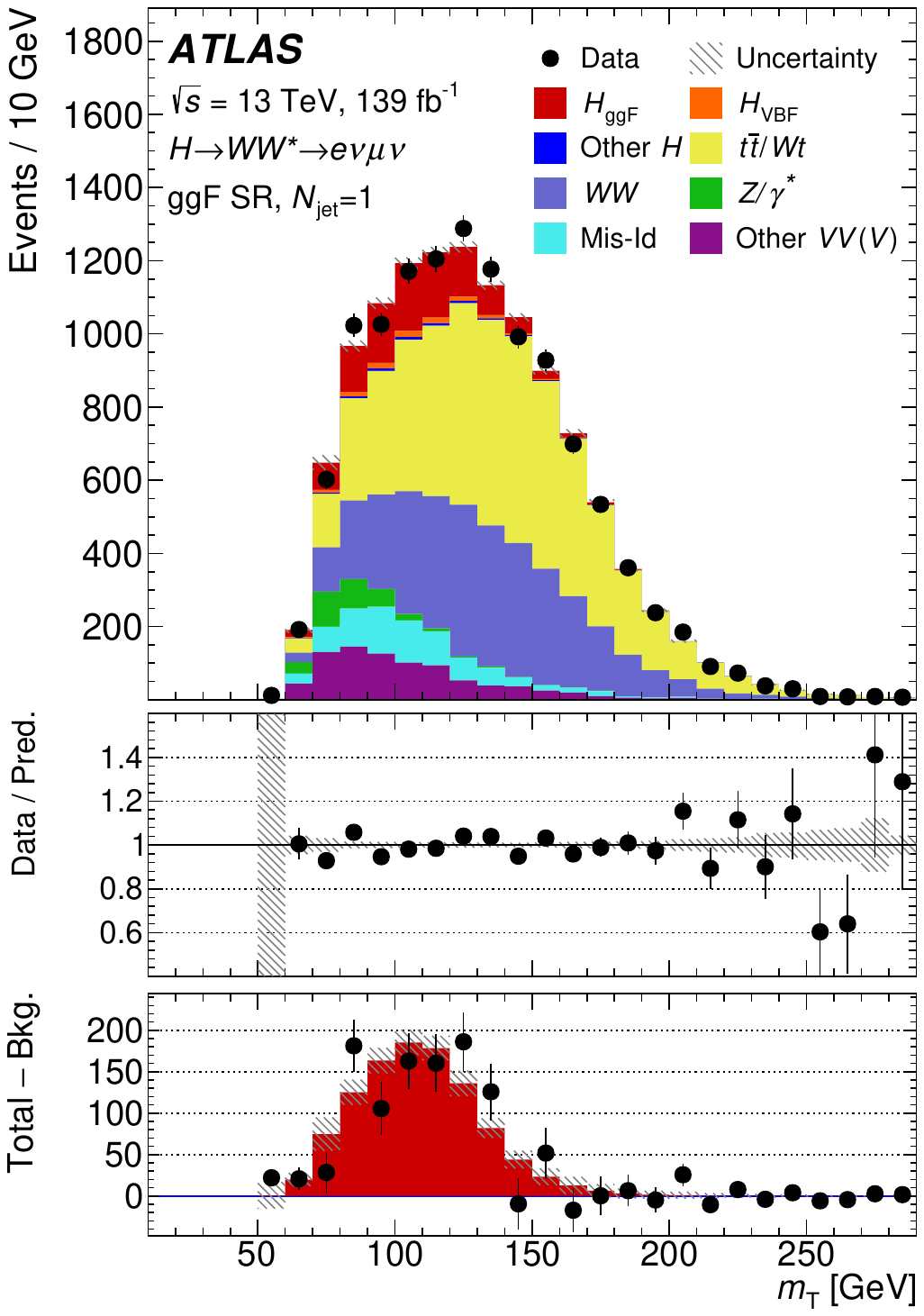}
} \\
\subfloat[]{
\includegraphics[width=0.355\textwidth]{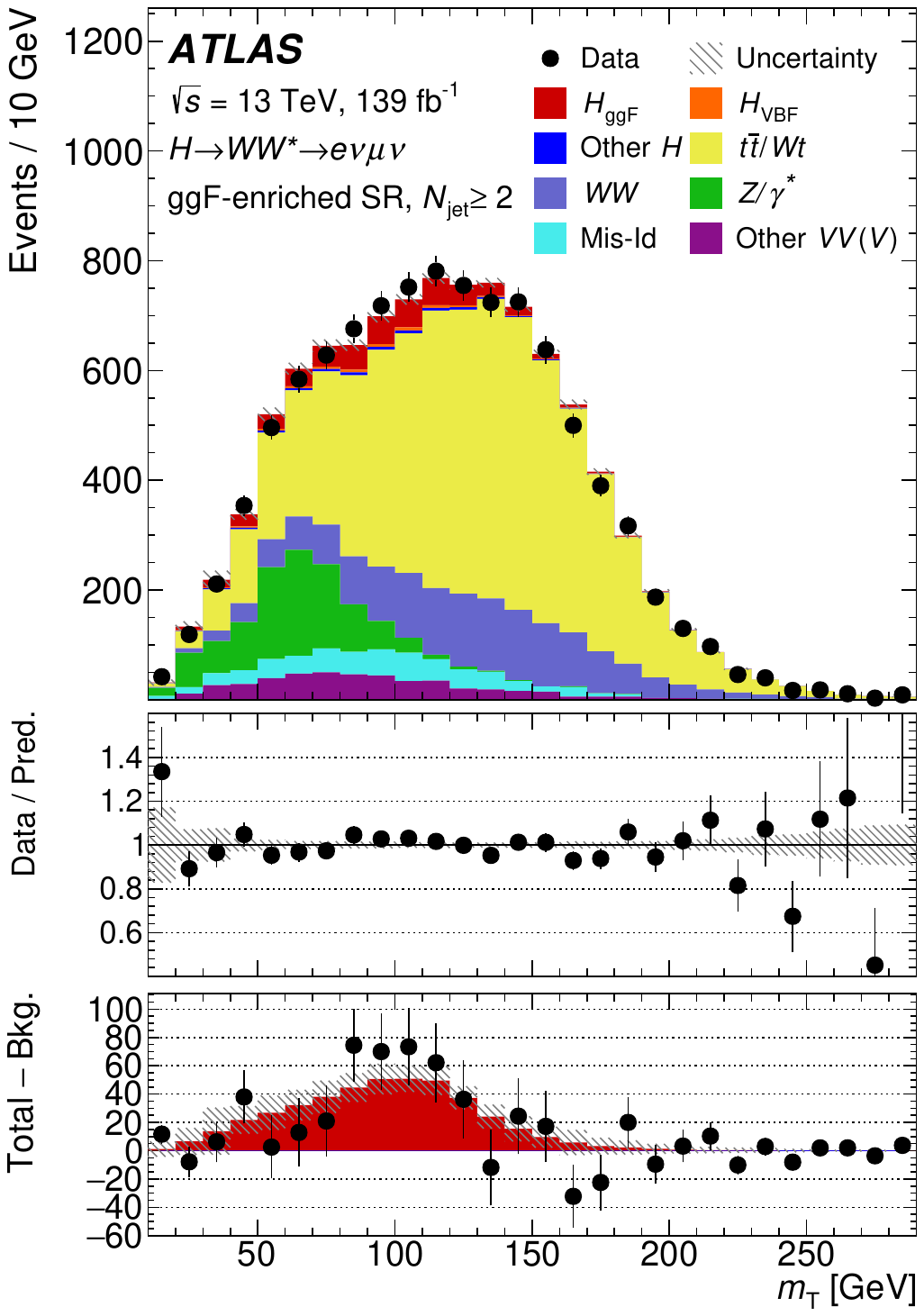}
}
\subfloat[]{
\includegraphics[width=0.355\textwidth]{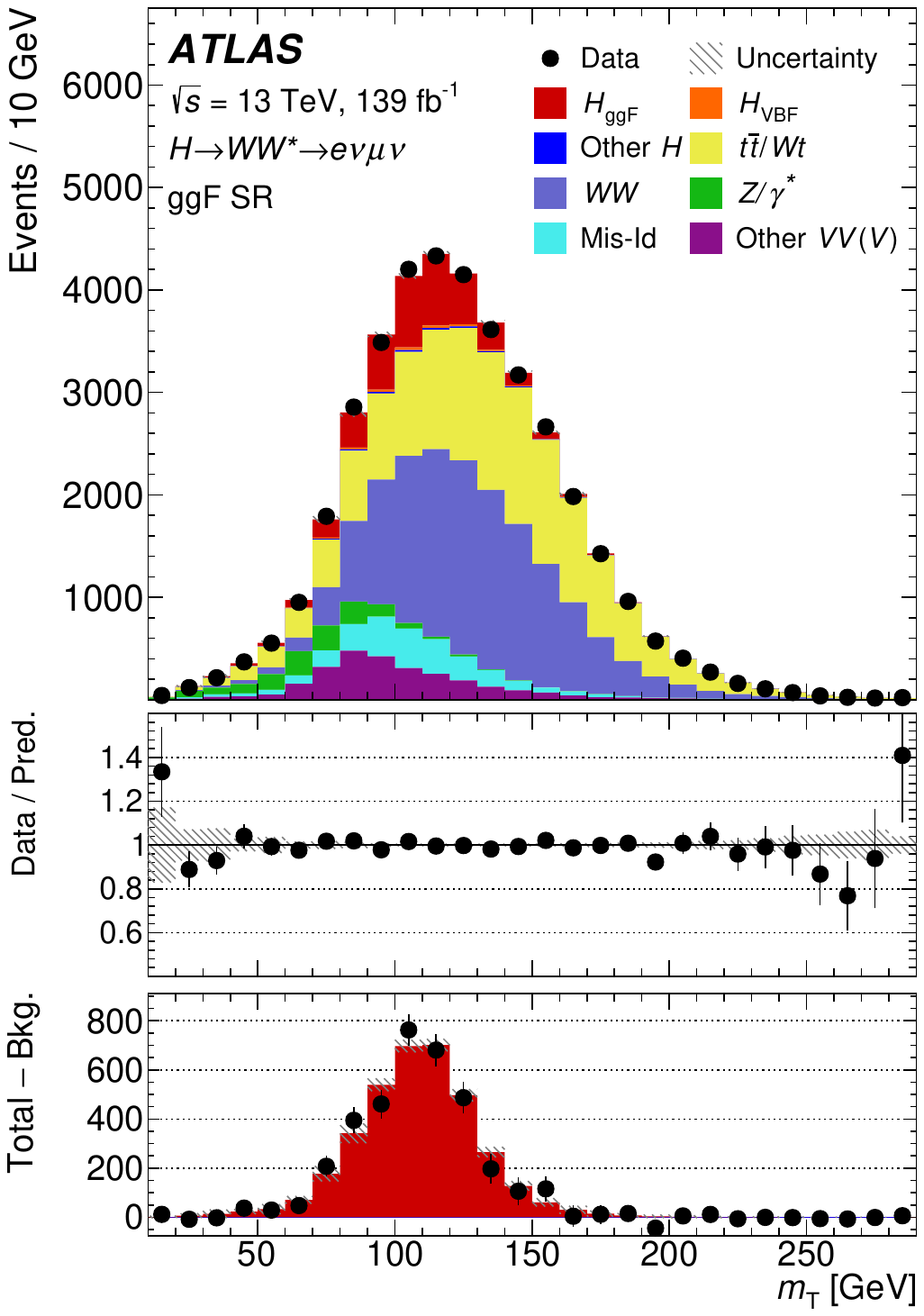}
}
\caption{
Postfit $\mT$ distributions with the signal and the background modeled contributions
in the (a) $\ZeroJet$, (b) $\OneJet$, (c) ggF-enriched \TwoJet, and (d) combined ggF signal regions. Underflow and overflow events are not included.
The hatched band shows the total uncertainty, assuming SM Higgs boson production.
The middle panel shows the ratio of the data to the sum of the fitted signal and background.
The bottom panel displays the difference between the data and the estimated background compared to the simulated signal distribution, where the hatched band indicates the combined statistical and systematic uncertainty for the fitted signal and background.
\label{fig:ggF_MT}
}
\end{figure}

The VBF DNN output distribution in the final signal region is presented in Fig.~\ref{fig:VBF_DNN}.
The observed (expected) VBF signal reaches a significance of $5.8\sigma$ ($6.2\sigma$) above the background expectation.

\begin{figure}[htb]
\centering
\includegraphics[width=0.7\textwidth]{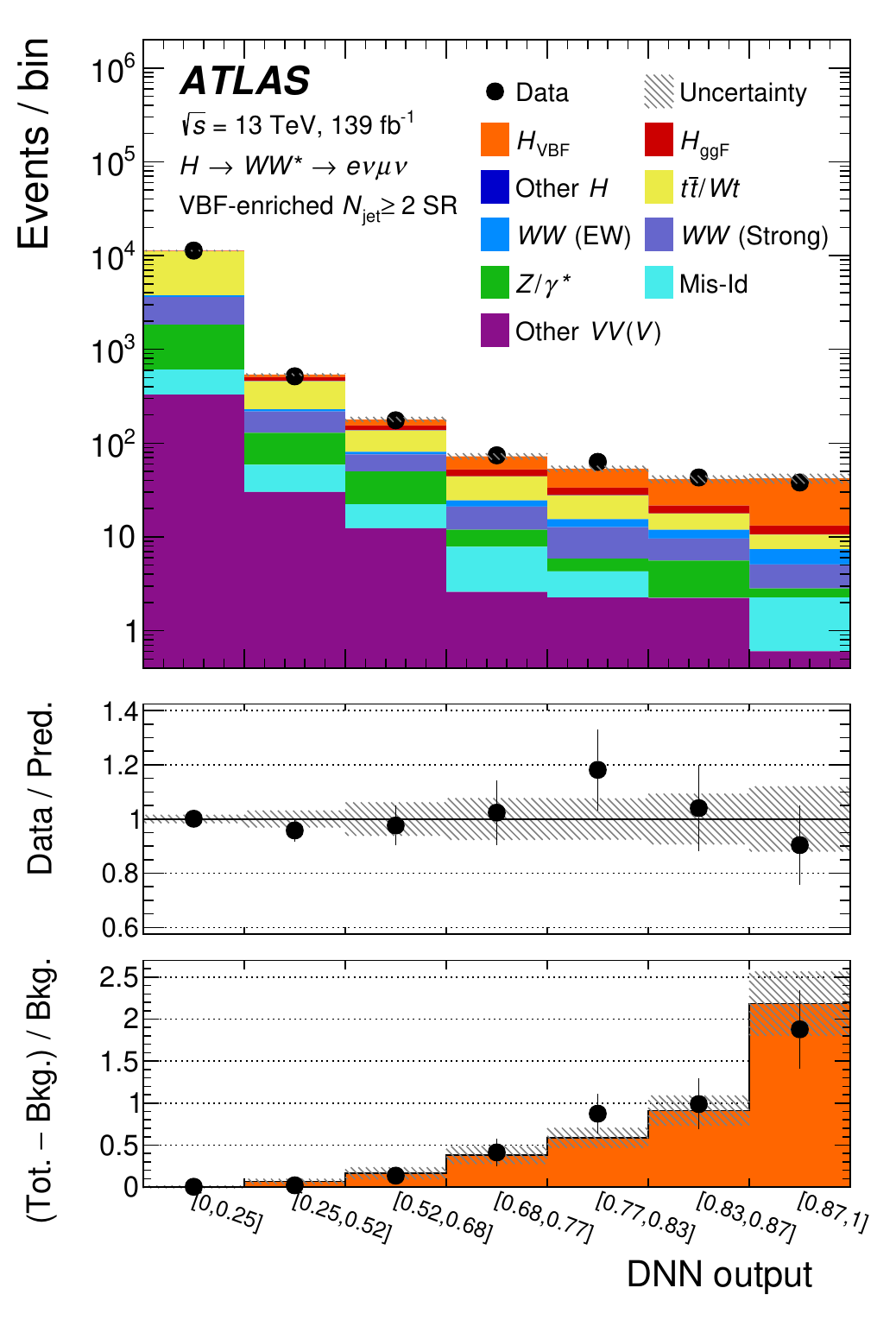}
\caption{
Postfit distribution of the DNN output in the VBF signal region.
The hatched band shows the total uncertainty, assuming SM Higgs boson production.
The middle panel shows the ratio of the data to the sum of the fitted signal and background.
The bottom panel displays the signal-to-background ratio, where the hatched band indicates the combined statistical and systematic uncertainty for the fitted signal and background.
}
\label{fig:VBF_DNN}
\end{figure}

Figure~\ref{fig:couplings-POIs} shows the best-fit values and uncertainties of the $\hww$ cross section for the ggF and VBF processes and their combination, normalized to the corresponding SM prediction.
The cross sections times branching ratio for the ggF and VBF production modes for a Higgs boson with mass $\mH=125.09$~\GeV\ in the $\hww$ decay channel, $\sigma_{\mathrm{ggF}} \cdot \mathcal{B}_{\hww}$ and $\sigma_{\mathrm{VBF}} \cdot \mathcal{B}_{\hww}$, are simultaneously measured to be
 
\begin{eqnarray*}
\sigma_{\mathrm{ggF}} \cdot \mathcal{B}_{H \to WW^{\ast}} &=& 12.0 \pm 1.4~\mathrm{pb} \\
&=& 12.0 \pm 0.6\;(\mathrm{stat.}) ^{+0.9}_{-0.8}\;(\mathrm{exp.~syst.})\;^{+0.6}_{-0.5}\;(\mathrm{sig.~theo.})\; \pm 0.8~(\mathrm{bkg.~theo.})~\mathrm{pb} \\
\sigma_{\mathrm{VBF}} \cdot \mathcal{B}_{H \to WW^{\ast}} &=& 0.75\;^{+0.19}_{-0.16}~\mathrm{pb} \\
&=& 0.75 \pm 0.11\;(\mathrm{stat.})\;^{+0.07}_{-0.06}\;(\mathrm{exp.~syst.})\;^{+0.12}_{-0.08}\;(\mathrm{sig.~theo.})\;^{+0.07}_{-0.06}\;(\mathrm{bkg.~theo.})~\mathrm{pb},
\end{eqnarray*}
 
compared to the SM predicted values of $10.4\pm 0.5$ and $0.81\pm 0.02$~pb for ggF and VBF~\cite{deFlorian:2016spz},\footnote{The uncertainties in the predicted cross sections include the uncertainty in the Higgs boson mass.} respectively.
The combined cross section times branching ratio, $\sigma_{\mathrm{ggF+VBF}} \cdot \mathcal{B}_{\hww}$, obtained from fitting a single POI, is measured to be
\begin{eqnarray*}
\sigma_{\mathrm{ggF+VBF}} \cdot \mathcal{B}_{H \to WW^{\ast}} &=& 12.3 \pm 1.3~\mathrm{pb} \\
&=& 12.3 \pm 0.6\;(\mathrm{stat.})\;^{+0.8}_{-0.7}\;(\mathrm{exp.\ syst.})\;\pm 0.6\;(\mathrm{sig.\ theo.})\;\pm 0.7\;(\mathrm{bkg.\ theo.})~\mathrm{pb},
\end{eqnarray*}
compared to the SM predicted value of $11.3\pm 0.5$~pb.
 
\begin{figure}[htb]
\centering
\includegraphics[width=0.8\textwidth]{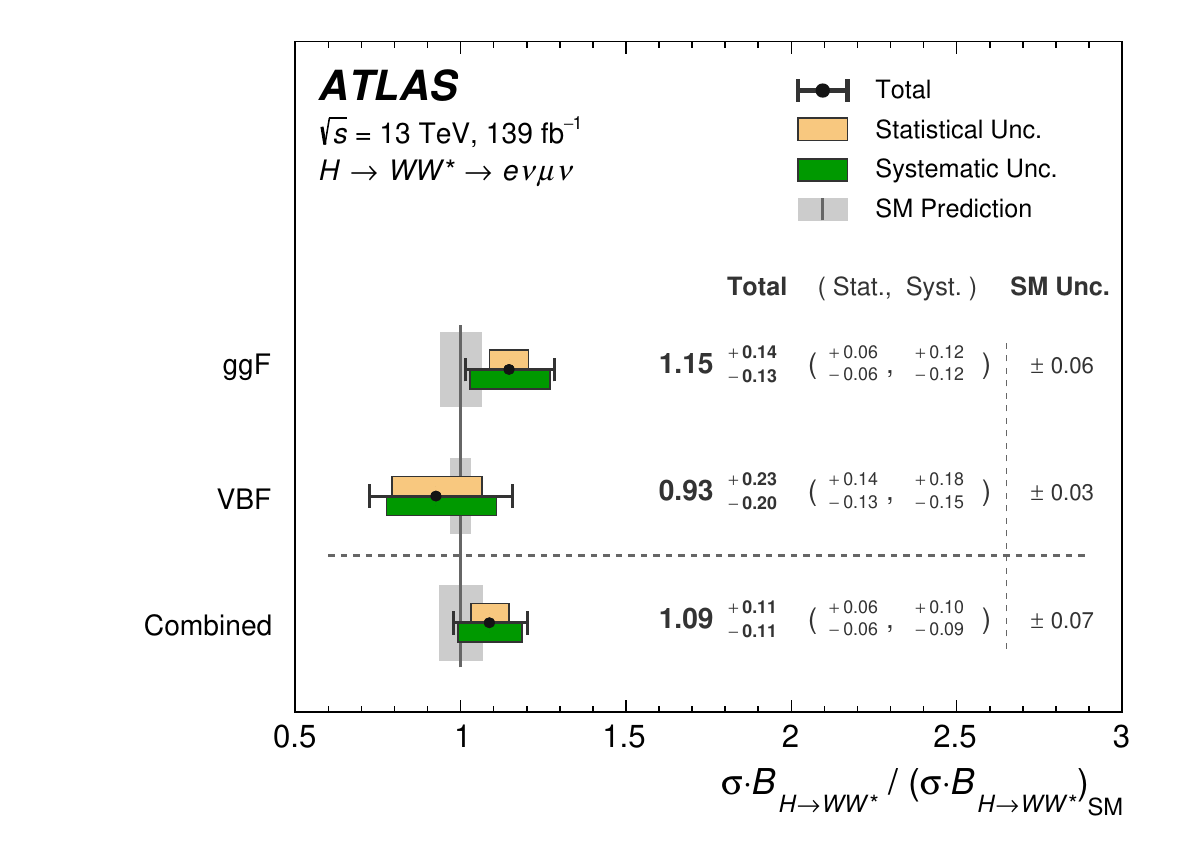}
\caption{
Best-fit values and uncertainties of the $\hww$ cross section for the ggF and VBF processes and their combination, normalized to the corresponding SM prediction.
The black error bars, green boxes, and tan boxes show the total, systematic, and statistical uncertainties in the measurements, respectively.
The gray bands represent the theory uncertainty of the corresponding Higgs production mode.
\label{fig:couplings-POIs}
}
\end{figure}

Table~\ref{tab:UncertaintyBreakdown_2-POI} shows the relative impact of the main uncertainties on the measured values for $\sigma_{\mathrm{ggF+VBF}} \cdot \mathcal{B}_{\hww}$, $\sigma_{\mathrm{ggF}} \cdot \mathcal{B}_{\hww}$, and $\sigma_{\mathrm{VBF}}  \cdot \mathcal{B}_{\hww}$.
The measurements are dominated by systematic uncertainties.
For the ggF measurement, uncertainties from experimental and theoretical sources are comparable.
For the VBF measurement, signal theory uncertainties make up the largest contribution and the dominant ones are those related to the modeling of potential jets in addition to the tagging jets.
 
\begin{table}[htb]
\centering
\caption{
Breakdown of the main contributions to the total uncertainty in $\sigma_{\mathrm{ggF+VBF}} \cdot \mathcal{B}_{\hww}$, $\sigma_{\mathrm{ggF}} \cdot \mathcal{B}_{\hww}$, and $\sigma_{\mathrm{VBF}} \cdot \mathcal{B}_{\hww}$, relative to the measured value.
The individual sources of systematic uncertainties are grouped together.
The sum in quadrature of the individual components differs from the total uncertainty due to correlations between the components.
}
\scalebox{0.8}{
\begin{tabular}{lrrr}
\hline\hline
\noalign{\vskip 1mm}
Source         & $\frac{\Delta\sigma_{\mathrm{ggF+VBF}} \cdot \mathcal{B}_{\hww}}{\sigma_{\mathrm{ggF+VBF}} \cdot \mathcal{B}_{\hww}}$ $[\%]$ & $\frac{\Delta\sigma_{\mathrm{ggF}} \cdot \mathcal{B}_{\hww}}{\sigma_{\mathrm{ggF}} \cdot \mathcal{B}_{\hww}}$ $[\%]$  & $\frac{\Delta\sigma_{\mathrm{VBF}} \cdot \mathcal{B}_{\hww}}{\sigma_{\mathrm{VBF}} \cdot \mathcal{B}_{\hww}}$ $[\%]$  \\
\noalign{\vskip 1mm}
\hline\hline
Data statistical uncertainties & 4.6 & 5.1 & 15\phantom{.0}\tabularnewline
Total systematic uncertainties & 9.5 & 11\phantom{.0} & 18\phantom{.0}\tabularnewline
\hline
MC statistical uncertainties & 3.0 & 3.8 & 4.9\tabularnewline
Experimental uncertainties & 5.2 & 6.3 & 6.7\tabularnewline
\hspace*{4mm} Flavor tagging & 2.3 & 2.7 & 1.0\tabularnewline
\hspace*{4mm} Jet energy scale & 0.9 & 1.1 & 3.7\tabularnewline
\hspace*{4mm} Jet energy resolution & 2.0 & 2.4 & 2.1\tabularnewline
\hspace*{4mm} \met & 0.7 & 2.2 & 4.9\tabularnewline
\hspace*{4mm} Muons & 1.8 & 2.1 & 0.8\tabularnewline
\hspace*{4mm} Electrons & 1.3 & 1.6 & 0.4\tabularnewline
\hspace*{4mm} Mis-Id extrapolation factors & 2.1 & 2.4 & 0.8\tabularnewline
\hspace*{4mm} Pileup & 2.4 & 2.5 & 1.3\tabularnewline
\hspace*{4mm} Luminosity & 2.1 & 2.0 & 2.2\tabularnewline
Theoretical uncertainties & 6.8 & 7.8 & 16\phantom{.0} \tabularnewline
\hspace*{4mm} ggF & 3.8 & 4.3 & 4.6\tabularnewline
\hspace*{4mm} VBF & 3.2 & 0.7 & 12\phantom{.0}\tabularnewline
\hspace*{4mm} $WW$ & 3.5 & 4.2 & 5.5\tabularnewline
\hspace*{4mm} Top & 2.9 & 3.8 & 6.4\tabularnewline
\hspace*{4mm} $Z\tau\tau$ & 1.8 & 2.3 & 1.0\tabularnewline
\hspace*{4mm} Other $VV$ & 2.3 & 2.9  & 1.5\tabularnewline
\hspace*{4mm} Other Higgs & 0.9 & 0.4 & 0.4\tabularnewline
Background normalizations & 3.6 & 4.5 & 4.9\tabularnewline
\hspace*{4mm} $WW$ & 2.2 & 2.8 & 0.6 \tabularnewline
\hspace*{4mm} Top & 1.9 & 2.3 & 3.4\tabularnewline
\hspace*{4mm} $Z\tau\tau$ & 2.7 & 3.1 & 3.4 \tabularnewline
\hline
\noalign{\vskip 1mm}
Total & 10\phantom{.0}       & 12\phantom{.0}            & 23\phantom{.0}    \\
\hline\hline
\end{tabular}
}
\label{tab:UncertaintyBreakdown_2-POI}
\end{table}

The 68\% and 95\% confidence level two-dimensional contours of $\sigma_{\mathrm{ggF}} \cdot \mathcal{B}_{H \to WW^{\ast}}$ and $\sigma_{\mathrm{VBF}} \cdot \mathcal{B}_{H \to WW^{\ast}}$ are shown in Fig.~\ref{fig:LL2D} and are consistent with the SM predictions.
 
\begin{figure}[htb]
\centering
\includegraphics[width=0.75\textwidth]{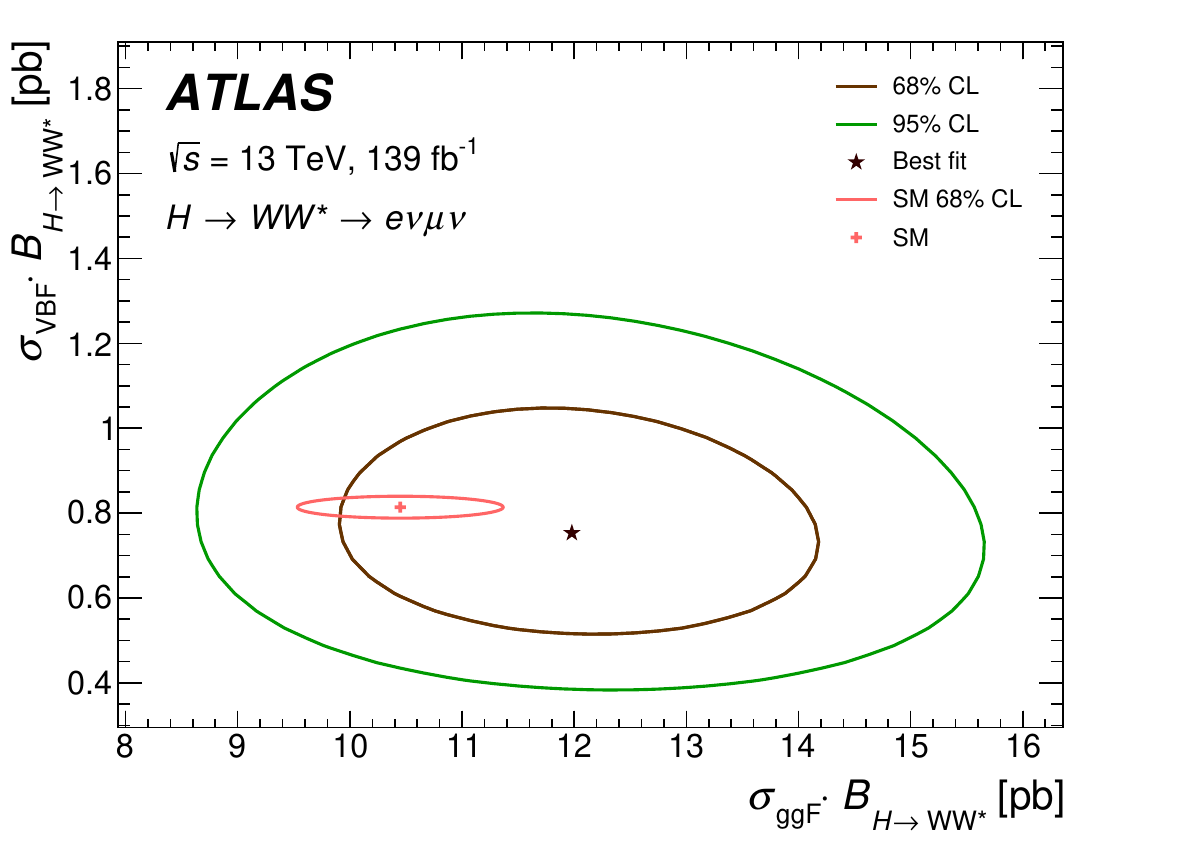}
\caption{
68\% and 95\% confidence level (C.L.) two-dimensional contours of $\sigma_{\mathrm{ggF}} \cdot \mathcal{B}_{\hww}$ vs \mbox{$\sigma_{\mathrm{VBF}}\cdot\mathcal{B}_{\hww}$}, compared to the SM prediction shown by the red marker.
The 68\% C.L. contour for the SM predictions of the ggF and VBF cross sections times branching ratio~\cite{deFlorian:2016spz} is indicated by the red ellipse.
\label{fig:LL2D}
}
\end{figure}

Figure~\ref{fig:11-POI_measurement} shows a summary of the $\hww$ cross sections measured in each of the 11 STXS bins, normalized to the corresponding SM prediction. The correlation matrix for the measured cross sections is shown in Fig.~\ref{fig:STXS-correlation}.
The largest correlations between the measured cross sections are around 30\% and are primarily caused by the migration of signal events between STXS bins and reconstructed signal regions.
The measured cross sections for the five STXS bins targeting EW~$qqH$ production are on average lower than the VBF cross section measured in the two-POI fit.
Events with high $\mjj$ or high $\pTH$ carry a larger statistical weight than events at low $\mjj$ in the two-POI fit, and in these STXS bins, the measured value is close to 1.
Table~\ref{tab:STXS-XSecs} provides the central value and uncertainties of each of the measured STXS cross sections, together with the SM predictions.
The results are compatible with the Standard Model predictions, with a $p$-value of 53\%.
 
\begin{figure}[htb]
\centering
\includegraphics[width=0.9\textwidth]{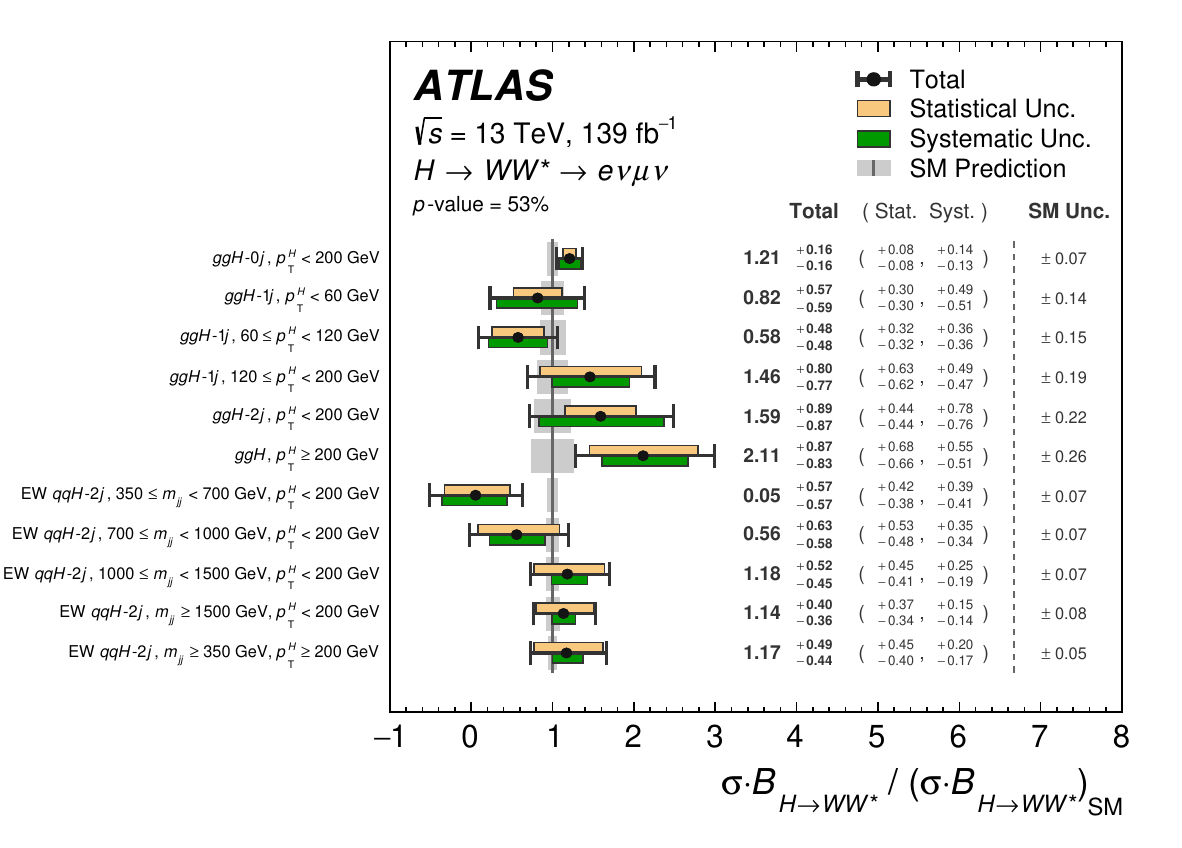}
\caption{
Best-fit value and uncertainties for the $\hww$ cross section measured in each of the STXS bins, normalized to the corresponding SM prediction.
The black error bars, green boxes, and tan boxes show the total, systematic, and statistical uncertainties in the measurements, respectively.
The gray bands represent the theory uncertainty of the signal yield in the corresponding STXS bin.
\label{fig:11-POI_measurement}
}
\end{figure}

\begin{figure}[htb]
\centering
\scalebox{0.9}{
\includegraphics[width=1.2\textwidth]{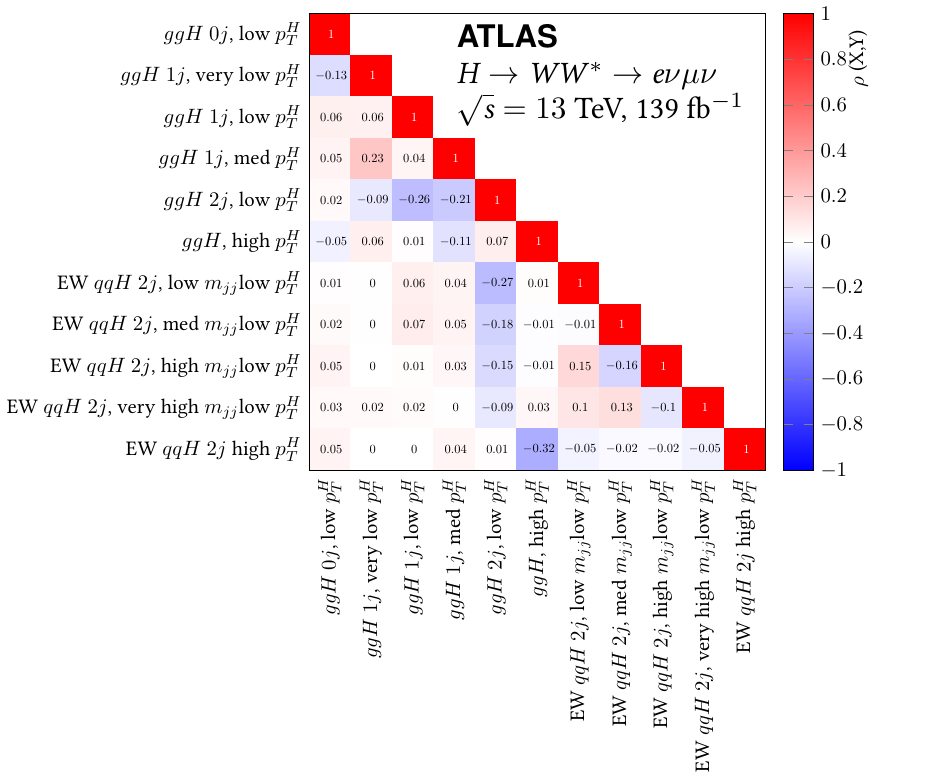} 
}
\caption{
Correlations between the cross-section measurements in the 11 STXS bins for the \hwwenmn analysis.
\label{fig:STXS-correlation}
}
\end{figure}

\begin{table}[htp]
\caption{
Best-fit values and uncertainties for the production cross section times $\hww$ branching ratio $({\sigma_i \cdot \mathcal{B}_{H \to WW^{\ast}}})$ in each STXS bin.
}
\begin{center}
\small
\renewcommand{\arraystretch}{1.5}
\scalebox{0.90}{
\begin{tabular}{l|rccccc|S[table-format=5,table-number-alignment=right]@{$\,\pm\,$}
S[table-format=3,table-number-alignment=left]}
\hline \hline
\multirow{2}{*}{STXS bin $({\sigma_i \cdot \mathcal{B}_{H \to WW^{\ast}}})$} & \multicolumn{1}{c}{Value} & \multicolumn{5}{c|}{ Uncertainty [fb]} & \multicolumn{2}{c}{SM prediction} \\
&  \multicolumn{1}{c}{[fb]} &     Total                 & Stat.                       & Exp.\ Syst.                  & Sig.\ Theo.                  &    Bkg.\ Theo.    & \multicolumn{2}{c}{[fb]}   \\
\hline
\makecell[l]{\noalign{\vskip 2mm} \ggHZeroJ  \\ {\scriptsize \ggHZeroJMath}}                  & $7100$                    & $^{+ 950}_{-910}$           & $^{+480}_{-470}$          & $^{+570}_{-530}$            & $^{+320}_{-260}$            & $^{+490}_{-480}$          & 5870  & 390           \\ [0.4cm]
\makecell[l]{\noalign{\vskip 1mm}\ggHOneJVLowPt  \\ {\scriptsize \ggHOneJVLowPtMath}}          & $1140$                    & $^{+ 800}_{-820}$           & $^{+420}_{-410}$          & $^{+380}_{-380}$            & $^{+80\phantom{0}}_{-70}$   & $^{+570}_{-600}$          & 1400 &  190            \\ [0.4cm]
\makecell[l]{\ggHOneJLowPt  \\ {\scriptsize \ggHOneJLowPtMath}}                 & $540$                     & $^{+ 470}_{-470}$           & $^{+310}_{-310}$          & $^{+230}_{-230}$            & $^{+42\phantom{0}}_{-47}$   & $^{+270}_{-280}$          & 970  & 150             \\ [0.4cm]
\makecell[l]{\ggHOneJMedPt  \\ {\scriptsize \ggHOneJMedPtMath}}                 & $230$                     & $^{+ 130}_{-120}$           & $^{+100}_{-100}$ & $^{+60\phantom{0}}_{-60}$   & $^{+10\phantom{0}}_{-10}$   & $^{+50\phantom{0}}_{-50}$ & 160  & 30             \\ [0.4cm]
\makecell[l]{\ggHTwoJ  \\ {\scriptsize \ggHTwoJMath}}                     & $1610$                    & $^{+ 900}_{-890}$           & $^{+440}_{-440}$          & $^{+430}_{-420}$            & $^{+300}_{-150}$   & $^{+640}_{-650}$          & 1010  & 220             \\ [0.4cm]
\makecell[l]{\ggHHighPt  \\ {\scriptsize \ggHHighPtMath}}                  & $260$                     & $^{+ 100}_{-100}$           & $^{+80\phantom{0}}_{-80}$ & $^{+40\phantom{0}}_{-40}$   & $^{+40\phantom{0}}_{-20}$   & $^{+40\phantom{0}}_{-40}$ & 122 & 31             \\ [0.4cm]
\makecell[l]{\qqHLowMjj \\ {\scriptsize \qqHLowMjjMath}}                  & $6$                     & $^{+ 63\phantom{0}}_{-62}$  & $^{+46\phantom{0}}_{-42}$ & $^{+31\phantom{0}}_{-34}$   & $^{+11\phantom{0}}_{-14}$   & $^{+24\phantom{0}}_{-26}$ & 109 & 7             \\ [0.4cm]
\makecell[l]{\qqHMedMjj \\ {\scriptsize \qqHMedMjjMath}}                   & $31$                      & $^{+ 35\phantom{0}}_{-33}$  & $^{+30\phantom{0}}_{-27}$ & $^{+15\phantom{0}}_{-14}$   & $^{+8\phantom{0}}_{-7}$    & $^{+11\phantom{0}}_{-10}$ & 56 & 4             \\ [0.4cm]
\makecell[l]{\qqHHighMjj \\ {\scriptsize \qqHHighMjjMath}}                 & $60$                      & $^{+ 26\phantom{0}}_{-23}$  & $^{+23\phantom{0}}_{-21}$ & $^{+7\phantom{00}}_{-7}$    & $^{+9\phantom{00}}_{-5}$    & $^{+5\phantom{00}}_{-5}$  & 51 & 4            \\ [0.4cm]
\makecell[l]{\qqHVHighMjj \\ {\scriptsize \qqHVHighMjjMath}}          & $57$                      & $^{+ 20\phantom{0}}_{-18}$  & $^{+18\phantom{0}}_{-17}$ & $^{+5\phantom{00}}_{-5}$    & $^{+3\phantom{00}}_{-3}$    & $^{+4\phantom{00}}_{-4}$  & 50 & 4             \\ [0.4cm]
\makecell[l]{\qqHHighPt \\ {\scriptsize   \qqHHighPtMath}}                & $37$                      & $^{+ 16\phantom{0}}_{-14}$  & $^{+14\phantom{0}}_{-13}$ & $^{+4\phantom{00}}_{-3}$    & $^{+4\phantom{00}}_{-3}$    & $^{+3\phantom{00}}_{-3}$  & 32 & 1            \\ [0.4cm]
\hline
\end{tabular}
}
\end{center}
\label{tab:STXS-XSecs}
\end{table}


\clearpage
\section{Conclusions}
\label{sec:Conclusions}

 
\sloppy
The \hwwenmn decay channel was used to measure Higgs boson production by gluon-gluon fusion and vector-boson fusion.
The measurements are based on a dataset of proton-proton collisions with an integrated luminosity of \RunTwoLumi\ recorded with the ATLAS detector at the LHC in 2015--2018 at a center-of-mass energy of 13~\TeV.
The ggF and VBF cross sections times the \hww branching ratio are simultaneously measured to be
$12.0\pm0.6\;(\mathrm{stat.})^{+0.9}_{-0.8}\;(\mathrm{exp.~syst.})\;^{+0.6}_{-0.5}\;(\mathrm{sig.~theo.})\pm0.8\;(\mathrm{bkg.~theo.})$
and
$0.75\pm0.11\;(\mathrm{stat.})\;^{+0.07}_{-0.06}\;(\mathrm{exp.~syst.})\;^{+0.12}_{-0.08}\;(\mathrm{sig.~theo.})\;^{+0.07}_{-0.06}\;(\mathrm{bkg.~theo.})~\mathrm{pb}$, in agreement with the Standard Model predictions of $10.4\pm 0.6$ and $0.81\pm 0.02$~pb, respectively. These measurements are significantly more precise than the previous Higgs boson cross sections times \hww branching ratio results from ATLAS because of several improvements to the analysis in addition to the larger dataset, most notably the inclusion of a dedicated signal region for the ggF production mode in conjunction with two or more reconstructed jets.
Higgs boson production in the \hww decay channel is further characterized through STXS measurements in a total of 11 categories. The STXS results are compatible with the Standard Model predictions, with a $p$-value of 53\%.



\section*{Acknowledgments}


We thank CERN for the very successful operation of the LHC, as well as the
support staff from our institutions without whom ATLAS could not be
operated efficiently.
 
We acknowledge the support of
ANPCyT, Argentina;
YerPhI, Armenia;
ARC, Australia;
BMWFW and FWF, Austria;
ANAS, Azerbaijan;
CNPq and FAPESP, Brazil;
NSERC, NRC and CFI, Canada;
CERN;
ANID, Chile;
CAS, MOST and NSFC, China;
Minciencias, Colombia;
MEYS CR, Czech Republic;
DNRF and DNSRC, Denmark;
IN2P3-CNRS and CEA-DRF/IRFU, France;
SRNSFG, Georgia;
BMBF, HGF and MPG, Germany;
GSRI, Greece;
RGC and Hong Kong SAR, China;
ISF and Benoziyo Center, Israel;
INFN, Italy;
MEXT and JSPS, Japan;
CNRST, Morocco;
NWO, Netherlands;
RCN, Norway;
MEiN, Poland;
FCT, Portugal;
MNE/IFA, Romania;
MESTD, Serbia;
MSSR, Slovakia;
ARRS and MIZ\v{S}, Slovenia;
DSI/NRF, South Africa;
MICINN, Spain;
SRC and Wallenberg Foundation, Sweden;
SERI, SNSF and Cantons of Bern and Geneva, Switzerland;
MOST, Taiwan;
TENMAK, T\"urkiye;
STFC, United Kingdom;
DOE and NSF, United States of America.
In addition, individual groups and members have received support from
BCKDF, CANARIE, Compute Canada and CRC, Canada;
PRIMUS 21/SCI/017 and UNCE SCI/013, Czech Republic;
COST, ERC, ERDF, Horizon 2020 and Marie Sk{\l}odowska-Curie Actions, European Union;
Investissements d'Avenir Labex, Investissements d'Avenir Idex and ANR, France;
DFG and AvH Foundation, Germany;
Herakleitos, Thales and Aristeia programmes co-financed by EU-ESF and the Greek NSRF, Greece;
BSF-NSF and MINERVA, Israel;
Norwegian Financial Mechanism 2014-2021, Norway;
NCN and NAWA, Poland;
La Caixa Banking Foundation, CERCA Programme Generalitat de Catalunya and PROMETEO and GenT Programmes Generalitat Valenciana, Spain;
G\"{o}ran Gustafssons Stiftelse, Sweden;
The Royal Society and Leverhulme Trust, United Kingdom.
 
The crucial computing support from all WLCG partners is acknowledged gratefully, in particular from CERN, the ATLAS Tier-1 facilities at TRIUMF (Canada), NDGF (Denmark, Norway, Sweden), CC-IN2P3 (France), KIT/GridKA (Germany), INFN-CNAF (Italy), NL-T1 (Netherlands), PIC (Spain), ASGC (Taiwan), RAL (UK) and BNL (USA), the Tier-2 facilities worldwide and large non-WLCG resource providers. Major contributors of computing resources are listed in Ref.~\cite{ATL-SOFT-PUB-2021-003}.


\printbibliography
 
\clearpage
 
\begin{flushleft}
\hypersetup{urlcolor=black}
{\Large The ATLAS Collaboration}

\bigskip

\AtlasOrcid[0000-0002-6665-4934]{G.~Aad}$^\textrm{\scriptsize 101}$,
\AtlasOrcid[0000-0002-5888-2734]{B.~Abbott}$^\textrm{\scriptsize 119}$,
\AtlasOrcid[0000-0002-7248-3203]{D.C.~Abbott}$^\textrm{\scriptsize 102}$,
\AtlasOrcid[0000-0002-2788-3822]{A.~Abed~Abud}$^\textrm{\scriptsize 36}$,
\AtlasOrcid[0000-0002-1002-1652]{K.~Abeling}$^\textrm{\scriptsize 55}$,
\AtlasOrcid[0000-0002-8496-9294]{S.H.~Abidi}$^\textrm{\scriptsize 29}$,
\AtlasOrcid[0000-0002-9987-2292]{A.~Aboulhorma}$^\textrm{\scriptsize 35e}$,
\AtlasOrcid[0000-0001-5329-6640]{H.~Abramowicz}$^\textrm{\scriptsize 150}$,
\AtlasOrcid[0000-0002-1599-2896]{H.~Abreu}$^\textrm{\scriptsize 149}$,
\AtlasOrcid[0000-0003-0403-3697]{Y.~Abulaiti}$^\textrm{\scriptsize 116}$
% ...
\bigskip
\\

% ...
$^{29}$Physics Department, Brookhaven National Laboratory, Upton NY; United States of America.\\
% ...
$^{35}$$^{(a)}$Facult\'e des Sciences Ain Chock, R\'eseau Universitaire de Physique des Hautes Energies - Universit\'e Hassan II, Casablanca;$^{(b)}$Facult\'{e} des Sciences, Universit\'{e} Ibn-Tofail, K\'{e}nitra;$^{(c)}$Facult\'e des Sciences Semlalia, Universit\'e Cadi Ayyad, LPHEA-Marrakech;$^{(d)}$LPMR, Facult\'e des Sciences, Universit\'e Mohamed Premier, Oujda;$^{(e)}$Facult\'e des sciences, Universit\'e Mohammed V, Rabat;$^{(f)}$Institute of Applied Physics, Mohammed VI Polytechnic University, Ben Guerir; Morocco.\\
$^{36}$CERN, Geneva; Switzerland.\\
% ...
$^{55}$II. Physikalisches Institut, Georg-August-Universit\"{a}t G\"ottingen, G\"ottingen; Germany.\\
% ...
$^{101}$CPPM, Aix-Marseille Universit\'e, CNRS/IN2P3, Marseille; France.\\
$^{102}$Department of Physics, University of Massachusetts, Amherst MA; United States of America.\\
% ...
$^{116}$Department of Physics, New York University, New York NY; United States of America.\\
% ...
$^{119}$Homer L. Dodge Department of Physics and Astronomy, University of Oklahoma, Norman OK; United States of America.\\
% ...
$^{149}$Department of Physics, Technion, Israel Institute of Technology, Haifa; Israel.\\
$^{150}$Raymond and Beverly Sackler School of Physics and Astronomy, Tel Aviv University, Tel Aviv; Israel.\\
% ...
$^{a}$ Also Affiliated with an institute covered by a cooperation agreement with CERN.\\
$^{b}$ Also at Borough of Manhattan Community College, City University of New York, New York NY; United States of America.\\
$^{c}$ Also at Bruno Kessler Foundation, Trento; Italy.\\
$^{d}$ Also at Center for High Energy Physics, Peking University; China.\\
$^{e}$ Also at Centro Studi e Ricerche Enrico Fermi; Italy.\\
$^{f}$ Also at CERN, Geneva; Switzerland.\\
% ...

\end{flushleft}


\end{document}